\documentclass[11pt]{article}  
 \pdfoutput=1 
 
 \usepackage{tabu}
 \usepackage{jheppub}
\usepackage[utf8]{inputenc} 
\usepackage{subcaption}
\usepackage{cancel}
\usepackage{float}
\usepackage{tikz}
\usepackage{titlesec}
\usepackage{fancyhdr}
\usepackage{verbatim}
\usepackage[section]{placeins}
\usetikzlibrary{positioning,arrows}
\usetikzlibrary{decorations.pathmorphing}
\usetikzlibrary{decorations.markings}
\usepackage{graphicx,rotating,colortbl}
\usepackage{hyperref,slashed,amsfonts}
\usepackage{color}
\usepackage{amsmath,amssymb,bbold}
\usepackage{mathtools}
\usepackage{multirow}

\def\circa#1{\,\raise.3ex\hbox{$#1$\kern-.75em\lower1ex\hbox{$\sim$}}\,}

\usepackage{mathtools}
\usepackage{multicol}
\usepackage{color}
\usepackage{blkarray}
\usepackage{lmodern}
\definecolor{rosso}{cmyk}{0,1,1,0.4}
\definecolor{rossos}{cmyk}{0,1,1,0.55}
\definecolor{rossoc}{cmyk}{0,1,1,0.2}
\definecolor{blu}{cmyk}{1,1,0,0.3}
\definecolor{blus}{cmyk}{1,1,0,0.6}
\definecolor{bluc}{cmyk}{1,1,0,0.1}
\definecolor{verde}{cmyk}{0.92,0,0.59,0.25}
\definecolor{verdec}{cmyk}{0.92,0,0.59,0.15}
\definecolor{verdes}{cmyk}{0.92,0,0.59,0.4}

\newcommand{\GeV}{\,{\rm GeV}}
\newcommand{\TeV}{\,{\rm TeV}}

\newcommand{\Tr}{\,{\rm Tr}}

\def\circa#1{\,\raise.3ex\hbox{$#1$\kern-.75em\lower1ex\hbox{$\sim$}}\,}

\newcommand{\beq}{\begin{equation}}
\newcommand{\eeq}{\end{equation}}

\newcommand{\bea}{\begin{eqnarray}}
\newcommand{\eea}{\end{eqnarray}}
\newcommand{\be}{\begin{equation}}
\newcommand{\ee}{\end{equation}}
\font\tenrsfs=rsfs10 at 12pt
\font\sevenrsfs=rsfs7
\font\fiversfs=rsfs5
\newfam\rsfsfam
\textfont\rsfsfam=\tenrsfs
\scriptfont\rsfsfam=\sevenrsfs
\scriptscriptfont\rsfsfam=\fiversfs
\def\mathscr#1{{\fam\rsfsfam\relax#1}}

\newcommand{\SO}{\,{\rm SO}}
\newcommand{\SU}{\,{\rm SU}}
\newcommand{\SP}{\,{\rm Sp}}
\newcommand{\U}{\,{\rm U}}

\newcommand{\ct}{c_\theta}
\newcommand{\st}{s_\theta}
\newcommand{\ctt}{c_{2\theta}}
\newcommand{\stt}{s_{2\theta}}
\newcommand{\xLL}{x_{\text{LL}}}
\newcommand{\xLR}{x_{\text{LR}}}
\title{Vacuum misalignment and pattern of scalar masses in the SU(5)/SO(5) composite Higgs model}

\author[a]{Alessandro~Agugliaro,}
\author[b]{Giacomo~Cacciapaglia,}
\author[b]{Aldo~Deandrea,}
\author[a]{Stefania~De~Curtis}

\affiliation[a]{INFN, Sezione di Firenze, and Department of Physics and Astronomy, University of Florence, via G.~Sansone 1, 50019 Sesto Fiorentino, Italy}
\affiliation[b]{Universit\'{e} de Lyon, France; Universit\'{e} Lyon 1, CNRS/IN2P3, UMR5822, IPNL F-69622 Villeurbanne Cedex, France}

\abstract{
Composite Higgs models based on $\SU(5)/\SO(5)$ are characterised by the presence of custodial triplets, like the Georgi-Machacek model. We classify 
all the operators giving rise to the top mass and Higgs potential in presence of fermion partial compositeness, with top partners in two-index representations
of $\SU(5)$. A detailed study of each operator allows us to find correlations in the couplings of Higgs and non-Higgs pseudo-Nambu-Goldstone bosons,
which depend only on one, two or three independent parameters. We also analyse the Higgs potential, finding that a misalignment along the custodial invariant triplet direction is forbidden by CP conservation.  Avoiding custodial breaking allows us to select a handful of feasible models, which feature universal patterns in the scalar masses related to their transformation properties under the custodial symmetry.
Finally, we briefly study the LHC phenomenology of the scalars, which are always below 1 TeV even for multi-TeV condensation scales, and find promising same-sign lepton final states enriched by hard photons.
}

\begin{document}

\begin{flushright}
LYCEN 2018-08
\end{flushright}
\maketitle
\flushbottom

\newpage 

\section{Introduction}

The discovery of a scalar particle with properties compatible with those of the Standard Model (SM)  Higgs boson -- announced by the ATLAS \cite{Harrington:2013cca} and 
CMS \cite{Chatrchyan:2013lba} collaborations in 2012 -- has crowned with success the efforts made by a large community of physicists during more than 
forty years. Although the SM explains with extremely high precision almost all the data collected at particle colliders, there are still many aspects that are  
poorly understood or that cannot find an explanation within the current established framework. Leaving aside issues related to Naturalness and the Hierarchy 
problem, we mention that the SM does not provide any dynamical explanation to the mechanism of electroweak symmetry breaking (EWSB), the latter being a feature that 
plays a major role in the theory of interactions between fundamental particles. Since a large part of the theoretical issues in the SM, including triviality and stability \cite{Degrassi:2012ry, Buttazzo:2013uya}, are related to the scalar sector, all such problems do not affect 
theories in which the scalar states are of composite nature. Of course, complying with the experimental data requires to account for the existence of a scalar 
mode compatible with the discovered $125 \GeV$ Higgs boson.

The Composite Higgs (CH) scenario represents a suitable framework for introducing a non-elementary scalar sector in the SM. We will consider the class of CH models in which 
the Higgs arises as a pseudo-Nambu-Goldstone boson (pNGB) of some spontaneously broken symmetry~\cite{Kaplan:1983fs}, related to a new confining sector. In particular, in CH 
models that admit an underlying fermionic theory, which we will denote as `realistic models', the symmetry is assumed to be broken by the formation of a 
bilinear condensate of the new ``hyperfermions'' (HF) $\psi$, interacting via a ``hypercolour'' (HC) dynamics that is responsible for the confinement. 
The underlying theories are selected in order to have a vacuum of  the HC dynamics that does not break the SM gauge symmetry. Its alignment, leading to the EWSB, is then determined by the low-energy dynamics~\cite{Kaplan:1983fs}. 
Remarkably, in realistic models, the Higgs is always accompanied by at least one additional pNGB~\cite{Cacciapaglia:2014uja}, implying that the potential of the scalar 
sector is richer than the minimal one used in the SM. For instance, it can exhibit mass mixing among the different pNGBs, depending on their quantum numbers 
and on the alignment direction of the vacuum. In fact, the vacuum is usually assumed to be misaligned only along the Higgs direction, even though this is not the most general scenario.
In this case, the relation between the vacuum misalignment angle $\theta$~\cite{Dugan:1984hq} and the condensate scale $f$ is simply given by $f \sin \theta = v$, where $v=246 \GeV$ 
is the electroweak vacuum expectation value (VEV) in the SM. Furthermore, the Higgs does not mix with other pNGBs and its couplings are modified in a universal way (see, for instance, examples in Refs~\cite{Ma:2015gra,Cai:2018tet}). \footnote{Mixing of the Higgs to other light scalars can be induced by adding ad-hoc couplings, like in Refs~\cite{Serra:2015xfa,Banerjee:2017qod}, or by misaligning the vacuum in non-minimal ways, as shown in Ref.~\cite{Cai:2018tet}. A mixing to a light composite spin-0 resonance, if present, is also inevitable~\cite{Arbey:2015exa}.}
However, especially in models where many neutral pNGBs are present, a more general vacuum structure cannot be 
excluded \emph{a priori} neither by electroweak precision data, specifically the bounds on the $T$ parameter, nor by the measurement of Higgs couplings. 
In fact, the vacuum misalignment is not arbitrary, but fully determined by the couplings of the SM fields to the strong dynamics, and non-minimal alignments are in some cases unavoidable (see Ref.~\cite{Cai:2018tet}).

All the features mentioned above are present in the $\SU(5)/\SO(5)$ CH model \cite{Dugan:1984hq,Katz:2005au,Ferretti:2014qta} that we consider in this work. The pNGBs can be conveniently classified on the (non-physical) vacuum that preserves the electroweak (EW) symmetry: in addition to a composite Higgs doublet, 
the other fields include a gauge singlet pseudo-scalar and three $\SU(2)_L$ triplets, of hypercharge $\pm 1$ and $0$. 
The low energy field content, therefore, resembles that of the Georgi-Machacek model~\cite{Georgi:1985nv,Chanowitz:1985ug}, with the addition of a singlet~\cite{Campbell:2016zbp}. \footnote{In the composite case, however, the singlet cannot be Dark Matter, as it decays via topological anomalies~\cite{Dugan:1984hq}.}
The physical eigenstates, however, due to the mixing, do not correspond to the gauge ones. In fact, a simpler way to describe 
the mass spectrum is to use the custodial $\SU(2)_C$  eigenstates \cite{Dugan:1984hq}, where the three triplets combine into a five-plet, a triplet and a singlet.
This approach is quite useful because the diagonal custodial symmetry is preserved by the vacuum misaligned along the Higgs direction and, therefore, in such vacuum every source of mixing between different $\SU(2)_C$ eigenstates can be traced back to an explicit breaking 
of this symmetry, as for instance the gauging of hypercharge.

A link between the composite and SM elementary fermion sector is provided via the Fermion Partial Compositeness (FPC) paradigm~\cite{Kaplan:1991dc}, which assumes the existence of linear mixing between the SM fermions and some composite spin-1/2 states that share the same quantum numbers.
In the realistic models we consider in this paper, FPC requires the presence 
of additional HFs $\chi$, which are charged under both QCD colour and hypercolour (while the HFs $\psi$ remain not charged under the QCD colour)~\cite{Barnard:2013zea,Ferretti:2013kya}. 
In this way, the heavy fermionic resonances needed by the FPC paradigm emerge as ``chimera'' bound 
states~\cite{Golterman:2015zwa} of three fermions, which can schematically be of the kind $\psi \chi \psi$ or $\chi \psi \chi$. It is worth noting that such resonances are  
vector-like, namely the HC dynamics generates a mass term for them without breaking the SM gauge symmetry, as opposed to what happens for SM elementary fields. 
As the heavy resonances, which naturally talk to the composite scalar sector, mix with the SM elementary fields,  the light physical fermions corresponding to the SM quarks and leptons are a superpositions between them. In particular, the heavier the fermion the higher its degree of compositeness.

 A classification of all the possible minimal realisations of FPC that give a fermionic partner with the same quantum numbers of the top (a top partner) 
 can be found in Ref.~\cite{Ferretti:2016upr}. 
 The strict requirement that the underlying theory at low energies (i.e., around the condensation scale) is not conformal leaves only 7 models with the $\SU(5)/\SO(5)$ symmetry breaking pattern~\cite{Ferretti:2016upr}. It is remarkable that some of these models are being studied on the lattice for HC groups $\SP(4)$~\cite{Bennett:2017ttu,Bennett:2017kga} and $\SU(4)$~\cite{Ayyar:2017qdf,Ayyar:2018zuk}.
Throughout this paper we will only consider chimera baryons of the form $\psi \chi \psi$, leading to composite fermionic partners 
transforming in two-index irreducible representations of $\SU(5)$. This case has also been studied in Ref.~\cite{Golterman:2017vdj} in connection with lattice studies, while some results for the case with top partners in the fundamental of $\SU(5)$ can be found in Refs~\cite{Ferretti:2014qta,Golterman:2015zwa}.

We will follow the approach of Refs~\cite{Golterman:2015zwa,Golterman:2017vdj,Alanne:2018wtp} where the 
 heavy resonances are integrated out, and we write down effective operators in terms of spurionic couplings and of the pNGB and SM fields only. Thus, we derive the most general basis of operators for the effective fermion (top) mass and the scalar pNGB potential. The latter also receives contributions from loops of gauge bosons and from the bare mass terms for the HFs. We 
consider all the possible embeddings of the third generation left- and right-handed quarks, respectively $q_L = ( t_L, b_L) ^\intercal$ and $t_R$, into 
different two-index irreducible representations of $\SU(5)$. These are given by the adjoint $ D = {\bf 24}$, the (anti-) symmetric $S \ (A) = {\bf 15} \ ({\bf 10})$, and the 
singlet $N = {\bf 1}$. 
We study in detail the pNGB masses and couplings expanding the operators around a vacuum misaligned only along the Higgs direction, thus extending the analysis of Ref.~\cite{Golterman:2017vdj}. A common issue of CH models based on $\SU(5)/\SO(5)$ is that the vacuum also tends to be misaligned along the custodial triplet thanks to the top Yukawas~\cite{Vecchi:2013bja}, and we therefore look for combinations of embeddings for the left- and right-handed tops that do not induce such a misalignment. We find that this is indeed possible in a few cases. Furthermore, a more general misalignment of the vacuum is also considered. Our goal here is to give the most general characterisation of viable pNGB spectra and couplings.

The paper is organised as follows: after summarising the main features of the model in Section~\ref{sec:model}, we give the most general form of the top 
mass and Yukawa couplings in Section~\ref{sec:gen}. Afterwards, Section~\ref{sec:cases} describes four phenomenologically viable cases at LO, while 
in Section~\ref{sec:antisym} we give the results for the anti-symmetric representation at NLO. 
In Section~\ref{sec:vacuum2}, we discuss the vacuum misaligned along the custodial triplet. Finally, in Section~\ref{sec:pheno} 
 a brief analysis of some interesting signatures that could be observed at the LHC and beyond is given, before presenting our conclusions in Section~\ref{sec:concl}.
 In the Appendix~\ref{app:models} we give some more details about the structure of the model.

\section{Description of the model} \label{sec:model}

The basic group-theory structure of composite models based on $\SU(5)/\SO(5)$ has been widely discussed in the literature~\cite{Dugan:1984hq,Katz:2005au,Ferretti:2014qta}, thus here we will limit ourselves to summarise the main properties.
For the embedding of the EW symmetry, we choose the generators of the custodial $\SU(2)_L \times \SU(2)_R$ to be identified with the following $\SU(5)$ generators:
\be\label{su5:gen}
T_L ^i = \frac{1}{2}
\left(
\begin{array}{c | c} 
\mbox{$ \mathbb{1} _2 \otimes \sigma^i $}  &   \\
\hline
  & 0 \end{array} \right)\,,
\quad 
T_R ^i = \frac{1}{2}
\left(
\begin{array}{c | c}
\mbox{$ \sigma^i \otimes \mathbb{1}_2 $ } & \\
\hline
& 0 \end{array} \right)\,.
\ee
With this choice, the bi-doublet of the custodial symmetry can be associated to the first four components of the fundamental of $\SU(5)$, while the fifth is a singlet. In a fermionic underlying theory, these would be the quantum numbers of the fermions $\psi$ generating the global symmetry, which can therefore be written as
\be
\psi = \left( \begin{array}{c}
\psi_d^{+} \\ \psi_d^- \\ \psi_s \end{array} \right)\,,
\ee
where $\psi_d^\pm$ are $\SU(2)_L$ doublets with hypercharge $\pm 1/2$ respectively and $\psi_s$ is a neutral singlet, while $(\psi_d^+, \psi_d^-)^T$ forms a doublet of $\SU(2)_R$.
The Lagrangian for the underlying theory would thus read:
\be
\mathcal{L}_{\rm UV} = i \bar{\psi} \sigma^\mu D_\mu \psi - \psi \mathcal{M}_\psi \psi + \mbox{h.c} 
\ee
where the covariant derivative contains both the HC generators $\lambda_{\rm HC}$ and the EW gauge interactions:
\be
D_\mu = \partial_\mu - i g W^i_\mu T_L^i - i g' B_\mu T_R^3 - i g_{\rm HC} G_\mu^a \lambda^a_{\rm HC}\,.
\ee
The matrix $\mathcal{M}_\psi$  gives  the most general gauge invariant mass term~\cite{Katz:2005au}:
\be
\psi \left( \begin{array}{ccc}
0 & \mu_d (i \sigma^2) & 0 \\
-\mu_d (i \sigma^2) & 0 & 0 \\
0 & 0 & \mu_s
\end{array} \right) \psi = \mu_d \psi_d^+ \psi_d^- + \mu_s \psi_s \psi_s\,,  \label{eq:Mpsi}
\ee
which thus contains a Dirac mass for the charged doublets and a Majorana mass for the neutral singlet. In general, this mass term breaks $\SU(5) \to \SO(4)\sim \SU(2)_L \times \SU(2)_R$, while for $\mu_d = \mu_s$ a larger $\SO(5)$ group will be preserved. Note, therefore, that the HF mass term cannot explicitly break the custodial symmetry in this model.

We also assume the presence of an additional U$(1)_X$ under which the top partners are charged, so that the hypercharge generator is defined as 
$Y = T_R^3 + X$. 
This charge is carried by the additional underlying states that form the so-called ``chimera'' baryons together with the $\psi$'s.  They can be either fermions~\cite{Barnard:2013zea,Ferretti:2013kya}, called $\chi$, or scalars~\cite{Sannino:2016sfx,Cacciapaglia:2017cdi}. The fermion multiplet 
$\psi$ carries zero $\U(1)_X$ charge, which implies that the condensate $\langle \psi \psi \rangle$ does not break $\U(1)_X$. As the $\chi$'s belong to a different representation of the $\psi$'s under HC, by gauge symmetry, the $\psi$ and $\chi$ fermions cannot have a mass mixing. The $\U(1)_X$ symmetry is therefore unaffected by the spontaneous breaking in the EW sector giving rise to the pNGBs under study.


Our reference vacuum matrix, $\Sigma_0$, that leaves the EW symmetry unbroken and transforms as a real two-index symmetric representation of $\SU(5)$,  
is chosen as
 \be \label{vacuum}
\Sigma_0=\left(
\begin{array}{cc|c}
 & i \sigma_2 &  \\
 -i \sigma_2 &  & \\ \hline
  & & 1 
\end{array} \right)\,.
\ee
Note that there is a second inequivalent choice for a real vacuum, defined by the same matrix with $1 \to -1$, as shown in Eq.~\eqref{vacuum2}. The two real vacua are related to each other via 
a $\SU(5)$ rotation and an overall phase redefinition, and the choice is mainly correlated with the sign of the masses in Eq.\eqref{eq:Mpsi}: 
for positive equal mass parameters, $\mu_s = \mu_d$, the mass term is proportional to the vacuum $\Sigma_0$ in Eq.\eqref{vacuum}, while for $\mu_s = - \mu_d$ 
the mass is proportional to the other vacuum.

The low-energy dynamics is described in terms of a linearly transforming matrix $\Sigma$, defined as
\be\label{goldstone}
\Sigma(x) = \Omega(\theta) \, \exp \left[4 i \Pi(x) /f \right] \Sigma_0 \, \Omega^\intercal (\theta)\,, 
\ee
where $\Pi (x) = \Pi^{\hat a} (x) X^{\hat a}$ is the pion matrix ($X^{\hat a}$ being the $\SU(5)$ generators broken by the vacuum $\Sigma_0$ and normalised as $\Tr [X^{\hat a} X^{\hat b}] = \frac{1}{2} \delta^{\hat{a} \hat{b}}$) and
$f$ is the physical scale generated by the strong dynamics (i.e., the pNGB decay constant).
The rotation $\Omega(\theta)$ describes the misalignment of the vacuum that breaks the EW symmetry, and is given explicitly in Eq.\eqref{eq:omega} of Appendix~\ref{app:vacuum}. Thus, we follow a different approach from most of the CH literature, including Ref.~\cite{Golterman:2017vdj}, as we define the pNGBs around the true vacuum of the theory, $\Sigma_\theta = \Omega(\theta) \Sigma_0 \Omega^\intercal (\theta)$, and we do not allow for any pNGB to develop a VEV. 
The explicit form of $\Omega(\theta)$ depends on the potential generated for the pNGBs: the minimal case is when the only misalignment happens along the direction of the Higgs boson and it can be described by a single parameter $\theta$ that encodes the breaking of the EW symmetry. The lowest order chiral Lagrangian reads:
\be
\mathcal{L}^{(2)} = \frac{f^2}{16} \mbox{Tr} \left[ (D_\mu \Sigma)^\dagger D^\mu \Sigma\right]\,,
\ee
where the normalisation is chosen such that the relation between the misalignment angle, $f$ and the EW vacuum energy $v$ is $v = f \sin \theta$ 
\cite{Belyaev:2016ftv}.

Besides the doublet $H$ that plays the role of the Higgs doublet, the pion matrix contains ten additional pNGBs: nine of them  constitute a bi-triplet of the custodial symmetry, 
$ {(3,3)} \equiv (\pi_+, \,\pi_0, \, \pi_-)$, while the remaining one is a singlet ${(1,1)}\equiv  \eta$ of $\SU(2)_L \times \SU(2)_R$. 
The pion matrix defined in Eq.\eqref{goldstone} is thus given by
 \begin{align} \label{eq:Pimatrix}
\Pi \ = \ \frac{1}{2} \left(
\begin{matrix}
\frac{\eta}{\sqrt{10}}\mathbb{1}_2 + \pi_0 & \pi_+ &  H \\
\pi_- & \frac{\eta}{\sqrt{10}} \mathbb{1}_2 - \pi_0  & - \tilde{H}\\
 H ^\dagger&- \tilde{H}^\dagger & -\frac{4}{\sqrt{10}} \eta  
\end{matrix} \right)
\end{align}
where $\tilde H = i \sigma^2 H^\ast$, and the triplet matrices are defined as $\pi_0  = \frac{1}{\sqrt{2}} \,\pi_0 ^i \, \sigma^i$ and $\pi_-  = \pi_- ^i \, \sigma ^i = (\pi_+)^\dagger$.
Note however that the identification of the triplets with eigenstates of the custodial symmetry can be strictly done on the EW preserving vacuum $\Sigma_0$: the misalignment parametrised by $\theta$ will also misalign the triplets with respect to the partially-gauged custodial symmetry. It is, therefore, more convenient to define a basis of eigenstates of the diagonal $\SU(2)_C$ symmetry which is left unbroken by the Higgs VEV.
The bi-triplet decomposes as
\be
{(3,3)} \rightarrow  {\bf 5} \oplus {\bf 3} \oplus {\bf 1}\,.
\ee
In the following, therefore, we will replace the fields $\pi^i_{0,\pm}$ with the following:
\be
{\bf 5} = ( \eta_5^{++},\ \eta_5^+,\ \eta_5^0,\ \eta_5^-,\ \eta_5^{--})\,, \quad {\bf 3} = (\eta_3^+,\ \eta_3^0,\ \eta_3^-)\,, \quad {\bf 1} = \eta_1^0\,.
\ee
More details about this change of basis can be found in Appendix~\ref{app:custodial}.

To complete the characterisation of the model, we now turn our attention to the explicit breaking of $\SU(5)$. Besides the gauging of the electroweak symmetry, encoded in the covariant derivative, and the HF mass term of Eq.~\eqref{eq:Mpsi}, we will consider here the couplings generating the top mass.
For the top quark, we work under the assumption of FPC~\cite{Kaplan:1991dc}: we consider top partners transforming under two-index representations of $\SU(5)$, as obtained in some underlying completions~\cite{Ferretti:2013kya}. More specifically, this situation occurs in the models M3 and M4 as defined in Table 1 of Ref.~\cite{Belyaev:2016ftv} (see also Ref.~\cite{Golterman:2017vdj} for an analysis of this class of models).

\subsection{Fermion partial compositeness}\label{spur}

\begin{table}[tb]
\centering
\begin{tabular}{l|c|c|}
 & $\SO(5)$ & $\SU(2)_L\times\SU(2)_R$ \\ \hline
$A = {\bf 10}$ & $\bf 10$ & $(2,2) \oplus (3,1) \oplus (1,3)$ \\ \hline
\multirow{2}{*}{$S = {\bf 15}$} & $\bf 14$ & $(1,1) \oplus (2,2) \oplus (3,3)$\\
                                                & $\bf 1$ & $(1,1)$ \\ \hline
\multirow{2}{*}{$D = {\bf 24}$} & $\bf 10$ & $(2,2) \oplus (3,1) \oplus (1,3)$ \\
                                                & $\bf 14$ & $(1,1) \oplus (2,2) \oplus (3,3)$ \\ \hline
$N = {\bf 1}$ & $\bf 1$ & $(1,1)$ \\ \hline
\end{tabular}
\caption{List of $\SU(5)$ spurions and their decomposition under $\SO(5)$ and the custodial $\SU(2)_L \times SU(2)_R$ symmetry. The left-handed doublet $q_L$ can be associated to any of the bi-doublets $(2,2)$, while the right-handed singlet $t_R^c$ can be associated to singlets $(1,1)$ or to the $SU(2)_R$ triplets $(1,3)$.} \label{tab:topspurions}
\end{table}

The form of the effective potential depends on the way the SM fermions are embedded into spurions transforming as complete representations of the $\SU(5)$ group. In the following, we consider two-index representations of $\SU(5)$, \emph{i.e.} the singlet $N =  {\bf 1}$, the (anti)-symmetric $(A={\bf 10}) \, S =  {\bf 15}$, and the adjoint $D=  {\bf 24}$.  These spurions are based on ``chimera'' baryons in the form $\langle \psi \psi \chi \rangle$, thus the $X$ charge assigned to the $\chi$ fermions is $X = 2/3$ in order to give the right hypercharge to the right-handed-top partners.~\footnote{This is straightforward in the underlying theories we are dealing with. At the effective level, this choice corresponds to assigning a $U(1)_X$ charge to the spurions the quarks belong to.}
The decompositions of the above representations of $\SU(5)$ are summarised in Table~\ref{tab:topspurions}.
Note that it is most convenient to write the SM fermion fields in Weyl notation, so that we will associate the spurions to the left-handed spinors $q_L = \left( \begin{matrix} t_L \\ b_L \end{matrix} \right)$ and $t_R^c$. 


The left-handed quark doublet $q_L$ has a unique embedding in the symmetric and anti-symmetric representations, given by the matrices below
\be
S_L =\frac{1}{\sqrt{2}}\ \left(
\begin{array}{cccc|c}
  &  &  &  &  \\
  &  &  &  &  \\
  &  &  &  & t_L \\
  &  &  &  & b_L \\ \hline
\phantom{t_L} &  \phantom{t_L} & t_L & b_L &  \\
\end{array}
\right), \quad 
A_L = \frac{1}{\sqrt{2}}\ \left(
\begin{array}{cccc|c}
 &  &  &  &  \\
  &  &  &  &  \\
  &  &  &  & -t_L \\
  &  &  &  & -b_L \\ \hline
\phantom{t_L} &  \phantom{t_L} & t_L & b_L &  \\
\end{array}
\right)\,.
\ee
The symmetric and anti-symmetric irreducible representations (\emph{irreps}), being complex, come with their conjugates, respectively $S^c _L = \Sigma_0^\dagger S_L \Sigma_0^\dagger$ and $A^c_L = \Sigma_0^\dagger A_L \Sigma_0^\dagger$. 
For the adjoint, there are two possible embeddings:
\be
D_L ^1 = \frac{1}{\sqrt{2}}\ \left(
\begin{array}{cccc|c}
  &  &  &  & \\
 &  &  &  & \\
  &  &  &  & t_L \\
  &  &  &  & b_L \\ \hline
 -b_L & t_L & \phantom{t_L}  & \phantom{t_L}  &  \\
\end{array}
\right), \quad
D_L ^2 = \frac{1}{\sqrt{2}}\ \left(
\begin{array}{cccc|c}
  &  &  &  &  \\
  &  &  &  & \\
  &  &  &  & t_L \\
  & &  &  & b_L \\ \hline
 b_L & -t_L &  \phantom{t_L}  & \phantom{t_L}   &  \\
\end{array}
\right)\,. 
\ee
The first embedding belongs to the anti-symmetric (adjoint) $\bf 10$ of $\SO(5)$, as it can be seen by the relation $D_L^1 \Sigma_0 = A_L$, while the second belongs to the symmetric $\bf 14$ of $\SO(5)$, as $D_L^2  \Sigma_0 = S_L$.
In general, therefore, the left-handed quarks will correspond to a linear combination of the two 
\be
D_L = c_{\alpha} D_L^1 + e^{i \varphi} s_{\alpha} D_L^2\,,
\ee
where $c_x = \cos x$ and $s_x = \sin x$ and we allow for a relative phase between the two coefficients. Note that the operators should be written in terms of this linear combination in order to 
better separate the parameters of the effective Lagrangian that originate from the underlying theory (like $\alpha$ and $\varphi$, in this case), and the 
coefficients generated by the strong dynamics. \footnote{Our approach differs from Ref.~\cite{Golterman:2017vdj}, where the two spurions $D_L^{1,2}$ 
are used to define separate operators. Of course, the physics consequences are not affected by this choice.} 

Similarly, the singlet $t_R^c$ can have the following embedding in the symmetric and anti-symmetric representations
\be
A_R =\frac{t_R^c}{2} \ \left(
\begin{array}{cccc|c}
  &  &  & 1 &   \\
  &  & -1 &  &  \\
  & 1 &  &  &  \\
 -1 &  &  &  &  \\ \hline
 &  &  &  & \phantom{0} \\
\end{array}
\right), \quad
S_R ^1 =\frac{t_R^c}{\sqrt{5}} \ \left(
\begin{array}{cccc|c}
  &  &  & 1 &  \\
  &  & -1 &  &  \\
  & -1 &  &  &  \\
 1 & &  &  &  \\ \hline
  &  &  &  & 1 \\
\end{array}
\right)\,, \quad
S_R ^2 = \frac{t_R^c}{2\sqrt{5}}\ \left(
\begin{array}{cccc|c}
  &  &  & 1 &  \\
  &  & -1 &  &  \\
  & -1 &  &  &  \\
1 & &  &  &  \\ \hline
  &  &  &  & -4 \\
\end{array}
\right)\,,
\ee
together with their conjugates, respectively $S^{i,c}_R= \Sigma_0^\dagger S^i_R \Sigma_0^\dagger \equiv S_R^i$ and $A^c_R = \Sigma_0^\dagger A_R \Sigma_0^\dagger \equiv - A_R$.
Note also that $S_R^1$, being aligned to the vacuum $\Sigma_0$, corresponds to the singlet of $\SO(5)$, while $S_R^2$ to the symmetric $\bf 14$.
For the two embeddings in the symmetric (and the conjugate), the linear combinations
\be
S_R = c_{\beta_{S}} S_R^1 + e^{i \gamma_{S}} s_{\beta_{S}} S_R^2\,, \quad S^c_R = c_{\beta^c_{S}} S_R^{1,c} + e^{i \gamma^c_{S}} s_{\beta^c_{S}} S_R^{2,c}\,,
\ee
need to be used to define the operators. 
The embeddings in the adjoint read 
\be
D_R^1 =\frac{t_R^c}{2}\ \left(
\begin{array}{cccc|c}
1  &  &  & &  \\
  & 1 &  &  &  \\
  &  & -1 &  &  \\
  &  &  & -1 &  \\ \hline
 &  &  &  & \phantom{0} \\
\end{array}
\right), \quad
D_R ^2 =\frac{t_R}{2\sqrt{5}}\ \left(
\begin{array}{cccc|c}
1  &  &  &  &  \\
  & 1 &  &  &  \\
  &  & 1 &  &  \\
  & &  & 1 &  \\ \hline
  &  &  &  & -4 \\
\end{array}
\right)\,,
\ee
and we define a linear combination
\be
D_R = c_{\beta} D_R^1 + e^{i \gamma} s_{\beta} D_R^2\,.
\ee
The first embedding can be associated to the anti-symmetric $\bf 10$ of $\SO(5)$ as $D_R^1 \Sigma_0 = A_R$, while the second to the symmetric $\bf 14$ as $D_R^1 \Sigma_0 = S_R^2$.
Finally, the singlet can be written in matrix form as
\be
N_R =\frac{t_R^c}{\sqrt{5}}\ \left(
\begin{array}{cccc|c}
1  &  &  & &  \\
  & 1 &  &  &  \\
  &  & 1 &  &  \\
  &  &  & 1 &  \\ \hline
 &  &  &  & 1 \\
\end{array}
\right).
\ee

The spurions we defined above contain the SM spinors, thus they are suitable to define the effective top mass operators. 
To define operators where the SM fermions are integrated in loops, like in the case of the pNGB potential, it is useful to define a corresponding set of 
field-independent spurions as follows:
\be
X_L = t_L \tilde{X}_L^1 + b_L \tilde{X}_L^2\,, \quad X_R = t_R^c \tilde{X}_R\,,
\ee
where $X = A, S, D, N$. Technically, the spurions $\tilde{X}_{L,R}$ carry indices of the SM gauge symmetries (EW and QCD colour) that are contracted with 
the ones of the SM fields, as well as indices of the global symmetries of the strong dynamics. For instance, $\tilde{X}_L^\alpha$ forms an anti-doublet of $\SU(2)_L$. 
In the following we will omit such indices and the tilde to simplify the notation, and we will leave understood that the spurions appearing in the pNGB potential 
are the tilded ones defined above.

\section{General form of the top mass term and pNGB potential} \label{sec:gen}

In this section, we will provide a classification of all the operators contributing to the mass of the top and to the pNGB potential. Our results match the classification in Refs~\cite{Golterman:2017vdj} and \cite{Alanne:2018wtp}. Furthermore, we provide the expansion of each operator around the vacuum misaligned along the Higgs direction.  This will allow us to study the couplings of the pNGBs, the stability of the vacuum alignment, the presence of tadpoles for other pNGBs (as a sign that a more general vacuum misalignment is needed) and the scalar spectra.

\subsection{Top Yukawa (mass) operators} \label{sec:topmass}

The mass for the top quark is generated by operators containing one $L$ and one $R$ field-dependent spurions. In general, the list of operators can be written as
\be \label{eq:topLO}
\mathcal{L}_{\rm top, LO} = - f\ \sum_{i_R,j_L} \frac{C_{t,i_R j_L}}{4 \pi} \mathcal{O}_{i_R j_L} - f\ \sum_{i_R,j_L} \frac{C'_{t,i_R j_L}}{4 \pi} \mathcal{O}'_{i_R j_L} + \mbox{h.c.}\,,
\ee
where the indices $i = A, A^c, S, S^c, D, N$ and $j = A, A^c, S, S^c, D$ cover all the allowed spurion representations, and the factor of $1/4\pi$ comes from na\"ive dimensional analysis (NDA)~\cite{Georgi:1992dw,Buchalla:2013eza}. As we will explicitly see below, the operators in the first sum are defined in terms of a single trace over the $\SU(5)$ indices, while the primed ones contain two traces. At leading order, operators with two spurions do not exist for all the combinations of representations. The list contains 5 operators with the same {\it irrep} for both, given by:
\bea
&  \mathcal{O}_{D_R D_L} = i\ \Tr [D_L^T \Sigma^\dagger D_R \Sigma]\,, \quad \mathcal{O}_{S_R S_L} = -i\ \Tr [S_L \Sigma^\dagger S_R \Sigma^\dagger]\,, \quad \mathcal{O}_{S^c_R S^c_L} = -i\ \Tr [S^c_L \Sigma S^c_R \Sigma]\,, & \nonumber \\
& \mathcal{O}_{A_R A_L} = i\ \Tr [A_L \Sigma^\dagger A_R \Sigma^\dagger]\,, \quad \mathcal{O}_{A^c_R A^c_L} = i\ \Tr [A^c_L \Sigma A^c_R \Sigma]\,;  & \label{eq:op2XX}
\eea
followed by 8 containing one adjoint and one (anti)-symmetric:
\bea
& \mathcal{O}_{D_R S_L} = i\ \Tr [S_L \Sigma^\dagger D_R]\,, \quad \mathcal{O}_{D_R S^c_L} = i\ \Tr [S^c_L \Sigma D_R^T ]\,, & \nonumber \\
& \mathcal{O}_{S_R D_L} = -i\ \Tr [D_L S_R \Sigma^\dagger]\,, \quad \mathcal{O}_{S^c_R D_L} = -i\ \Tr [D_L^T S^c_R \Sigma]\,, & \nonumber \\
& \mathcal{O}_{D_R A_L} = i\ \Tr [A_L \Sigma^\dagger D_R]\,, \quad \mathcal{O}_{D_R A^c_L} = i\ \Tr [A^c_L \Sigma D_R^T ]\,, & \nonumber \\
& \mathcal{O}_{A_R D_L} = i\ \Tr [D_L A_R \Sigma^\dagger]\,, \quad \mathcal{O}_{A^c_R D_L} = i\ \Tr [D_L^T A^c_R \Sigma]\,; & \label{eq:op1Sigma}
\eea
and 2 containing the singlet {\it irrep}:
\be\label{eq:op1SigmaN}
\mathcal{O}_{N_R S_L} = -i\ \Tr [S_L \Sigma^\dagger N_R]\,, \quad \mathcal{O}_{N_R S^c_L} = -i\ \Tr [S^c_L \Sigma N_R^T]\,.
\ee
Finally, the primed operators contain a double-trace, and count 4 operators only involving the symmetric {\it irrep}:
\bea
& \mathcal{O}'_{S_R S_L} = -i\ \Tr [S_L \Sigma^\dagger] \Tr [S_R \Sigma^\dagger]\,, \quad \mathcal{O}'_{S^c_R S^c_L} = -i\ \Tr [S^c_L \Sigma] \Tr [S^c_R \Sigma]\,, & \nonumber \\ 
& \mathcal{O}'_{S_R S^c_L} = -i\ \Tr [S^c_L \Sigma] \Tr [S_R \Sigma^\dagger]\,, \quad \mathcal{O}'_{S^c_R S_L} = -i\ \Tr [S_L \Sigma^\dagger] \Tr [S^c_R \Sigma]\,. &  \label{eq:op2Trace}
\eea
Note that we have built a basis of operators that is explicitly invariant under the unbroken group $\SU(5)$. In principle, a more appropriate procedure would consist in constructing operators that are only invariant under the unbroken $\SO(5)$, following for instance the procedure highlighted in Refs~\cite{Contino:2010rs,Mrazek:2011iu}. This fact may rise the question if some operators are missing, especially for spurions that contain more than one $\SO(5)$ irrep, like the adjoint and symmetric (see Table~\ref{tab:topspurions}). However, we explicitly checked that our basis of operators is complete and equivalent to a basis constructed in terms of $\SO(5)$ invariants (see Appendix~\ref{app:equivalence} for a detailed proof). In particular, the two operators constructed in terms of the symmetric and anti-symmetric $\SO(5)$ components of the adjoint are equivalent to $\mathcal{O}_{D_R D_L}$ up to a trivial pNGB-independent operator. For the symmetric, the double-trace operator contains the singlet component, while the single trace one is a linear combination of the symmetric and singlet $\SO(5)$ components.


In total, therefore, there are 19 independent operators contributing to the top mass. Note that the coefficients defined in Eq.~\eqref{eq:topLO} contain a form factor and the couplings (aka pre-Yukawa) of the elementary fields to the strong dynamics operators. The former is only determined by the strong dynamics and can be, in principle, computed on the lattice once the underlying theory is defined. The latter are determined by the UV physics generating the linear mixing {\it \`a la} partial compositeness.
Each coefficient can thus be factorised as
\be
C^{(\prime)}_{t,i_R j_L} = K^{(')}_{t,i_R j_L} y_{i_R} y_{j_L}\,,
\ee
where it is clear that an unambiguous separation of the two is not possible (the pre-Yukawas $y_x$ originate, for instance, from four-fermion operators in the UV). We will thus keep the most compact notation of the coefficients $C^{(\prime)}$ and leave as understood their scaling with the various pre-Yukawa couplings. It is noteworthy that some ratios of the coefficients are independent of the pre-Yukawas, thus uniquely determined by the strong dynamics. For example:
\be \label{eq:ratios}
R_{S} = \frac{C^{(\prime)}_{t, S_R S_L}}{C_{t, S_R S_L}}\,, \quad R'_{S} = \frac{C^{(\prime)}_{t, S^c_R S_L}}{\sqrt{C_{t, S_R S_L} C_{t, S^c_R S^c_L}}}\,,
\ee
and similarly for $S \leftrightarrow S^c$. Naively, all the ratios, like the ones above, are $\mathcal{O} (1)$ numbers, thus they can be used as indicators of violations of the NDA counting if they are phenomenologically needed to be much larger or smaller than unity. Finally, it is only the strong dynamics that determines their size, thus they cannot be freely tuned to the desired value. On the contrary, the value of the pre-Yukawas (or, pre-Yukawa dependent coefficients) can be tuned to the desired value.

In general, the contribution of the above operators to the top mass has the form
\be \label{eq:mtop}
m_{\rm top} = (\lambda^1 + \lambda^2 c_{2\theta}) f s_{2\theta}\,,
\ee
with
\be \label{eq:lambdak}
\lambda^{k} = \sum_{i_R, j_L} \frac{C_{t,i_R j_L}}{4 \pi} \tilde{\lambda}^{k}_{i_R j_L} + \sum_{i_R, j_L} \frac{C'_{t,i_R j_L}}{4 \pi} \tilde{\lambda}^{\prime,k}_{i_R j_L}\,, \quad k = 1,2\,.
\ee
The coupling of the would-be Higgs to the top can be extracted from the form of the top mass:
\be
y_{ht\bar{t}} \equiv \frac{1}{f} \frac{\partial m_{\rm top}}{\partial \theta} = 2 (\lambda^1 c_{2\theta} + \lambda^2 c_{4\theta}) = \frac{m_{\rm top}}{v} \frac{c_{2\theta}}{c_\theta} - 2 \lambda^2 s_{2\theta}^2\,,
\ee
where $m_{\rm top}/v$ is the SM value.
Note that the operators will, in general, also generate couplings of the other pNGBs to the top (and left-handed bottom).
The 19 operators can be grouped in 4 classes, depending on how many parametrically independent coefficients appear in the linear couplings of the pNGBs.

\subsubsection*{Class A: Single-$\Sigma$ operators and anti-symmetric spurions}

This first class is populated by operators that have a single coupling determining both the top mass and the couplings of all the other pNGBs: this coupling is the $\lambda^1$ defined in Eq.\eqref{eq:mtop} as the contribution to $\lambda^2$ vanishes. Furthermore, couplings of the singlets and of the custodial triplet are also generated, proportional to the top mass coupling: all operators with a single-$\Sigma$ insertion in Eqs \eqref{eq:op1Sigma} and \eqref{eq:op1SigmaN} belong to this class.
The couplings of the non-Higgs pNGBs can be written as:
\be \label{eq:classA}
\mathcal{L}_{A_\pm} = - \lambda^1\ t_R^c t_L \left[ \pm i \eta\ \frac{3}{\sqrt{10}} s_{2\theta} \pm i \eta_1^0\ \sqrt{\frac{3}{2}} s_{2\theta} \mp \eta_3^0\ 2 s_\theta \right] 
- \lambda^1\ t_R^c b_L \left[ \pm \eta_3^+\ 2 \sqrt{2} s_\theta \right]\,.
\ee
The upper signs ($A_+$) belong to the 6 operators $D_R A_L$, $A_R D_L$, $D_R S_L$, $S_R D_L$, $S^c_R D_L$ and $N_R S_L$.
The respective coefficients read:
\bea
& \tilde{\lambda}^1_{D_R A_L} = \frac{c_{\beta} - \sqrt{5} e^{i \gamma} s_{\beta}}{4}\,, \quad \tilde{\lambda}^1_{A_R D_L} = \frac{c_{\alpha} - e^{i \varphi} s_{\alpha}}{4}\,, & \nonumber \\
& \tilde{\lambda}^1_{D_R S_L} = \frac{\sqrt{5} c_{\beta} + 3 e^{i \gamma} s_{\beta}}{4 \sqrt{5}}\,, \quad \tilde{\lambda}^1_{S_R D_L} = \frac{4 e^{i \varphi} c_{\beta_{S}} s_{\alpha} - 5 e^{i \gamma_{S}} s_{\beta_{S}} c_{\alpha} - 3 e^{i (\gamma_{S} + \varphi)} s_{\beta_{S}} s_{\alpha}}{4\sqrt{5}}\,, & \nonumber \\
& \tilde{\lambda}^1_{S^c_R D_L} = - \frac{4 e^{i \varphi} c_{\beta^c_{S}} s_{\alpha} + 5 e^{i \gamma^c_{S}} s_{\beta^c_{S}} c_{\alpha} - 3 e^{i (\gamma^c_{S} + \varphi)} s_{\beta^c_{S}} s_{\alpha}}{4 \sqrt{5}}\,, \quad \tilde{\lambda}^1_{N_R S_L} = \frac{1}{\sqrt{5}}\,.  & 
\eea
The lower signs ($A_-$) belong to the operators $D_R A^c_L$, $A^c_R D_L$, $D_R S_L^c$ and $N_R S^c_L$, with coefficients
\bea
& \tilde{\lambda}^1_{D_R A^c_L} = \frac{c_{\beta} + \sqrt{5} e^{i \gamma} s_{\beta}}{4}\,, \quad \tilde{\lambda}^1_{A^c_R D_L} = \frac{c_{\alpha} + e^{i \varphi} s_{\alpha}}{4}\,, & \nonumber \\
& \tilde{\lambda}^1_{D_R S^c_L} = \frac{\sqrt{5} c_{\beta} - 3 e^{i \gamma} s_{\beta}}{4 \sqrt{5}}\,, \quad \tilde{\lambda}^1_{N_R S^c_L} = - \frac{1}{\sqrt{5}}\,.  &
\eea

The 2 operators containing only anti-symmetric spurions in Eq.\eqref{eq:op2XX} also belong to this class, with the difference that couplings to the custodial 5-plet are also generated. The pNGB couplings read
\begin{multline}
\mathcal{L}_{A_R A_L/A^c_R A^c_L} = - \lambda^1\ t_R^c t_L \left[ \pm i \eta\ \frac{1}{\sqrt{10}} s_{2\theta} \pm i \eta_1^0\ \frac{1}{\sqrt{6}} s_{2\theta} \pm \eta_3^0\ 2 s_\theta \mp i \eta_5^0\ \frac{4}{\sqrt{3}} s_{2\theta} \right] \\
- \lambda^1\ t_R^c b_L \left[ \pm \eta_3^+\ 2 \sqrt{2} s_\theta \mp i \eta_5^+\ 2 \sqrt{2} s_{2\theta} \right] \,,
\end{multline}
with the upper signs belonging to $A_R A_L$ and the lower ones to $A^c_R A^c_L$, and coefficients
\be
\tilde{\lambda}^1_{A_R A_L} = - \tilde{\lambda}^1_{A^c_R A^c_L} = \frac{1}{2}\,.
\ee

\subsubsection*{Class B: Double-trace operators with symmetric spurions}

This second class comprises operators that generate only the two couplings in the top mass, i.e. $\lambda^1\neq 0$ and $\lambda^2 \neq 0$, which also 
determine the linear couplings to the singlets and custodial triplets. It is populated by the double-trace operators with the symmetric spurion. 
For operators $S_R S_L$ and $S^c_R S^c_L$, the couplings read
\begin{multline}
\mathcal{L}_{S_R S_L/S^c_R S^c_L} = -  t_R^c t_L \left[ \pm i \eta\ \left( \frac{-7 \lambda^1 + 15 \lambda^2}{3\sqrt{10}} + 3 \sqrt{\frac{2}{5}} \lambda^2 c_{2\theta} \right) s_{2\theta} \pm i \eta_1^0\ \left( \sqrt{\frac{3}{2}} (\lambda^1 - \lambda^2) + \right. \right. \\
\left. \left. \phantom{\sqrt{\frac{3}{2}} } + \sqrt{6} \lambda^2 c_{2\theta} \right) s_{2\theta} 
\mp \eta_3^0\ \left( 2\lambda^1  + 2 \lambda^2 c_{2\theta} \right) s_\theta\right] -  t_R^c b_L \left[ \pm \eta_3^+\ \left( 2 \sqrt{2} \lambda^1 + 2 \sqrt{2} \lambda^2 c_{2\theta} \right) s_\theta \right] \,,
\end{multline}
with
\bea
& \tilde{\lambda}^{\prime,1}_{S_R S_L} = \frac{6 c_{\beta_{S}} + 3 e^{i \gamma_{S}} s_{\beta_{S}}}{2 \sqrt{5}}\,, \quad \tilde{\lambda}^{\prime,2}_{S_R S_L} = \frac{4 c_{\beta_{S}} - 3 e^{i \gamma_{S}} s_{\beta_{S}}}{2 \sqrt{5}}\,, & \nonumber \\
& \tilde{\lambda}^{\prime,1}_{S^c_R S^c_L} = - \frac{6 c_{\beta^c_{S}} + 3 e^{i \gamma^c_{S}} s_{\beta^c_{S}}}{2 \sqrt{5}}\,, \quad \tilde{\lambda}^{\prime,2}_{S^c_R S^c_L} = - \frac{4 c_{\beta^c_{S}} - 3 e^{i \gamma^c_{S}} s_{\alpha^c_{S}}}{2 \sqrt{5}}\,. &
\eea

For operators $S^c_R S_L$ and $S_R S^c_L$, the couplings read
\begin{multline}
\mathcal{L}_{S^c_R S_L/S_R S^c_L} = -  t_R^c t_L \left[ \pm i \eta\ \sqrt{\frac{5}{2}} \frac{5\lambda^1 -3 \lambda^2}{3}  s_{2\theta} \pm i \eta_1^0\ \sqrt{\frac{3}{2}} (\lambda^1 + \lambda^2)  s_{2\theta} \mp \eta_3^0\ \left( 2\lambda^1  + 2 \lambda^2 c_{2\theta} \right) s_\theta\right] \\
-  t_R^c b_L \left[ \pm \eta_3^+\ \left( 2 \sqrt{2} \lambda^1 + 2 \sqrt{2} \lambda^2 c_{2\theta} \right) s_\theta \right] \,,
\end{multline}
with
\bea
& \tilde{\lambda}^{\prime,1}_{S^c_R S_L} = - \tilde{\lambda}^{\prime,1}_{S^c_R S^c_L}\,, \quad \tilde{\lambda}^{\prime,2}_{S^c_R S_L} = - \tilde{\lambda}^{\prime,2}_{S^c_R S^c_L}\,, & \nonumber \\
& \tilde{\lambda}^{\prime,1}_{S_R S^c_L} = - \tilde{\lambda}^{\prime,1}_{S_R S_L}\,, \quad \tilde{\lambda}^{\prime,2}_{S_R S^c_L} = - \tilde{\lambda}^{\prime,2}_{S_R S_L}\,. &
\eea

\subsubsection*{Class C: Single-trace operators with symmetric spurions}

The third class also comprises contributions with two independent couplings, however only one appears in the top mass ($\lambda^2 \neq 0$, while $\lambda^1=0$), 
and another independent one appears in the pNGB couplings. The two operators with a single trace and symmetric spurions populate this class.
The couplings of the pNGBs read
\begin{multline}\label{eq:classC}
\mathcal{L}_{S_R S_L/S^c_R S^c_L} = -  t_R^c t_L \left[ i \eta\ \left(  \lambda^3 \pm 3 \sqrt{\frac{2}{5}} \lambda^2 c_{2\theta} \right) s_{2\theta} + i \eta_1^0\ \left( -\sqrt{\frac{3}{5}} \lambda^3 \pm \sqrt{6} \lambda^2 c_{2\theta} \right) s_{2\theta} \right. \\
\left. + \eta_3^0\ \left( \mp 2\lambda^2 + 2 \sqrt{\frac{2}{5}} \lambda^3 \mp 2 \lambda^2 c_{2\theta} \right) s_\theta\right] -  t_R^c b_L \left[ + \eta_3^+\ \left( \pm 2 \sqrt{2} \lambda^2 + \frac{4}{\sqrt{5}} \lambda_3 \pm 2 \sqrt{2} \lambda^2 c_{2\theta} \right) s_\theta \right] \,,
\end{multline}
with
\bea
& \tilde{\lambda}^2_{S_R S_L} = \frac{4 c_{\beta_{S}} - 3 e^{i \gamma_{S}} s_{\beta_{S}}}{2 \sqrt{5}}\,, \quad \tilde{\lambda}^3_{S_R S_L} = - \frac{5 e^{i \gamma_{S}} s_{\beta_{S}}}{2 \sqrt{2}}\,, & \nonumber \\
& \tilde{\lambda}^2_{S^c_R S^c_L} = -\frac{4 c_{\beta^c_{S}} - 3 e^{i \gamma^c_{S}} s_{\beta^c_{S}}}{2 \sqrt{5}}\,, \quad \tilde{\lambda}^3_{S^c_R S^c_L} = - \frac{5 e^{i \gamma^c_{S}} s_{\beta^c_{S}}}{2 \sqrt{2}}\,. &
\eea

\subsubsection*{Class D: Adjoint spurions}

Finally, this class comprises operators generating three independent couplings: the only operator in this class is the one with the adjoint spurions.
We can define the following couplings:
\bea
& \tilde{\lambda}^1_{D_R D_L} = \frac{1}{2} e^{i \varphi} s_{\alpha} c_{\beta}\,, \quad \tilde{\lambda}^2_{D_R D_L} = \frac{\sqrt{5}}{2} e^{i \gamma} c_{\alpha} s_{\beta}\, & \nonumber \\
&  \tilde{\lambda}^3_{D_R D_L} = \frac{1}{2} c_{\alpha} c_{\beta}\,, \quad \tilde{\lambda}^4_{D_R D_L} = \frac{\sqrt{5}}{2} e^{i (\varphi + \gamma)} s_{\alpha} s_{\beta}\,, & 
\eea
which are related by
\be
\tilde{\lambda}^1_{D_R D_L} \tilde{\lambda}^2_{D_R D_L} = \tilde{\lambda}^3_{D_R D_L} \tilde{\lambda}^4_{D_R D_L}\,.
\ee

The pNGB couplings also include the five-plet, and are given by the following Lagrangian:
\begin{multline} \label{eq:lambda}
\mathcal{L}_{D_R D_L} = -  t_R^c t_L \left[ - i \eta\  \frac{5}{\sqrt{2}} \left( \lambda^3 + \lambda^4 \right) s_{2\theta}  -  i \eta_1^0\ \sqrt{\frac{3}{2}} \frac{5 \lambda^3 - 3\lambda^4}{3} s_{2\theta}  \right. \\
\left.  \phantom{\sqrt{\frac{3}{2}}} + \eta_3^0\ \left( 2\lambda^3 + 2 \left( 2 \lambda^3- \lambda^4\right) c_{2\theta} \right) s_\theta - i \eta_5^0 \frac{4}{\sqrt{3}} \lambda^3 s_{2\theta} \right] \\
 -  t_R^c b_L \left[  \eta_3^+\ \left( - 2 \sqrt{2} \lambda^3 + 2 \sqrt{2} \lambda^4 c_{2\theta} \right) s_\theta - i \eta_5^+ \ 2 \sqrt{2} \lambda^3 s_{2\theta}\right] \,.
\end{multline}
It is noteworthy	 that the pNGB couplings can be set to zero by choosing the embedding of the top left- and right-handed components into different $\SO(5)$ {\it irreps} of the adjoint: $t_L$ in the anti-symmetric and $t_R^c$ in the symmetric (i.e., $\alpha = 0$, $\beta = \pi/2$) for which $\lambda^1=0$, and  $t_L$ in the symmetric and $t_R^c$ in the anti-symmetric (i.e., $\alpha = \pi/2$, $\beta = 0$) for which $\lambda^2=0$.
This is the only scenario where the pNGBs can decouple from the SM fermions, barring cancellations between different operators.

\subsection{Discrete symmetries and the topological term}\label{disc}

As we have seen in the previous section for the Yukawa Lagrangian, there are some features that are specific to the coset under consideration. Among these, we mention that the custodial triplet is the only field -- except the Higgs --  exhibiting linear couplings to scalar currents, while all the other pNGBs couple to the pseudo-scalar combination of the top/bottom spinors. Of course, this is true only if no CP-violating phases appear in the underlying pre-Yukawa sector. For instance, in the case described by Eq.~\eqref{eq:classA}, making the top mass real via a suitable choice of basis, uniquely determines the CP-parity of all the pNGBs. However, when the couplings to fermions are sufficiently uncorrelated from the top mass coefficients $\lambda^1$ and $\lambda^2$ -- such is the case of the adjoint described by Eq.~\eqref{eq:lambda} -- the Yukawa Lagrangian alone cannot determine the CP-parity of the pNGBs. On the other hand, the Wess-Zumino-Witten (WZW) term~\cite{Wess:1971yu,Witten:1983tw}, when present, gives linear couplings only to CP-odd pNGBs, therefore fixing the CP properties of all the pNGBs.
In realistic models with a fermionic underlying description the WZW term is always present, and it allows us to determine that the custodial triplet $\eta_3$ is the only CP-even state, while all the others are CP-odd. Explicit coefficients of the WZW term in the custodial basis can be found in Ref.~\cite{Dugan:1984hq}. This fact has important consequences, as it implies that the vacuum cannot be misaligned along the singlet direction of the triplets without violating CP. 

\subsection{General pNGB potential} \label{sec:pot}

\begin{figure}[tbh]
\centering
\hspace{-35pt} \includegraphics[width=0.7\textwidth]{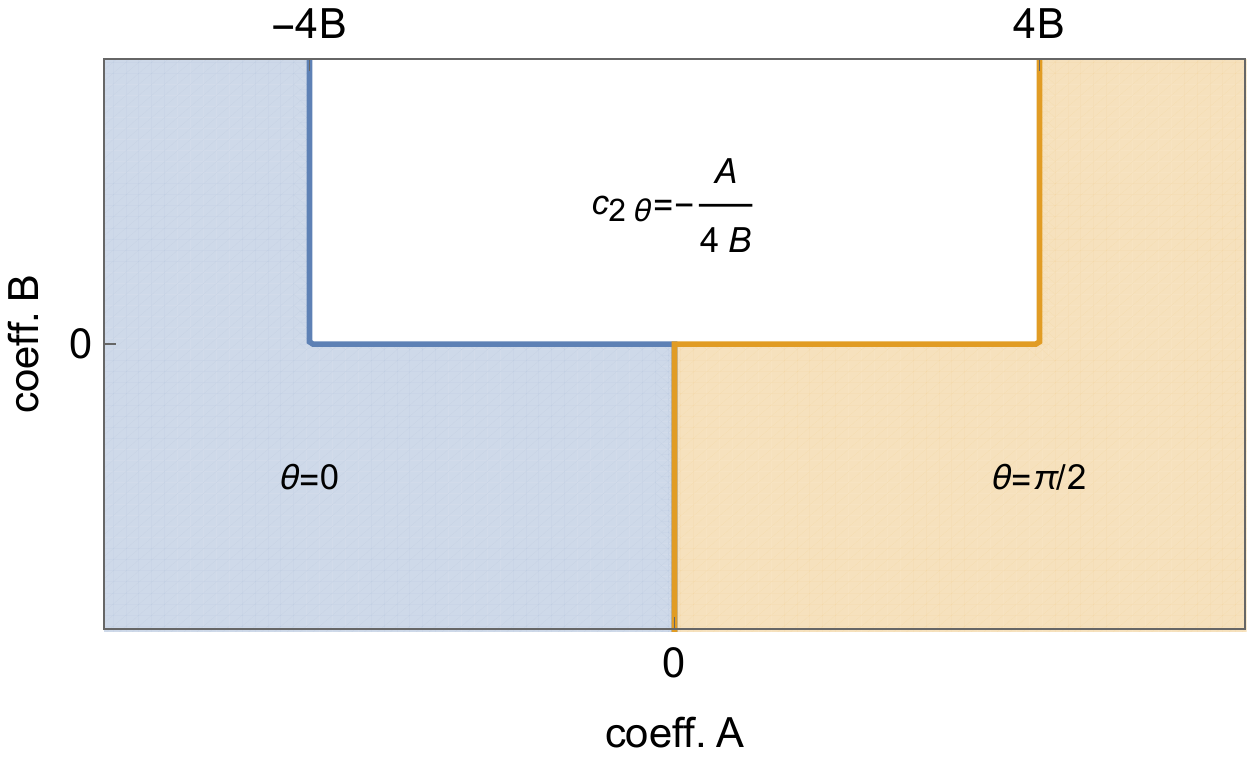}
\caption{Phase space of the vacuum misalignment as a function of the two coefficients of the potential $A$ and $B$. In the blue region the EW symmetry is unbroken, in the orange one the theory is in the Technicolor vacuum, while non-trivial values of $\theta$ are obtained in the white region.}
\label{fig:vacuumstructure}
\end{figure}

The potential for the pNGBs, that determines both the misalignment of the vacuum (i.e. the angle $\theta$) and the scalar spectrum, 
is generated by loops of the SM fields. Both tops and gauge fields will contribute. At LO, we include all the operators that are quadratic in the spurions. To this 
class it belongs also the LO operator generated by the underlying fermion mass.

In general the potential can be expanded in powers of $s_\theta$, and at LO it only contains terms $s_\theta^2$ and $s_\theta^4$: this can be easily 
understood as all the operators contain at most 2 insertions of the vacuum $\Sigma_\theta$, which is at most quadratic in $s_\theta$ and $c_\theta$. We find convenient 
to write the potential in the equivalent form:
\be \label{eq:Vtheta}
V (\theta) = f^4 (A\ c_{2 \theta} + B\ c_{4\theta})\,.
\ee
The phase space of the theory as a function of the two coefficients is shown in Fig.~\ref{fig:vacuumstructure}. The region where a small misalignment angle $\theta$ can be obtained is for $B>0$ and a negative $A$ close to the boundary $A \gtrsim -4 B$.
In that region, the mass of the Higgs can be computed from the second derivative of the above potential, and it can be expressed as
\be \label{eq:higgsmass}
m_h^2 = 64\, B \, v^2 c_\theta^2\,.
\ee
This simple analysis unveils the two tunings that are needed in the potential: on the one hand, obtaining a small misalignment requires $A \sim -4 B$; on the other hand, the correct value of the Higgs mass requires a small coefficient $B$:
\be
B = \frac{1}{c_\theta^2} \frac{m_h^2}{64 v^2} \sim 0.004\ \frac{1}{c_\theta^2}\,.
\ee
While the above issues are common to all composite pNGB Higgs models, we anticipate that, analysing the contribution of the top to the potential, another 
issue arises in the specific $\SU(5)/\SO(5)$ model, namely one needs to avoid a misalignment of the vacuum along the custodial triplet, which would generate a tree level 
contribution to the $\rho$ parameter.

In the rest of the section, we will analyse in detail the contribution to the potential from the gauge, underlying HF mass and top couplings.

\subsubsection{Gauge loops} \label{sec:gaugeloops}

The contribution of gauge loops to the pNGB potential can be written as~\cite{Peskin:1980gc,Preskill:1980mz}:
\be
V_{\rm gauge} = C_g f^4\ \left( g^2\ \Tr [T_L^a \Sigma (T_L^a \Sigma)^\ast] + {g'}^2\ \Tr [T_R^3 \Sigma (T_R^3 \Sigma)^\ast] \right)\,,
\ee
where $C_g$ is an undetermined low energy constant. Expanding up to linear terms in the pNGB fields, we get
\be \label{eq:vgauge}
V_{\rm gauge} = C_g f^4 \frac{3 g^2 + {g'}^2}{2} \left( - (c_{2\theta} + 1) + 2 s_{2\theta} \frac{h}{f} + \dots \right)\,.
\ee
Thus, if this were the only contribution to the potential, the minimum would be at $\theta=0$ (as $B=0$, $A<0$) for $C_g>0$: the assumption on the sign 
of the coefficient relies on the fact that gauge loops typically tend not to break the gauge symmetry itself. Note that the tadpole for the Higgs $h$ vanishes 
at the minimum.

\subsubsection{Mass term} \label{sec:HFmass}

The contribution of the HF mass term to the potential can be written following the standard chiral Lagrangian formalism as
\be
V_{\rm mass} = - C_m f^3\  \Tr[M_\psi \Sigma + \Sigma^\dagger M_\psi^\dagger]\,.
\ee
Expanding up to linear terms in the pNGB fields:
\be
\label{eq:cm}
V_{\rm mass} = 2 C_m f^3 \left( - 3 \mu_d  - (\mu_d + \mu_s) c_{2\theta} + 2 (\mu_d + \mu_s) s_{2\theta} \frac{h}{f} + \dots \right)\,,
\ee
where we have chosen the masses to be real via an appropriate definition of the HF phases.
Similarly to the gauge loops, if $C_m>0$ and for positive masses, the minimum from this term is also at $\theta=0$: this is again expected, as the mass of the 
underlying fermions should simply give a mass to the pNGBs, if defined around the correct vacuum.

It is tantalising that, by changing the sign of the mass terms, the alignment of the theory may change: for instance, if we keep $\mu_d>0$, turning the other mass negative will change the sign of the potential for $\mu_s < -\mu_d$ and thus push the minimum at $\theta=\pi/2$. The breaking of the EW symmetry by HF mass terms alone is, however, only a consequence of the inappropriate choice of the EW preserving vacuum, as we  prove in Appendix~\ref{app:massvacuum}. It is in fact enough to define the theory around the second inequivalent vacuum, defined in Eq.\eqref{vacuum2}, to flip all the signs in front of $\mu_s$. Thus, the theory becomes equivalent to the one with positive $\mu_s$ defined around $\Sigma_0$, with minimum at $\theta=0$. In the Appendix we also show that for negative $\mu_s$ the singlet $\eta$ acquires a tachyonic mass around the wrong vacuum, and that the connection between the two vacua is given by a misalignment along the singlet direction.

This example shows how important it is to study the theory on the correct vacuum, and that the presence of tachyonic mass terms cannot simply be cured by assuming a VEV for the corresponding pNGB but needs a change of vacuum. Examples of this sort have also been pointed out in the $\SU(4)/\SP(4)$ CH models in Ref.~\cite{Alanne:2018wtp}.

\subsubsection{Top couplings}

At LO, we can construct operators that are bilinear in the top spurions and contribute to the pNGB potential. The templates for the operators are the same as for the top mass, with the difference that we need to use the spurions that do not contain external fields (see discussion at the end of Sec.~\ref{spur}) and that they  need to contain only spurions either from $Q_L$ or from $t_R^c$ in order to preserve the SM gauge invariance.
We will employ the following notation for the operators:
\be
\mathcal{O}_{V,X Y} = \mathcal{O} (\bar{X}, Y)\,,
\ee
where $X,Y$ are generic spurions and the bar indicates the hermitian conjugate spurion. The most general form of the potential can thus be written as
\be \label{eq:Vtopgen}
V_{\rm top} = f^4 \sum_{i_L, j_L} \frac{C_{V,i_L j_L}}{4\pi}\ \mathcal{O}_{V,i_L j_L} + f^4 \sum_{i_L, j_L} \frac{C'_{V,i_L j_L}}{4\pi}\ \mathcal{O}'_{V,i_L j_L}  + (L \rightarrow R)\,,
\ee
where the $1/4\pi$ factor follows from NDA, and the prime indicates double-trace operators.
We remark that, by construction, operators with reversed indices are related by the hermitian conjugate:
\be
\mathcal{O}^{(\prime)}_{V, X Y} = \bar{\mathcal{O}}^{(\prime)}_{V, Y X}\,, \quad \mbox{and} \;\; C^{(\prime)}_{V,X Y} = C^{(\prime), \ast}_{V, Y X}\,.
\ee
It follows that operators containing the same spurions, i.e. $\mathcal{O}^{(\prime)}_{V, X X}$ are self-hermitian.
As in the case of the top mass, the pre-Yukawas are absorbed in the coefficients and the fact that each coefficient scales with powers of them allows to define 
pre-Yukawa independent ratios.

The only single-trace operators that are self-hermitian are generated by the adjoint spurions:
\be
\mathcal{O}_{V, D_x D_x} = \Tr [\bar{D}_x^T \Sigma^\dagger D_x \Sigma ]\,, \quad x = L, R\,.
\ee
It follows two single-trace operators containing (anti-)symmetric spurions:
\be
\mathcal{O}_{V, S_x S^c_x} = \Tr [\bar{S}_x \Sigma  S^c_x \Sigma]\,, \quad \mathcal{O}_{V, A_x A^c_x} = \Tr [\bar{A}_x \Sigma  A^c_x \Sigma]\,;
\ee
four containing one adjoint
\bea
& \mathcal{O}_{V, S_x D_x} = \Tr [\bar{S}_x \Sigma D_x]\,, \quad \mathcal{O}_{V, S^c_x D_x} = \Tr [\bar{S}^c_x \Sigma^\dagger D_x]\,, & \nonumber \\
& \mathcal{O}_{V, A_x D_x} = \Tr [\bar{A}_x \Sigma D_x]\,, \quad \mathcal{O}_{V, A^c_x D_x} = \Tr [\bar{A}^c_x \Sigma^\dagger D_x]\,; &
\eea
ad two right-handed ones containing the singlet
\be
\mathcal{O}_{V, S_R N_R} = \Tr [\bar{S}_R \Sigma N_R]\,, \quad \mathcal{O}_{V, S^c_R N_R} = \Tr [\bar{S}^c_R \Sigma^\dagger N_R^T]\,.
\ee
Finally, four double-trace operators can also be built with the symmetric spurions:
\bea
& \mathcal{O}^\prime_{V,S_x S_x} = \Tr [\bar{S}_x \Sigma]\ \Tr [S_x \Sigma^\dagger]\,, \quad  \mathcal{O}^\prime_{V,S^c_x S^c_x} = \Tr [\bar{S}^c_x \Sigma^\dagger]\ \Tr [S^c_x \Sigma]\,, & \nonumber \\
& \mathcal{O}^\prime_{V,S^c_x S_x} = \Tr [\bar{S}^c_x \Sigma^\dagger]\ \Tr [S_x \Sigma^\dagger]\,, \quad  \mathcal{O}^\prime_{V,S_x S^c_x} = \Tr [\bar{S}_x \Sigma]\ \Tr [S^c_x \Sigma]\,. &  
\eea
Similarly to the case of the top mass operators in Section~\ref{sec:topmass}, it can be shown that our basis is equivalent to one constructed in terms of $\SO(5)$ invariants (see Appendix~\ref{app:equivalence} for an explicit proof).
Therefore, there are in total 11 operators associated to the left-handed spurions, and 13 associated to the right-handed ones.
The contribution of the operators to the potential in Eq.~(\ref{eq:Vtheta}) can be written, in analogy to Eq.~(\ref{eq:lambdak}), as
\be
A = \sum_{i_L, j_L} \frac{C_{V,i_L j_L}}{4\pi}\ \tilde{a}_{V,i_L j_L} +  \sum_{i_L, j_L} \frac{C'_{V,i_L j_L}}{4\pi}\ \tilde{a}'_{V,i_L j_L}  + (L \rightarrow R)\,,
\ee
and analogously for $B$. The result for the 24 independent operators are listed in Tables~\ref{tab:potLL} and~\ref{tab:potRR}.

\begin{table} \centering
\begin{tabular}{|l|c|c|c|c|c|c|l|}
 \hline
\multirow{2}{*}{Operator}   & \multicolumn{2}{c|}{potential} & \multicolumn{4}{c|}{tadpoles} &  \\
& $\tilde{a}$ & $\tilde{b}$ & $\eta_3^0$ & $\eta$ & $\eta_1^0$ & $\eta_5^0$ & Coefficients \\
\hline
\hline
$\mathcal{O}_{V,D_L D_L}$ & $\surd$ & $\surd$ &  $\surd$ & $\surd$ & $\surd$ & $-$ & $\begin{array}{l} \tilde{a} = - \frac{3}{2} c_{2\alpha} \\ \tilde{b} = - \frac{1}{2} c_\alpha^2 \end{array}$ \\ \hline
$\mathcal{O}_{V,S_L S^c_L}$ & $\surd$ & $\surd$ &  $-$ & $\surd$ & $\surd$ & $-$ & $\begin{array}{l} \tilde{a} = \frac{3}{2} \\ \tilde{b} = \frac{1}{2} \end{array}$ \\ \hline
$\mathcal{O}_{V,A_L A^c_L}$ & $\surd$ & $-$ &  $-$ & $\surd$ & $\surd$ & $-$ & $\begin{array}{l} \tilde{a} = - \frac{3}{2} \end{array}$ \\ \hline \hline
$\mathcal{O}_{V,S_L D_L}$ & $\surd$ & $-$ &  $-$ & $\surd$ & $\surd$ & $-$ & $\begin{array}{l} \tilde{a} = \frac{1}{4} (3 c_{\alpha} + 5 e^{i \varphi} s_\alpha) \end{array}$ \\ \hline
$\mathcal{O}_{V,S^c_L D_L}$ & $\surd$ & $-$ &  $-$ & $\surd$ & $\surd$ & $-$ & $\begin{array}{l} \tilde{a} = \frac{1}{4} (- 3 c_{\alpha} + 5 e^{i \varphi} s_\alpha) \end{array}$ \\ \hline
$\mathcal{O}_{V,A_L D_L}$ & $\surd$ & $-$ &  $-$ & $\surd$ & $\surd$ & $-$ & $\begin{array}{l} \tilde{a} = \frac{1}{4} (-5 c_{\alpha} - 3 e^{i \varphi} s_\alpha) \end{array}$ \\ \hline
$\mathcal{O}_{V,A^c_L D_L}$ & $\surd$ & $-$ &  $-$ & $\surd$ & $\surd$ & $-$ & $\begin{array}{l} \tilde{a} = \frac{1}{4} (5 c_{\alpha} - 3 e^{i \varphi} s_\alpha) \end{array}$ \\ \hline\hline
$\mathcal{O}^\prime_{V,S_L S_L}$ & $-$ & $\surd$ &  $\surd$ & $-$ & $-$ & $-$ & $\begin{array}{l} \tilde{b} = -\frac{1}{2} \end{array}$ \\ \hline
$\mathcal{O}^\prime_{V,S^c_L S^c_L}$ & $-$ & $\surd$ &  $\surd$ & $-$ & $-$ & $-$ & $\begin{array}{l} \tilde{b} = -\frac{1}{2} \end{array}$ \\ \hline
$\mathcal{O}^\prime_{V,S_L S^c_L}$ & $-$ & $\surd$ &  $-$ & $\surd$ & $\surd$ & $-$ & $\begin{array}{l} \tilde{b} = \frac{1}{2} \end{array}$ \\ \hline
$\mathcal{O}^\prime_{V,S^c_L S_L}$ & $-$ & $\surd$ &  $-$ & $\surd$ & $\surd$ & $-$ & $\begin{array}{l} \tilde{b} = \frac{1}{2} \end{array}$ \\ \hline
\end{tabular}
\caption{Contribution of the operators containing left-handed spurions to the pNGB potential, and presence of tadpoles for the neutral pNGBs except the Higgs.}\label{tab:potLL}
\end{table}

\begin{table} \centering
\begin{tabular}{|l|c|c|c|c|c|c|l|}
 \hline
\multirow{2}{*}{Operator}   & \multicolumn{2}{c|}{potential} & \multicolumn{4}{c|}{tadpoles} &  \\
& $\tilde{a}$ & $\tilde{b}$ & $\eta_3^0$ & $\eta$ & $\eta_1^0$ & $\eta_5^0$ & Coefficients \\
\hline
\hline
$\mathcal{O}_{V,D_R D_R}$ & $\surd$ & $\surd$ &  $\surd$ & $-$ & $-$ & $-$ & $\begin{array}{l} \tilde{a} = - \frac{1}{2} c_{\beta} ^2 \\ \tilde{b} = \frac{5}{8} s_\beta^2 \end{array}$ \\ \hline
$\mathcal{O}_{V,S_R S^c_R}$ & $-$ & $\surd$ &  $-$ & $\surd$ & $\surd$ & $-$ & $\begin{array}{l}  \tilde{b} = \frac{1}{40} (4 c_{\beta_S} - 3 e^{-i \gamma_S} s_{\beta_S}) (4 c_{\beta^c_S} - 3 e^{i \gamma^c_S} s_{\beta^c_S}) \end{array}$ \\ \hline
$\mathcal{O}_{V,A_R A^c_R}$ & $\surd$ & $-$ &  $-$ & $\surd$ & $\surd$ & $\surd$ & $\begin{array}{l} \tilde{a} = - \frac{1}{2} \end{array}$ \\ \hline\hline
$\mathcal{O}_{V,S_R D_R}$ & $\surd$ & $-$ &  $-$ & $\surd$ & $\surd$ & $-$ & $\begin{array}{l} \tilde{a} = \frac{e^{i \gamma} s_{\beta}}{20} (-6 c_{\alpha_S} + 17 e^{-i \gamma_S} s_{\alpha_S}) \end{array}$ \\ \hline
$\mathcal{O}_{V,S^c_R D_R}$ & $\surd$ & $-$ &  $-$ & $\surd$ & $\surd$ & $-$ & $\begin{array}{l} \tilde{a} = \frac{e^{i \gamma} s_{\beta}}{20} (-6 c_{\alpha^c_S} + 17 e^{-i \gamma^c_S} s_{\alpha^c_S}) \end{array}$ \\ \hline
$\mathcal{O}_{V,A_R D_R}$ & $\surd$ & $-$ &  $-$ & $\surd$ & $\surd$ & $-$ & $\begin{array}{l} \tilde{a} = - \frac{1}{4} c_{\beta} \end{array}$ \\ \hline
$\mathcal{O}_{V,A^c_R D_R}$ & $\surd$ & $-$ &  $-$ & $\surd$ & $\surd$ & $-$ & $\begin{array}{l} \tilde{a} = \frac{1}{4} c_{\beta} \end{array}$ \\ \hline\hline
$\mathcal{O}_{V,S_R N_R}$ & $\surd$ & $-$ &  $-$ & $\surd$ & $\surd$ & $-$ & $\begin{array}{l} \tilde{a} = \frac{1}{10} (4 c_{\alpha_S} - 3 e^{-i \gamma_S} s_{\alpha_S}) \end{array}$ \\ \hline
$\mathcal{O}_{V,S^c_R N_R}$ & $\surd$ & $-$ &  $-$ & $\surd$ & $\surd$ & $-$ & $\begin{array}{l} \tilde{a} = \frac{1}{10} (4 c_{\alpha^c_S} - 3 e^{-i \gamma^c_S} s_{\alpha^c_S}) \end{array}$ \\ \hline\hline
$\mathcal{O}^\prime_{V,S_R S_R}$ & $\surd$ & $\surd$ &  $-$ & $\surd$ & $\surd$ & $-$ & $\begin{array}{l} \tilde{a} = \frac{3}{20} (5 + 11 c_{2\beta_S} - 2 s_{2 \beta_S} c_{\gamma_S}) \\ \tilde{b} = \frac{1}{80} (25 + 7 c_{2 \beta_S} - 24 s_{2\beta_S} c_{\gamma_S}) \end{array}$ \\ \hline
$\mathcal{O}^\prime_{V,S^c_R S^c_R}$ & $\surd$ & $\surd$ &  $-$ & $\surd$ & $\surd$ & $-$ & $\begin{array}{l} \tilde{a} = \frac{3}{20} (5 + 11 c_{2\beta^c_S} - 2 s_{2 \beta^c_S} c_{\gamma^c_S}) \\ \tilde{b} = \frac{1}{80} (25 + 7 c_{2 \beta^c_S} - 24 s_{2\beta^c_S} c_{\gamma^c_S}) \end{array}$ \\ \hline
$\mathcal{O}^\prime_{V,S_R S^c_R}$ & $\surd$ & $\surd$ &  $-$ & $\surd$ & $\surd$ & $-$ & {\footnotesize $\begin{array}{l} \tilde{a} = \frac{3}{10} ( c_{\beta_S} (8 c_{\beta_S^c} - e^{i \gamma_S^c} s_{\beta_S^c}) - 3 e^{-i \gamma_S} s_{\beta_S} (c_{\beta_S^c} + 3 e^{i \gamma_S^c} s_{\beta_S^c})) \\ \tilde{b} = \frac{1}{40} (4 c_{\beta_S} - 3 e^{-i \gamma_S} s_{\beta_S}) (4 c_{\beta^c_S} - 3 e^{i \gamma^c_S} s_{\beta^c_S}) \end{array}$} \\ \hline
$\mathcal{O}^\prime_{V,S^c_R S_R}$ & $\surd$ & $\surd$ &  $-$ & $\surd$ & $\surd$ & $-$ & {\footnotesize $\begin{array}{l} \tilde{a} = \frac{3}{10} ( c_{\beta_S^c} (8 c_{\beta_S} - e^{i \gamma_S} s_{\beta_S}) - 3 e^{-i \gamma_S^c} s_{\beta_S^c} (c_{\beta_S} + 3 e^{i \gamma_S} s_{\beta_S})) \\ \tilde{b} = \frac{1}{40} (4 c_{\beta_S} - 3 e^{i \gamma_S} s_{\beta_S}) (4 c_{\beta^c_S} - 3 e^{-i \gamma^c_S} s_{\beta^c_S}) \end{array}$} \\ \hline
\end{tabular}
\caption{Same as Table~\ref{tab:potLL}, but for operators containing right-handed spurions.}\label{tab:potRR}
\end{table}

The operators in Eq.~(\ref{eq:Vtopgen}) also contain tadpoles for the Higgs (that vanish at the minimum) and for other neutral pNGBs. The presence of such tadpoles for each individual operator is also indicated in Tables~\ref{tab:potLL} and ~\ref{tab:potRR}.
For the pseudo-scalars $\eta$, $\eta_1^0$ and $\eta_5^0$, the tadpoles are always proportional to a phase which is present in the underlying theory: either the phase between two embeddings of the tops into the same spurion, or the phase of the overall coefficient (i.e., the phase of the pre-Yukawas). Thus, such spurions are naturally zero if no overall CP-violating phases are present in the pre-Yukawas. \footnote{Note that the CP violating phase in the CKM does not contribute here. In fact, only relative phases between the pre-Yukawas of the same generation count.}
On the other hand, the tadpole for the CP-even scalar $\eta_3^0$ is proportional to the real part of the coefficients, and its presence is dangerous as it would lead to a VEV for the custodial triplet, i.e. a misalignment of the vacuum along its direction, thus breaking the custodial invariance. This tadpole is generated by only 4 operators: $\mathcal{O}_{V, D_L D_L}$, $\mathcal{O}_{V, D_R D_R}$, $\mathcal{O}^\prime_{V, S_L S_L}$ and $\mathcal{O}^\prime_{V, S^c_L S^c_L}$. The most general expression is
\be  \label{eq:tadpole}
V_{\rm top} \supset - f^3 \eta_3^0 \ c_\theta s_\theta^2\ \left( 4 \frac{C_{V,D_L D_L}}{4 \pi} s_{2\alpha} c_\varphi + 2 \sqrt{5} \frac{C_{V,D_R D_R}}{4 \pi} s_{2\beta} c_\gamma + 8 \frac{C^\prime_{V, S_L S_L}}{4\pi} - 8 \frac{C^\prime_{V, S^c_L S^c_L}}{4\pi} \right)\,.
\ee
It is essential, for the phenomenological viability of the model, to suppress such tadpole in order to avoid large contributions to the $\rho$ parameter. In the case of the adjoint spurions, this can be done by choosing $\alpha = 0, \pi/2$ and $\beta = 0, \pi/2$. Interestingly, in these cases the couplings of the non-Higgs pNGBs to top and bottom vanish. For the symmetric spurion, it would be needed to suppress the double-trace operators with respect to the single trace ones or require a cancellation between the operators with $S_L$ and $S^c_L$.

From the general results we found in this section, we can thus identify the following scenarios:

\begin{itemize}

\item[-] Adjoint: When both spurions are in the adjoint representation, it is possible to generate the mass of the top and avoid the triplet tadpole by choosing the $Q_L$ and $t_R$ in two different $\SO(5)$ {\it irreps}. However, only the choice $\alpha = 0$ ($Q_L$ in the anti-symmetric) and $\beta=\pi/2$ ($t_R^c$ in the symmetric) allows for a term $\sim c_{4\theta}$ in the potential, which is essential in order to achieve a minimum at small $\theta$. No linear couplings of the non-Higgs pNGBs to tops (and bottom) are generated.

\item[-] Anti-symmetric: When both spurions are in the anti-symmetric (or both in the conjugate), the LO potential for the vacuum is not generated. Thus, one needs to study operators at NLO. The additional suppression of $1/4\pi$ from NDA helps in achieving a naturally light Higgs mass (see Ref.~\cite{Alanne:2018wtp} for an analogous observation for  $\SU(4)/\SP(4)$ CH models).

\item[-] Symmetric: The left-handed top in the symmetric should be avoided as it generates tadpoles for the triplet via the double-trace operators. One possibility is that the double-trace operator is suppressed by some effects of the strong dynamics (for instance, at large number of hypercolours). The potential would then be generated dominantly by the right-handed spurions or single-trace operators. Alternatively, a cancellation may occur, needing the presence of both the symmetric and its conjugate.

\item[-] The remaining combinations of spurions without triplet tadpoles are: $D_L$--$S_R^{(c)}$, $D_L$--$A_R^{(c)}$ and $A_L^{(c)}$--$D_R$. As the anti-symmetric does not generate the coefficient $B$ (term $\sim c_{4\theta}$), then the adjoints are forced to be in the situation $\alpha = 0$ for $D_L$ (i.e., anti-symmetric of $\SO(5)$) and $\beta=\pi/2$ for $D_R$ (i.e., symmetric of $\SO(5)$).
 In the case $D_L$--$S_R^{(c)}$, both limits $\alpha = 0$, $\pi/2$ are allowed.

\end{itemize}

In the next section we will study in detail the cases highlighted above where a tadpole for the custodial triplet can be avoided. In section~\ref{sec:antisym} we will study the NLO potential in the case of the anti-symmetric spurion, while in section~\ref{sec:vacuum2} we will show some results for the general vacuum misaligned along the custodial triplet.

\section{Special cases with LO potential} \label{sec:cases}

We start with the four cases where the potential from the top spurions is generated at LO.

\subsection{Adjoint in the absence of tadpoles} \label{sec:adj}

We consider here the case where both the left- and right-handed fermions are embedded in the adjoint representation of $\SU(5)$. \footnote{This case has also been studied  in Ref.\cite{Ferretti:2016upr}, with which our results are in agreement.}
The minimal LO potential 
contains the two operators $\mathcal{O}_{V,D_R D_R}$ and $\mathcal{O}_{V,D_L D_L}$, plus the gauge contribution. To simplify the notation we 
call $C_R$ and $C_L$ the coefficients of the two operators respectively.
As already discussed in the previous section, the absence of tadpoles for pNGBs other than the Higgs plus the presence of the coefficient $B$ in 
the potential of Eq.~(\ref{eq:Vtheta}) requires that we choose $\alpha = 0$ and $\beta = \pi/2$ (and real coefficients). We recall that this corresponds 
to embedding the left-handed doublet in the anti-symmetric of $\SO(5)$ and the right handed top in the symmetric, and that this choice guarantees 
that the couplings of the pNGBs (except for the Higgs) to top and bottom vanish.
The coefficients of the potential can be read off from Tables~\ref{tab:potLL} and~\ref{tab:potRR}:
\begin{equation} \label{eq:potDD}
 A= \frac{1}{4\pi}\left(-\frac{3}{2} C_L \right) - \frac{1}{2} C_g (3g^2 + g'^2)\,, \qquad
 B=\frac{1}{4\pi} \left( -\frac{1}{2}  C_L +\frac{5}{8}  C_R\right) \,.
\end{equation}
We can immediately see that, to obtain the correct vacuum, the coefficient $C_{R}$ needs to be positive and that the following inequalities must hold:
\begin{equation}
5 C_R > 4 C_L\,, \qquad \frac{3}{4\pi} C_L > - C_g (3 g^2 + g'^2)\,. 
\end{equation}

Assuming that the correct vacuum is attained, the two coefficients $C_{L/R}$ can be fixed by imposing the minimisation condition and the value of the Higgs mass $m_h$:
\begin{align}
C_L  \ = \ & \pi \frac{c_{2\theta}}{6 c_\theta^2} \frac{m_h^2}{v^2} -  \frac{4 \pi}{3} C_g (3 g^2 + g'^2)\,, \nonumber  \\
 C_R \ = \ &  \pi \frac{8 c_{\theta}^2 -1}{30 c_\theta^2} \frac{m_h^2}{v^2} -  \frac{16 \pi}{15} C_g (3 g^2 + g'^2)\,.
\end{align}
The expressions above can be used to define Yukawa-free ratios of coefficients, as explained in the previous section, which in the present case amounts to a single ratio
\be \label{eq:ratioDD}
 R_\text{D} \equiv \frac{ C_t ^2}{  C_L C_R } = \ \frac{576 v^2 c_{\theta }^2 m_t^2}{c_{2 \theta }^3 \left(4 c_{2 \theta }+3\right) m_h^4} \ \approx   \frac{576}{7} \frac{v^2 m_t^2}{m_h^4} \left( 1 + \frac{43}{7} s_\theta^2 + \mathcal{O} (s_\theta^4) \right)\,.
 \ee
Numerically $R_D \sim 600$, thus showing that the form factors associated to the potential operators are required to be significantly smaller than the NDA estimate.

\begin{figure}
\centering
\includegraphics[width=.48\textwidth]{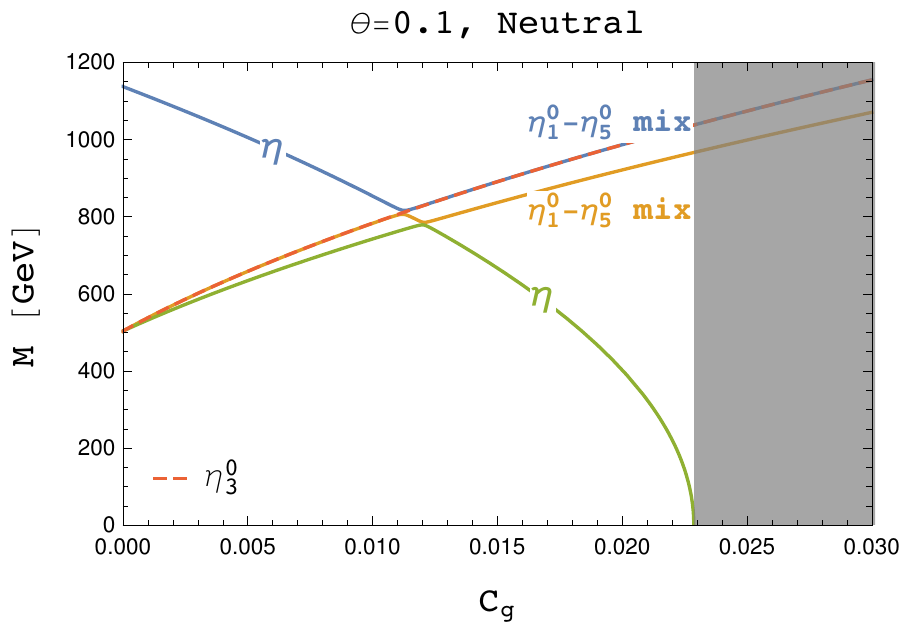}
\includegraphics[width=.48\textwidth]{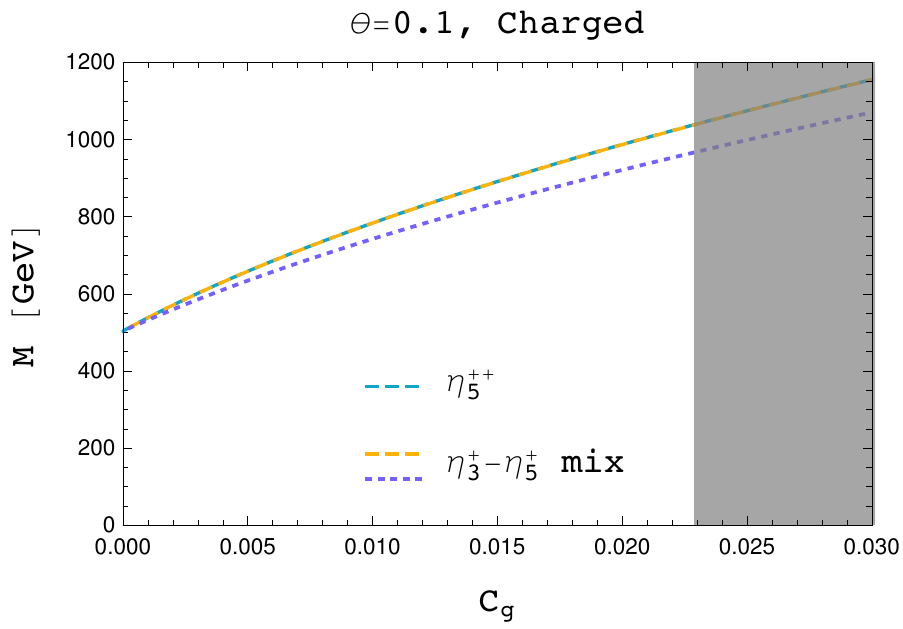}
\caption{Masses of the neutral (left) and charged (right) pNGBs in the case of the adjoint representation for $\theta=0.1$, as a function of the coefficient $C_g$. A tachyonic state appears for $C_g \gtrsim 0.023$, almost aligned in the direction of the gauge singlet $\eta$. }
\label{fig:spectrumDD}
\end{figure}

We are now ready to study the spectrum of the scalar pNGB sector of the theory. The Higgs candidate, as expected, does not mix with the other pNGBs. 
Similarly, the only other CP-even scalar, $\eta_3^0$, does not mix and has mass
\begin{align} \label{eta}
m_{\eta_3^0}^2 & \ =\  \frac{2}{3} m_h ^2 \cot^2 2\theta +4 \frac{C_g v^2}{\st^2} \left( (3g^2+ g'^2) + \frac{2}{3} g'^2 \ctt \right)\,.
\end{align}
The three pseudo-scalars, $\eta$, $\eta_1^0$ and $\eta_5^0$ enjoy non-zero mixing. The $3\times 3$ mass matrix can be parametrised as
\begin{equation} \label{eq:neutmix}
M_{\rm PS}^2 = \frac{m_h^2}{\stt^2}\ \mathcal{M}_1^2 + C_g (3 g^2 + g'^2) \frac{v^2}{\st^2}\ \mathcal{M}_2^2  + C_g g'^2 \frac{v^2}{\st^2}\ \mathcal{M}_3^2\,,
\end{equation}
with
\begin{eqnarray}
\mathcal{M}_1^2 &=& \left(
\begin{array}{ccc}
 \frac{5}{48} \left(20 c_{2 \theta }+3 c_{4 \theta }+9\right) & -\frac{1}{4} \sqrt{\frac{5}{3}} \left(5 c_{2 \theta }+3\right) s_{\theta }^2 & 0 \\
 -\frac{1}{4} \sqrt{\frac{5}{3}} \left(5 c_{2 \theta }+3\right) s_{\theta }^2 & \frac{1}{48} \left(-36 c_{2 \theta }+25 c_{4 \theta }+43\right) & 0 \\
 0 & 0 & \frac{2 c_{2 \theta }^2}{3} \\
\end{array}
\right) \,, \nonumber  \\
 \mathcal{M}_2^2 &=&\left(
\begin{array}{ccc}
 -\frac{20}{3}  & 0 & 0 \\
 0 & 4 & 0 \\
0 &0 & 4 \\
\end{array}
\right), \ 
\mathcal{M}_3^2 =\left(
\begin{array}{ccc}
 0 & 0 &-2 \sqrt{\frac{10}{3}} s_{\theta }^2 \\
 0 & 0 & \frac{1}{3} \sqrt{2} \left(5 c_{2 \theta }+3\right) \\
-2 \sqrt{\frac{10}{3}} s_{\theta }^2 &\frac{1}{3} \sqrt{2} \left(5 c_{2 \theta }+3\right) & \frac{4}{3}(\ctt-3) \\
\end{array}
\right)\,.
\end{eqnarray}
The term $\mathcal{M}_2^2$ encodes custodial invariant corrections generated by gauge loops to the mass term proportional to the Higgs mass, i.e. 
$\mathcal{M}_1^2$. As such, it cannot contain mixing between the three states, while a mixing between the two singlets is contained in $\mathcal{M}_1^2$, 
though suppressed by $\st^2$. Finally, the third term $\mathcal{M}_3^2$ contains custodial violating contributions, thus proportional to $g'^2$, and it's the 
only source of mixing for $\eta_5^0$.

The structure of the spectrum becomes clearer in an expansion for small $\theta$, where mixing terms are negligible except for the mixing between $\eta_1^0$ and $\eta_5^0$ in the third term of Eq.~\eqref{eq:neutmix}. The mass of the singlet is thus equal to
\be
m_\eta ^2 \ \approx \ \frac{5}{6} \frac{m_h^2}{\st^2}-\frac{20}{3} C_g (3g^2+g'^2) \frac{v^2}{\st^2} + \mathcal{O} (\st^2)\,, \\
\label{eq:custsing}
\ee
while the other two give mass eigenstates
\begin{eqnarray}
m_{\eta_1} ^2\ &\approx & \ \frac{1}{6}\frac{m_h^2}{\st^2}+ \frac{4}{3} C_g (9g^2-g'^2) \frac{v^2}{\st^2} + \mathcal{O} (\st^2)\,, \\
m_{\eta_2}^2 \ &\approx & \ m_{\eta_1}^2 + 8 C_g g'^2 \frac{v^2}{\st^2} + \mathcal{O} (\st^2)\,.
\end{eqnarray}
Note that, for $C_g \approx 0$, the singlet $\eta$ is heavier than the other two states by a factor $\sqrt{5}$, whereas it receives a negative correction from the gauge sector. There will therefore be a value of $C_g$ where the singlet state becomes tachyonic, so that consistency of the minimum is only maintained for 
\be
C_g \ < \ \frac{m_h^2}{8 v^2(3 g^2+g'^2) \ct^2} \sim \frac{0.023}{\ct^2}\,.
\ee
The other two states receive positive corrections from $C_g$ and they are lighter than the singlet for small gauge loops. We see all these features in the left plot of Figure~\ref{fig:spectrumDD}.
The singlet $\eta$ is clearly recognisable as the line that decreases with increasing $C_g$ and significantly mixes with the other two only when the masses are very close.
The other two states are a mixture of $\eta_1^0$ and $\eta_5^0$. Finally, the dashed red line   represents the scalar triplet $\eta_3^0$, which does not mix with any of the other states. 
We  remark that, even though we chose $\theta = 0.1$, for which the compositeness scale $f \approx 2.5$~TeV, the masses are below a TeV in most of the parameter space, with the singlet becoming increasingly light while approaching the tachyonic bound.

The spectrum of the charged scalars features one exact mass eigenstate, namely the doubly charged field $\eta_5^{\pm \pm}$, with mass 
\be
m_{\eta_5^{\pm \pm} } ^2 \ = \  \frac{2}{3} m_h^2 \cot  ^2  2\theta + 4 \frac{C_g v^2 }{\st^2}(3 g^2+ g'^2) + \frac{4}{3} \frac{C_g v^2}{\st^2}  \left(3-\ctt\right) g'^2\,.
\ee
The remaining two singly-charged custodial eigenstates, $\eta_3^\pm$ and $\eta_5^\pm$, have a mixing matrix that can be parametrised, similarly to the neutral sector:
\be
\mathcal M_+ ^2 \ = \   \frac{2}{3}m_h^2\cot ^2 2 \theta \ \mathcal M _{1,+}^2  +  4C_g (3 g^2 + g'^2) \frac{v^2}{\st^2} \ \mathcal{M}_{2,+}^2  + \frac{2}{3} C_g g'^2 \frac{v^2}{\st^2}\ \mathcal{M}_{3,+}^2\,,
\ee
where the matrices $\mathcal M_{i,+}^2$ are given by
\begin{eqnarray}
\mathcal M _{1,+}^2 = \mathcal{M}_{2,+}^2 = \mathbb{1}_2\, , \quad \mathcal M_{3,+} ^2 \ = \  \left( \begin{array}{cc} 
-2 \ctt & 6 \ct \\
6 \ct &  -3 + \ctt 
\end{array} \right)\,.
\end{eqnarray}
In a small $\theta$ expansion we remark that, at leading order, 
\be
m_{\eta_5^{\pm \pm} }  \approx m_{\eta_1^+} \approx m_{\eta_1}\,, \quad m_{\eta_2^+} \approx m_{\eta_2}\,,
\ee
where $m_{\eta_{1/2}^+}$ are the two mass eigenvalues in the singly charged sector.
Note, also, that the charged triplet and five-plet are exact mass eigenstates only when the hypercharge coupling $g'$ vanishes, and that in the same limit all the fields belonging to the triplet and five-plet are completely degenerate. 

 The charged spectrum is illustrated in the right panel of Figure \ref{fig:spectrumDD}. The doubly-charged field, represented by the dashed-blue line, is the heaviest state, while the other two fields are almost equal mixtures (the mixing is large and the difference between the diagonal terms is $6 \st^2$) of the singly-charged triplet and five-plet fields.

\subsection{Other cases with only one adjoint}

We now list the remaining cases that are free of triplet tadpoles. Note that the cases with the irreps conjugate to the (anti-)symmetric give similar results.

\subsubsection{Anti-symmetric $A_L$ plus adjoint $D_R$}

The anti-symmetric does not induce any potential at LO, thus the potential for this model is similar to the $D_L$--$D_R$ case in Eq.~\eqref{eq:potDD}, with $C_L = 0$ ($\beta = \pi/2$ remains  the only choice that keeps $B\neq 0$). As long as $C_R >0$, the potential has the correct minimum. However, in the absence of HF masses,  there is always a tachyonic state in the pNGB spectrum, meaning that the vacuum misalignment is not stable.
Therefore, in this case it is necessary to include the contribution of the HF masses, in such a way to give a positive contribution to the pNGB masses and remove tachyons. We remark that this is not an ad-hoc choice, as the HF masses are always present in all models. For simplicity, we also set the coefficient $C_m$ to one, as this is equivalent to reabsorbing it into the definition of the mass parameters $\mu_{d,s}$. We also define the quantity $\displaystyle \delta \equiv \frac{\mu_s - \mu_d}{f}$, which measures the amount of explicit $\SO(5)$ breaking due to the HF mass term, and an average mass $\displaystyle \mu \equiv \frac{ \mu_d + \mu_s}{2}$. The minimum and Higgs mass conditions are easily enforced by solving for two of the free parameters: in the following, we chose to solve for $C_R$--$\mu$ and $C_R$--$C_g$, alternatively, in order to give a broader picture of the available parameter space. 

\begin{figure}
\centering
\includegraphics[width=.4\textwidth]{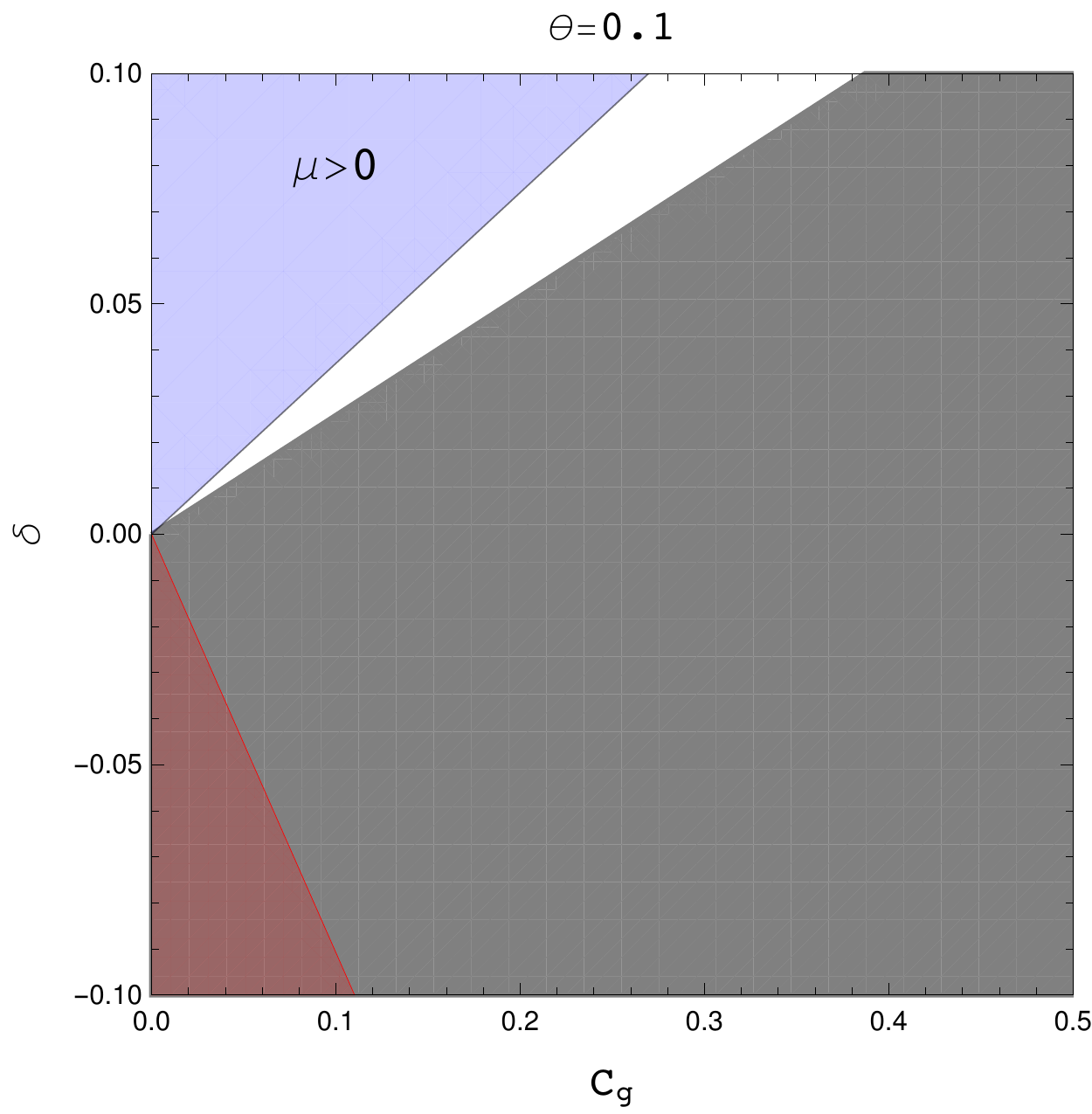} 
\includegraphics[width=.4\textwidth]{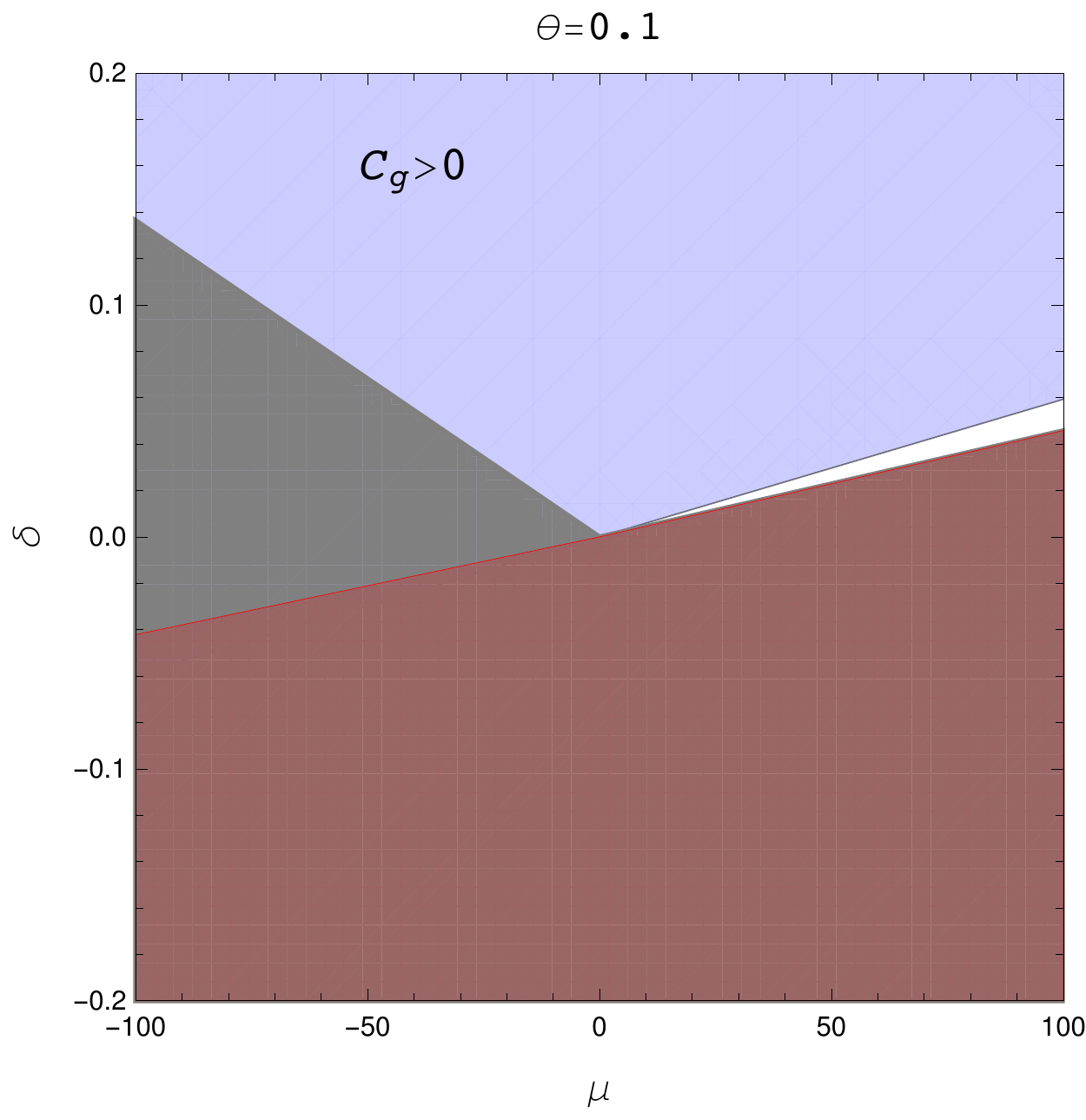}\\ 
\caption{Allowed region in the $A_L$-$D_R$ model for $\theta = 0.1$. The unaccessible grey area is determined by one neutral pseudo-scalar state becoming tachyonic. In a subset of this region, highlighted in red,  one charged scalar also has negative mass squared. The plots are given in the $C_g$\,--\,$\delta$  and $\mu$\,--\,$\delta$  planes. For each value of $\delta$, there is a maximum allowed value of $C_g>0$ (left-panel) and a range of of $\mu$ (right-panel) in order to avoid tachyons.}
\label{fig:ALDRexcl}
\end{figure}

In Fig.~\ref{fig:ALDRexcl} we show the allowed parameter space for the two choices, thus as a function of the remaining free parameters $C_g$--\,$\delta$ and $\mu$--$\delta$, and for fixed $\theta = 0.1$. The grey region is not accessible due to the presence of (at least) one tachyon, while the red shading indicates the presence of a charged tachyonic state. 
While we impose $C_g>0$ as a necessary condition (see discussion in Section~\ref{sec:gaugeloops}), the absence of tachyons implies that $\delta > 0$ and shows that, for each value of $\delta$, there is an upper limit on $C_g$, as we already observed in the $D_R$--$D_L$ case. The blue shading in the left panel indicates $\mu>0$, thus showing that negative values of the average mass are also allowed. The right panel of Fig.~\ref{fig:ALDRexcl} shows the same parameter space in terms of $\mu$--$\delta$, where $C_g>0$ corresponds to the blue shaded area.

\subsubsection{Adjoint $D_L$ plus anti-symmetric $A_R$}

This case is similar to the previous one, but with the left-handed $C_L$ being non zero ($\alpha = 0$ to avoid the triplet tadpole while guaranteeing $B \neq 0$).
However, even when adding the contribution of the fermion masses, there is always a tachyon in the spectrum, meaning that the simple vacuum misalignment is not consistent. We will therefore discard this case.

\subsubsection{Adjoint $D_L$ plus symmetric $S_R$}

 \begin{figure}
\centering
\includegraphics[width=.9\textwidth]{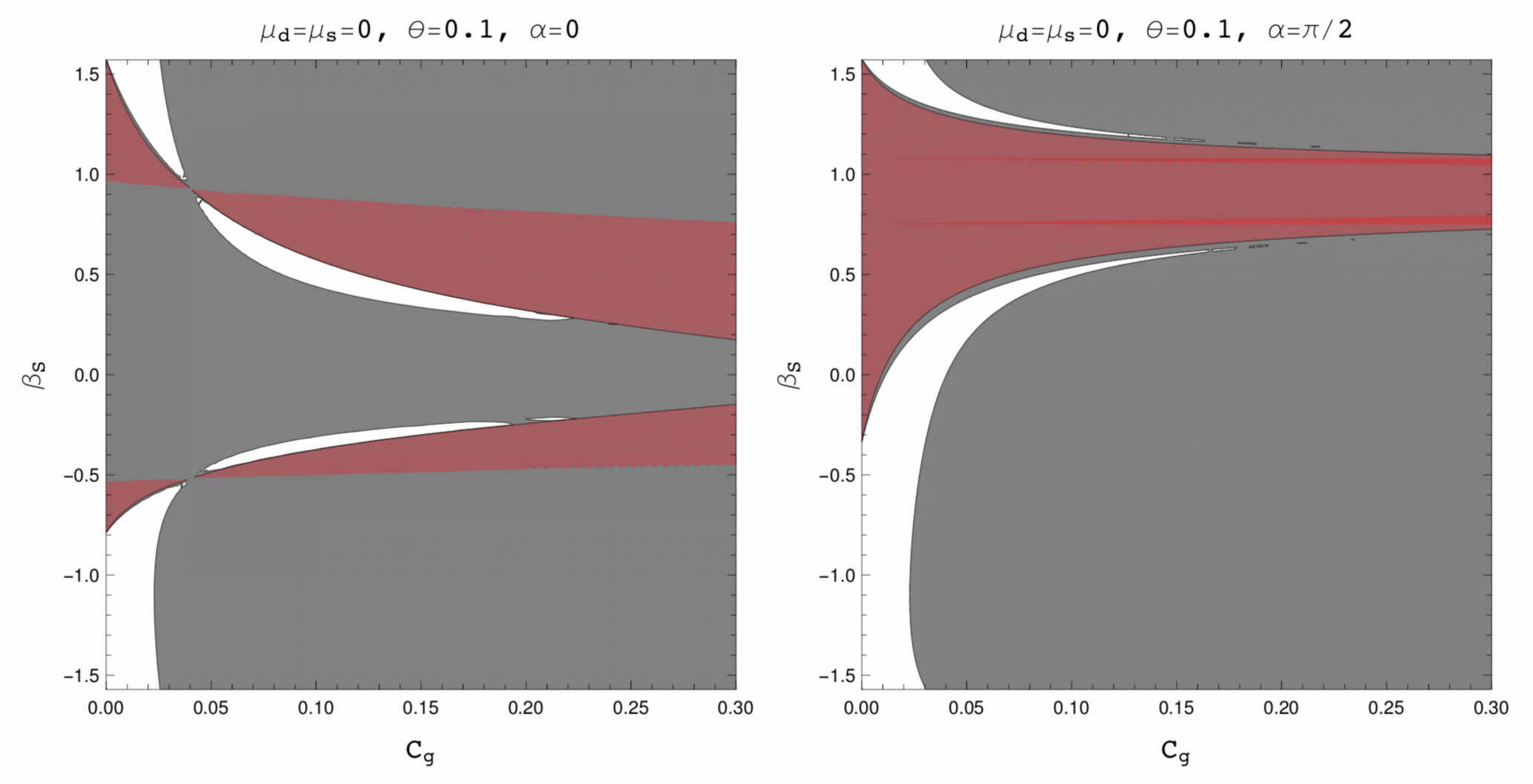}
\caption{
Allowed regions in the $D_L$--$S_R$ case, for $\mu_d = \mu_s =0$ and $\theta = 0.1$.  We consider both  $\alpha=0$ (left)  and $\alpha = \pi/2$ (right).  As before, the grey regions indicate that some neutral scalar becomes tachyonic, while the red shading signals the presence of charged tachyons.}
\label{fig:DLSR}
\end{figure}

This case is a bit more involved, because the potential features two operators: besides the L adjoint, there is the double-trace operator $\mathcal O' _{V, S_R, S_R}$. Furthermore, both choices $\alpha = 0,\pi/2$ are feasible, as the double-trace operator always induces $B \neq 0$. The coefficients in the potential read
\begin{eqnarray} \label{eq:potDS}
 A&= &\frac{1}{4\pi}\left(-\frac{3}{2} c_{2\alpha} C_L + \frac{3}{20} (5 + 11 c_{2\beta_S} - 2 s_{2\beta_S}) C_R \right) - \frac{1}{2} C_g (3g^2 + g'^2)\,, \nonumber \\
 B&=&\frac{1}{4\pi} \left( -\frac{1}{2} c_{\alpha}^2 C_L + \frac{1}{80} (25 + 7 c_{2\beta_S} - 24 s_{2\beta_S}) C_R \right) \,.
\end{eqnarray}
We choose to solve the minimum and Higgs mass conditions in terms of $C_L$ and $C_R$.
Thus, for either choice of $\alpha$, the two free parameters are $\beta_S$ and $C_g$ (note that we have fixed $\gamma_S = 0$ to preserve CP).
The allowed parameter space in the two cases $\alpha = 0, \pi/2$ is shown in Fig.~\ref{fig:DLSR} for $\theta= 0.1$. For simplicity and to reduce the number of free parameters, we will neglect the HF masses in this case.

\subsection{Spectra} \label{sec:spectra}

We now turn our attention to the mass spectra in the two cases, $A_L$--$D_R$ and $D_L$--$S_R$.

\begin{figure}
\centering
\includegraphics[width=.45\textwidth]{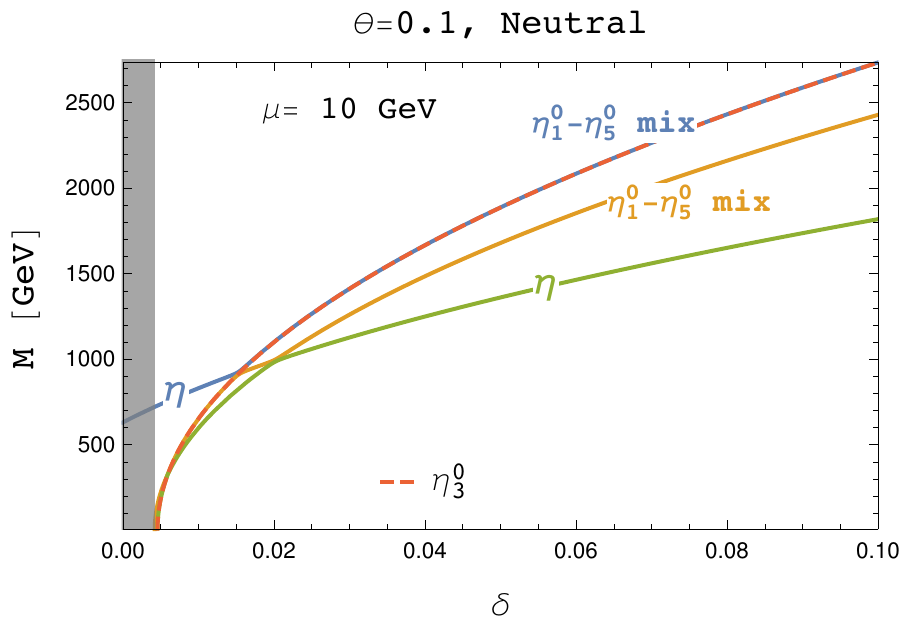}
\includegraphics[width=.45\textwidth]{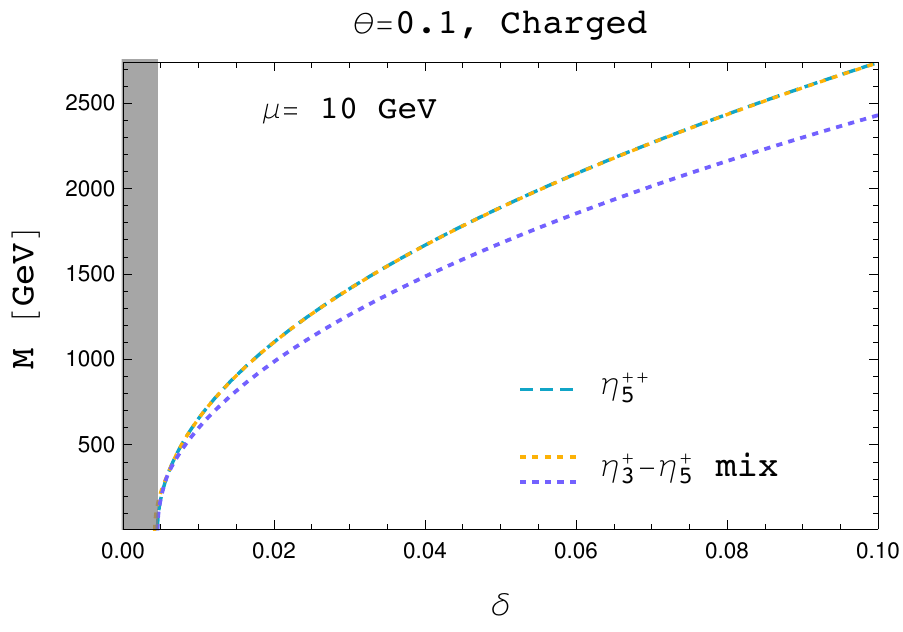} \\
\includegraphics[width=.45\textwidth]{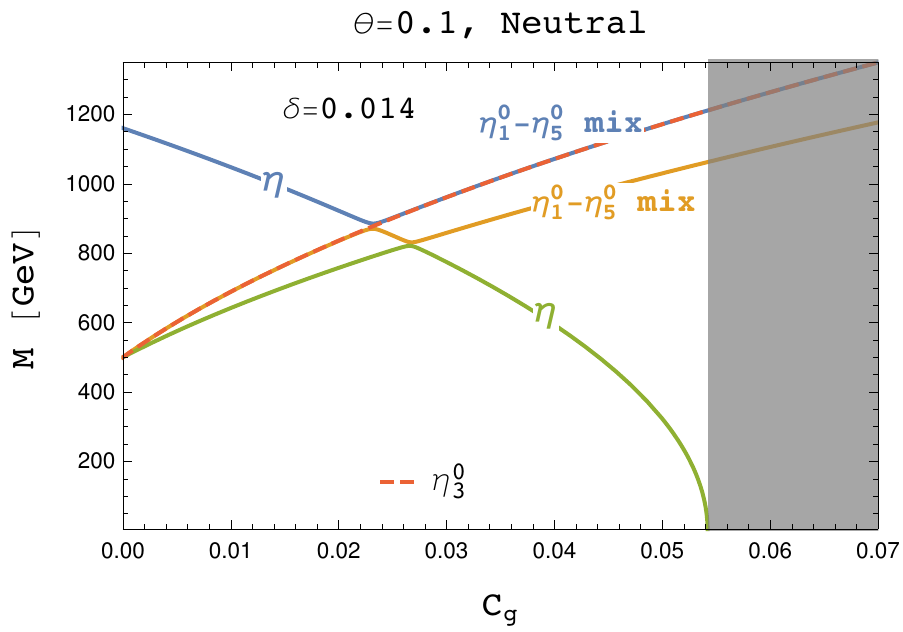}
\includegraphics[width=.45\textwidth]{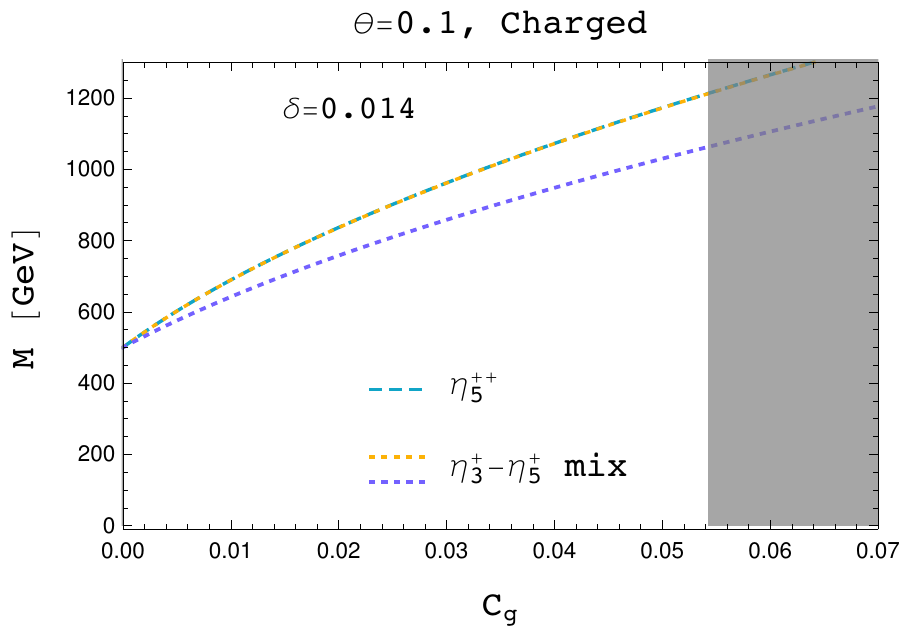} \\
\includegraphics[width=.45\textwidth]{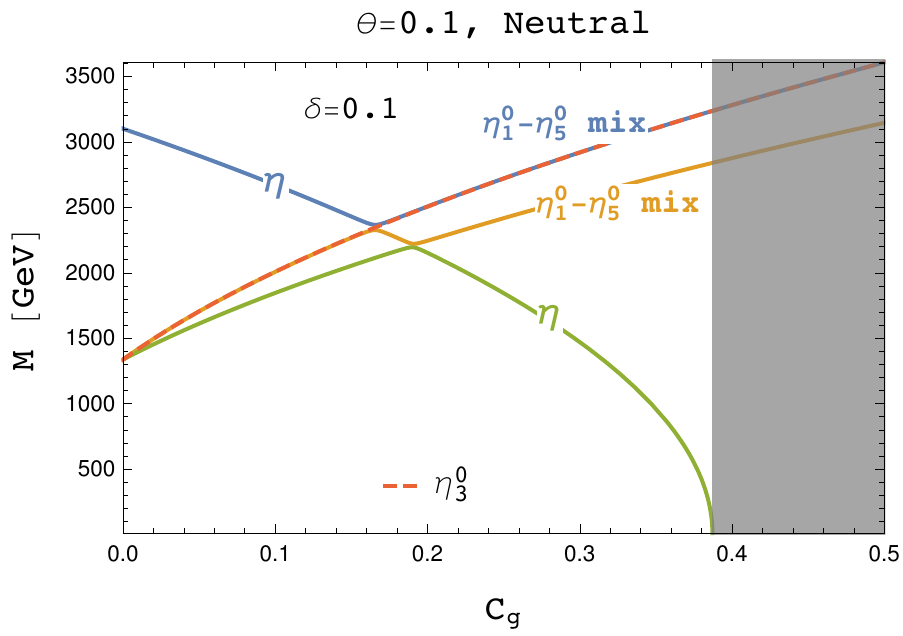} 
\includegraphics[width=.45\textwidth]{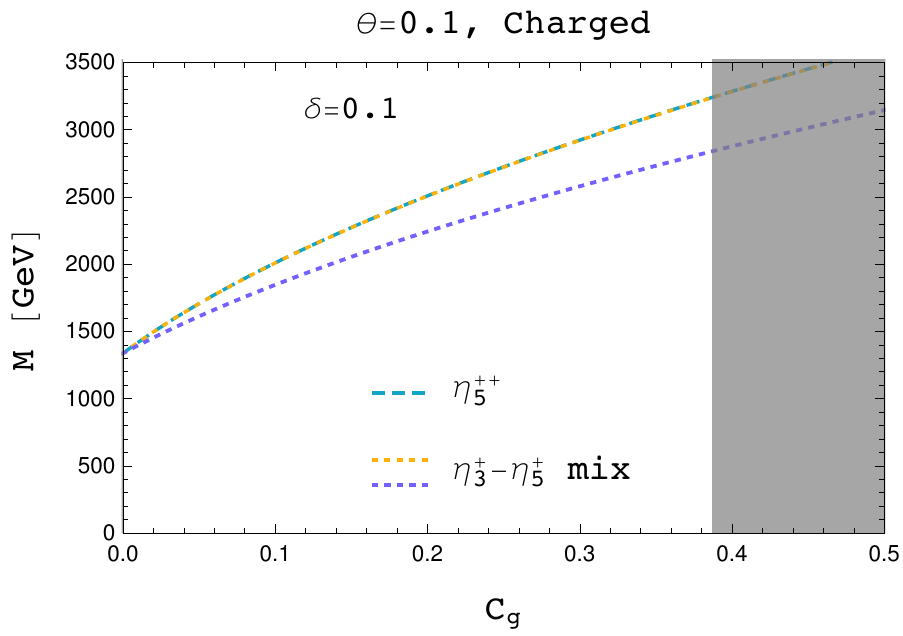} \\
\caption{
Masses of the neutral and charged pNGBs in the $A_L$--$D_R$ case. In the top panels $\mu = 10 \GeV$, and we plot the masses against $\delta$, while in the middle and bottom panels  we  fix $\delta = 0.014$ and  $0.1$, and plot the masses against $C_g$. Note that the pattern remains unchanged for different values of $\delta$.}
\label{fig:ALDR}
\end{figure}

In the former case, we first consider the dependence on the HF mass parameters $\mu$ and $\delta$. In the top row in Fig.~\ref{fig:ALDR} we show the masses as a function of $\delta$ for fixed $\mu=10$~GeV (thus, we solved for $C_R$--$C_g$). We see that the masses grow with $\delta$, quickly reaching the few TeV scale. Thus we can conclude that the pNGB masses tend to be larger than in the adjoint $D_L$--$D_R$ case, unless $\delta$ is close to the minimum allowed value for each $\mu$. Interestingly, the plots feature the same ordering and patterns as we saw in Fig.~\ref{fig:spectrumDD}.  The similarities are more clear when we plot the spectra as a function of $C_g$ for fixed value of $\delta$: the result is shown in the middle row of Fig.~\ref{fig:ALDR} for $\delta = 0.014$ and in the bottom one for $\delta = 0.1$. The value of $\delta=0.014$ has been chosen such that the value of the masses nearly coincide with those in the adjoint case in Fig.~\ref{fig:spectrumDD}. For different values of $\delta$, the pattern remains unchanged with the general rule that the allowed range of $C_g$ increases and the masses grow with growing $\delta$. 

\begin{figure}[!t]
\centering
\includegraphics[width=.45\textwidth]{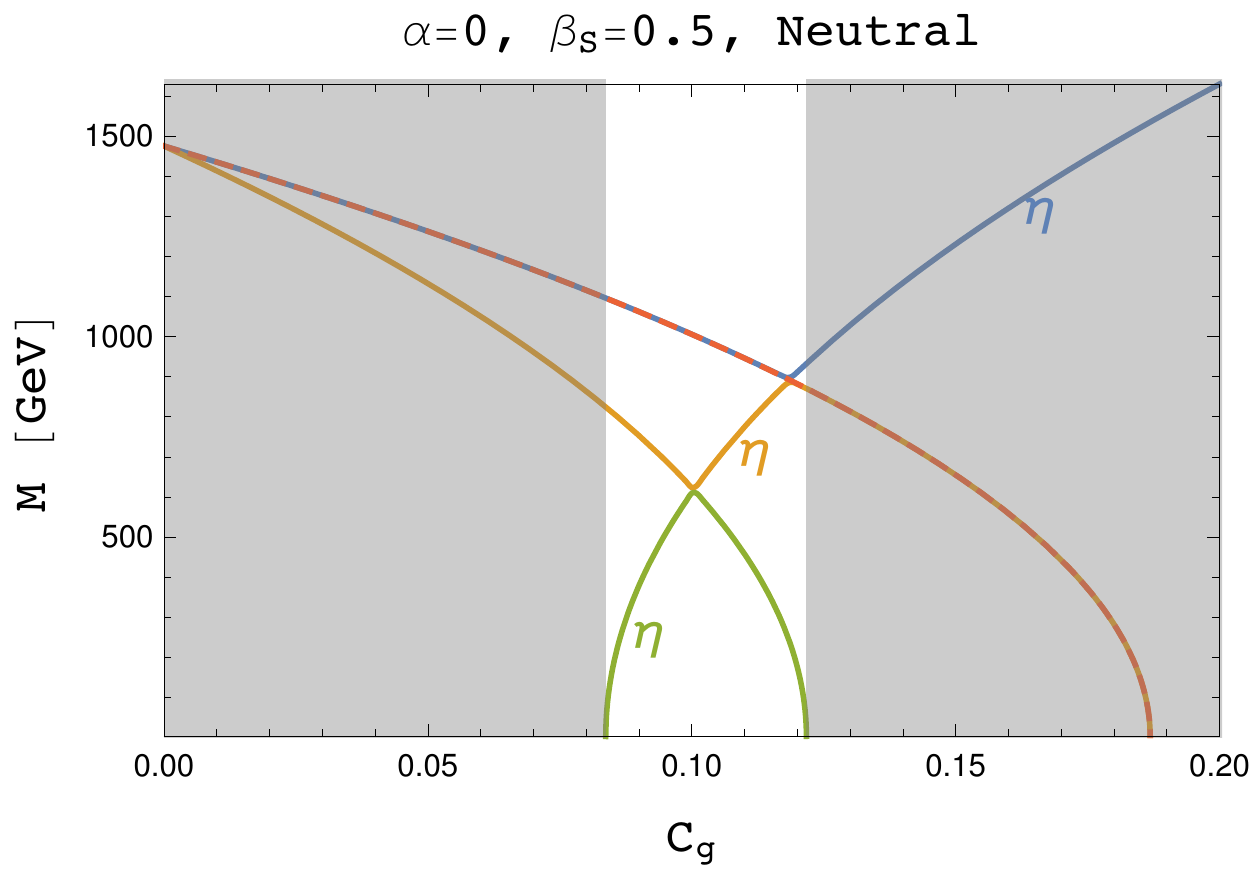}
\includegraphics[width=.45 \textwidth]{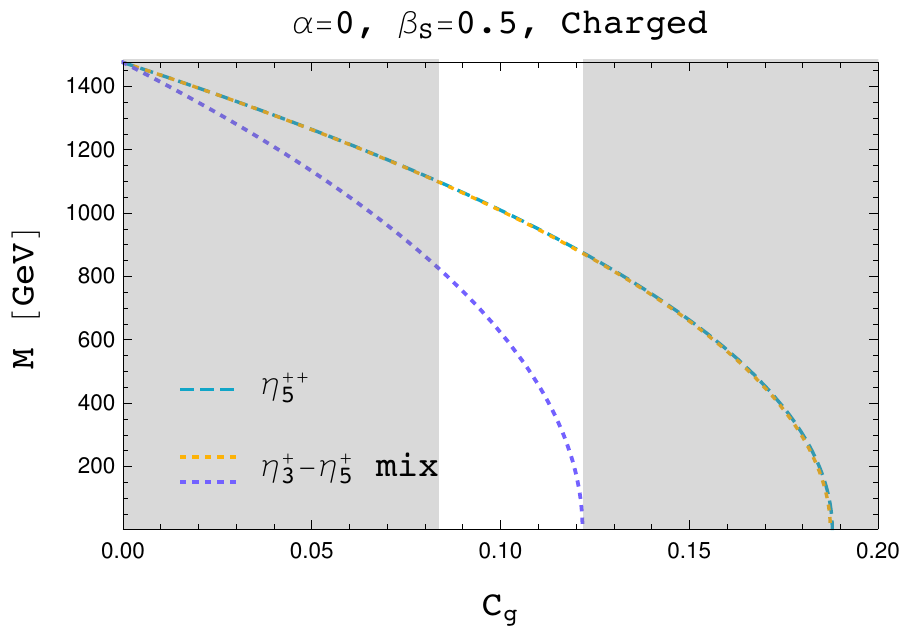} \\
\includegraphics[width=.45\textwidth]{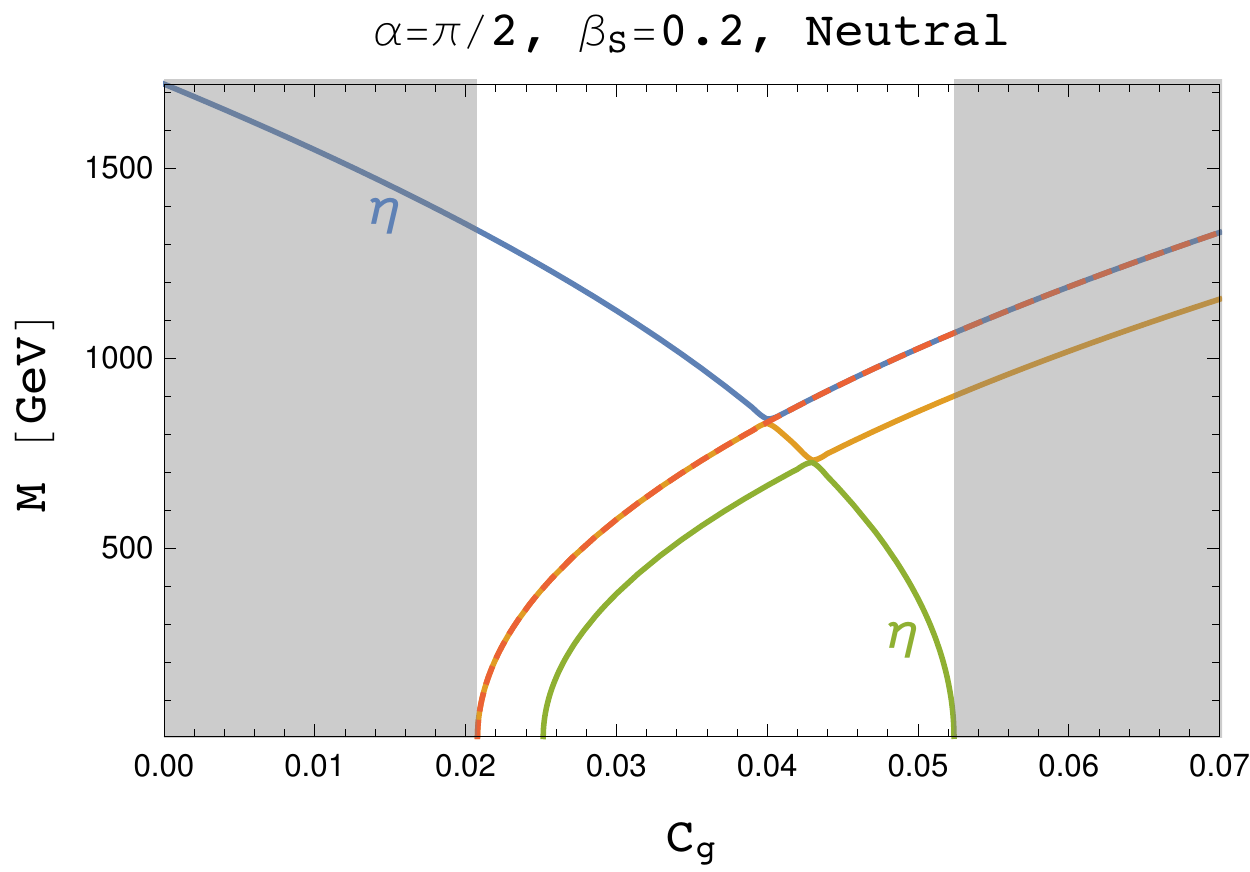}
\includegraphics[width=.45 \textwidth]{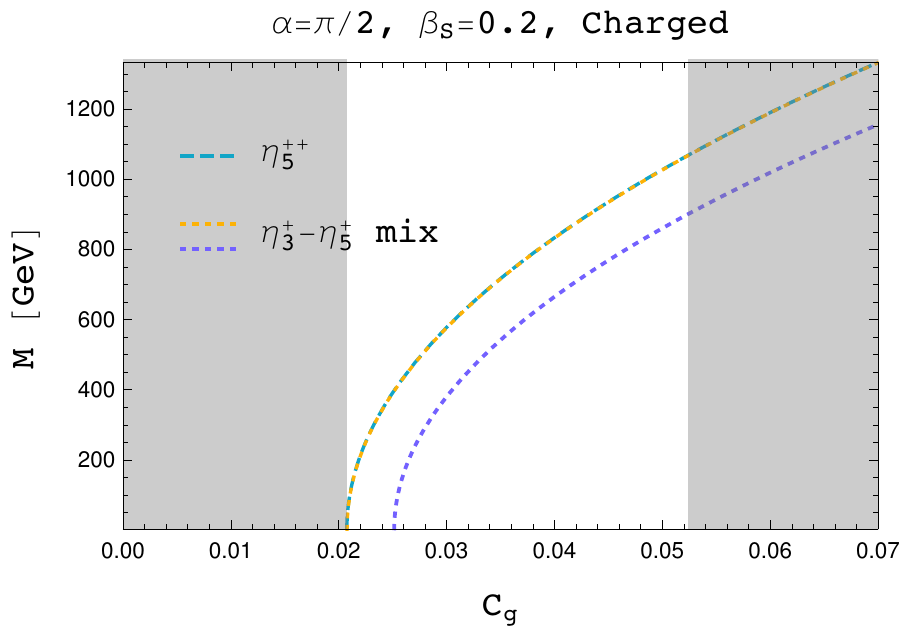}
\caption{Neutral and charged spectra in the $D_L$-$S_R$ case for two benchmark points corresponding to  $\alpha=0$ and $\beta_S=0.5$ (top row), and $\alpha=\pi/2$ and $\beta_S=0.2$ (bottom row). For the first benchmark we see that the role of the gauge corrections is inverted, i.e. the correction to the singlet increases for larger $C_g$ and the tachyon imposes a lower limit, while the triplets become tachyonic for large $C_g$. The second benchmark shows the same pattern as before, but shifted to a range of $C_g$ not starting at zero.}
\label{fig:benchDLSR1}
\end{figure}

A similar pattern of masses emerges in the $D_L$--$S_R$ case (for negligible HF masses). To test this, we fixed $\beta_S = -1.1$ (both for $\alpha=0$ and $\pi/2$) in order to match the value of the $\eta_3^0$ mass at $C_g=0$ with the adjoint case, and studied the spectrum as a function of $C_g$. We found that the mass differences between this case and the adjoint one in Fig.~\ref{fig:spectrumDD} are always of the order of $10^{-6} \GeV$ to $10^{-5} \GeV$ for both $\alpha = 0, \pi/2$. The two spectra, thus, perfectly overlap with each other.  We have verified that, with $\alpha=0$, for $\beta_S \lesssim-0.8$ the spectra are qualitatively unchanged, the overall effect being a shift of all the pNGBs along the $C_g$ axis. The same holds with $\alpha=\pi/2$, for $\beta_S \lesssim -0.4$. On the contrary, a qualitative change in the pattern of masses can be observed for some values of $\beta_S$, especially in the internal allowed (white) slices in Fig.~\ref{fig:DLSR} ($\alpha=0$), or for small positive values of $\beta_S$ ($\alpha=\pi/2$). Two such benchmark points are shown in Fig.~\ref{fig:benchDLSR1}, and are illustrated in the caption. We remark that there are cases where the corrections of the gauge loops to the $\eta$ and triplets have inverted signs, as it can be seen in the top row of the figure for a benchmark at $\alpha=0$.

To summarise, we found that the spectra of the pNGBs follow an approximately universal pattern, illustrated by Fig.~\ref{fig:spectrumDD} for the adjoint case, independently of the specific representations of the top partners. Also, in the $D_L$-$S_R$ case, for $\alpha=0$, we found an interesting alternative pattern, described by Fig.~\ref{fig:benchDLSR1}, where the gauge singlet mass increases with $C_g$, unlike it did in all the other cases.  This leads to a ``mirror image'' of the universal pattern, which retains its distinctive features.
 Such universality remains true for the cases where the potential is generated at LO, and in absence of tadpole for the custodial triplet. Nevertheless, some features of the model will not be the same: for instance, the couplings of the non-Higgs pNGBs to fermions (top and bottom) are different in the 3 models. This in turn will affect the phenomenology of the new states at the LHC, as we will discuss in Section~\ref{sec:pheno}, and allow to distinguish the various cases.

\section{Anti-symmetric: a case with NLO potential} \label{sec:antisym}

When both the left- and right-handed fermions are embedded into the anti-symmetric representation of $\SU(5)$ or its conjugate, we have shown that no potential is generated 
from the top sector at LO. Thus, no vacuum misalignment away from the extremes $\theta = 0, \pi/2$ can be achieved. It is therefore needed to include the effect 
of the top spurions at  NLO. 
This has a double advantage. On the one hand, the violation of NDA that we observed in the cases with LO operators, such as in Eq.~\eqref{eq:ratioDD}, is eased because now the coefficients in the potential are
 naturally suppressed compared to the top mass ones. On the other hand, more operators and structures arise, thus giving us more handles that can help misaligning the vacuum. We 
stress here that this advantageous situation is only obtained for the anti-symmetric representation of both spurions. 
At NLO, the operator coefficients scale with the fourth power of the pre-Yukawas, and one can identify three classes of operators scaling as 
$y_L^4$, $y_R^4$ and $y_L^2 y_R^2$. We will also include in the analysis the LO operators from gauge and fermion masses: this is justified, numerically, 
by the fact that the pre-Yukawas are needed to be fairly large in order to obtain the correct top mass, thus they will enhance the contribution of the 
top operators. We will, thus, discard NLO operators containing gauge couplings and mass insertions, including the ones belonging to the classes 
$g^2\, y_{L,R}^2$ and $\mu_{d/s}\,  y_{L,R}^2$, where $g$ is a generic gauge coupling. A complete list of NLO operators can be found in Ref.~\cite{Alanne:2018wtp}, from which 
we extract the following 8 independent operators:
\begin{subequations}
\begin{align} \label{eq:NLO}
 \text{LL}: \hspace{.7cm} & \mathcal O_1 ^{LL} =  \Tr[A_L^i \Sigma^\dagger A_L^j \Sigma^\dagger] \Tr[\Sigma \bar A_L ^i \Sigma \bar A_L ^j] \,,  \\
& \mathcal O_2 ^{LL}  =  \Tr[A_L ^i \Sigma^\dagger A_L ^j  \bar  A_{L,i}  \Sigma \bar A_{L,j}  ]\,, \\
& \mathcal O_3 ^{LL}  =  \Tr[A_L ^i \Sigma^\dagger A_L ^j  \bar A_{L,j}  \Sigma \bar A_{L,i}  ]   \\
 \text{RR}: \hspace{.7cm} & \mathcal O_1 ^{RR} =  \Tr[A_R  \Sigma ^\dagger A_R \Sigma^\dagger] \Tr[\Sigma \bar A_R \Sigma \bar A_R]  
\, , \\
& \mathcal O_2 ^{RR} =  \Tr[A_R  \Sigma^\dagger A_R  \bar A_R  \Sigma \bar A_R  ]  \\
 \text{LR}: \hspace{.7cm} &  \mathcal O_1^{LR} = \Tr[ A_{L} ^i \Sigma^\dagger A_R\Sigma^\dagger] \Tr[\bar A_L ^i \Sigma \bar A_R \Sigma] \,,  \\
 & \mathcal O _2^{LR} = \Tr[\bar A_{L} ^i \Sigma \bar A_R  A_L ^i \Sigma^\dagger A_R ] \,, \\
 & \mathcal O_3 ^{LR} =   \Tr[\bar A_L ^i A_L ^i \Sigma^\dagger A_R \bar A_R \Sigma] \,.
\end{align}
\end{subequations}
The resulting potential originating from the top quark contribution is:
\begin{align} \label{potAA}
V_{\text{top}} =\frac{1}{(4\pi)^2}\sum_{i} \left( C_{LL,i} \mathcal O_i ^{LL}+C_{RR,i} \mathcal O_i ^{RR}+C_{LR,i} \mathcal O_i ^{LR}\right)\,,
\end{align}
and, setting the fields to zero, the $\theta$ dependence of the potential is still the same  as in Eq.~(\ref{eq:Vtheta}) with:
\begin{align}
A&=  -\frac{1}{8(4 \pi)^2 } \left( 4 C_{LL,1} + C_{LL,2} + C_{LL,3} - 4 C_{RR,1}\right) \,,  \\
 B &= \frac{1}{32(4 \pi)^2 }\left( 4 C_{LL,1} + 5 C_{LL,2} + 9 C_{LL,3} - 4 C_{LR,1} - C_{LR,2} - 3 C_{LR,3} + 4 C_{RR,1} +C_{RR,2}\right) \,. \nonumber
\end{align}
An important point is that the tadpole for the $\eta_3^0$ field does not vanish, and is given by:
\begin{align}\label{eq:tadpole2}
V_{\text{top }}\ \supset  \  \eta_3^0\,\frac{f^3 \ct \, \st^2}{2(4\pi)^2}\left(- 8 \,C_{LL,1} - 2\, C_{LL,2} -6\, C_{LL,3} + 4 \,C_{LR,1} +C_{LR,2} + C_{LR,3}  \right)\,.  
\end{align}
In order to avoid the breaking of custodial invariance, therefore, a cancellation between the coefficients of the operators in the classes 
$y_L^4$ and $y_L^2 y_R^2$ needs to occur. The other possibility, that we will investigate in  Section~\ref{sec:vacuum2}, 
is that the vacuum is finally misaligned along the custodial triplet thus violating custodial invariance.

In the remaining of this section, for simplicity of notation, we replace the 8 coefficients of the top operators with five pre-Yukawa independent ratios, 
$\alpha$, $\beta_{1,2}$ and $\gamma_{1,2}$, defined as
\begin{equation} \label{eq:beta}
\alpha = \frac{C_{RR,2}}{C_{RR,1}}\,, \quad \beta_1 = \frac{C_{LL,2}}{C_{LL,1}}\,, \quad \beta_2 = \frac{C_{LL,3}}{C_{LL,1}}\,, \quad \gamma_1 = \frac{C_{LR,2}}{C_{LR,1}} \,, \quad \gamma_2 = \frac{C_{LR,3}}{C_{LR,1}}\,,
\end{equation}
and two ratios that scale with the pre-Yukawas
\be
\xLL = \frac{C_{LL}}{C_{RR}} \propto \frac{y_L^4}{y_R^4} \,, \quad \xLR = \frac{C_{LR}}{C_{RR}} \propto \frac{y_L^2}{y_R^2}\,,
\ee
and leave $C_{RR} \equiv C_{RR,1}$ as an overall coefficient.
Note that the first 5 parameters are of the same type of the pre-Yukawa independent ratios defined in Eq.~\eqref{eq:ratios}, which we expect to be of order one,  following NDA. Two more pre-Yukawa independent ratios can be defined, namely
\begin{equation}\label{eq:ratioAA}
R_{V}  \ =  \ \frac{ \lvert C_{LR} \rvert}{\sqrt{ \lvert C_{LL} C_{RR} \rvert}} \ = \  \frac{\xLR}{\sqrt{\lvert \xLL \rvert}}\,, \qquad
 R_t\  =   \ \frac{C_t^2}{\sqrt{\lvert C_{LL} C_{RR} \rvert}} \ = \frac{C_t ^2}{\sqrt{ \lvert \xLL \rvert } C_{RR}} \,, 
\end{equation}
where $\displaystyle C_t = \frac{m_t}{v \ct}$ in this case.

In general, to find the correct minimum, we need to ensure the vanishing of the triplet tadpole \eqref{eq:tadpole2}, the minimisation of the $\theta$--potential and match the value of the Higgs mass to the measured value. This allows to fix 3 parameters as a function of the misalignment angle $\theta$.
Let us  first discuss the vanishing of the tadpole. One possible way to achieve it is to tune the values of the pre-Yukawas, leading to the following relation between the ratios defined above:
\begin{align} \label{eq:ratioxLLxLR}
\frac{\xLL}{\xLR }&\ = \ \frac{\gamma _1+\gamma _2+4}{2 \beta _1+6 \beta _2+8}  \ \equiv  \ \frac{1}{\delta}  \,,
\end{align}
from which it follows
\be \label{eq:RV}
R_V =\delta \,  \sqrt{\lvert \xLL \rvert}\,.
\ee
Thus a value of $R_V$ natural in terms of NDA can be easily obtained. Note that this solution requires that the contributions of two classes of operators 
compensate each other by tuning parameters external to the strong dynamics, i.e. the pre-Yukawas.

Another interesting possibility is that the contributions of each class of operators vanish individually, thus leading to the relations below:
\be \label{weak}
\beta_1 + 3 \beta_2 = - 4\,, \quad \gamma_1 + \gamma_ 2 = -4 \,.
\ee
Note that this solution is qualitatively different from the one above, as the cancellation would be due uniquely to the strong dynamics. We also remark that the latter choice cannot be described by Eq.~\eqref{eq:ratioxLLxLR} as the ratio $\delta$ would be undetermined.

\subsection{Spectrum in the tuned case} \label{spectrum}

We now attempt to characterise the spectrum of the pNGBs when the triplet tadpole is tuned away by the condition in Eq.~\eqref{eq:ratioxLLxLR}. 
We complement the top potential by adding the contribution of the fermion masses, while neglecting the contribution of the gauge loops.
As already mentioned, the conditions of the minimum and the Higgs mass allow to determine 3 parameters in terms of $\theta$ 
(we choose  $x_{LR}$, $C_{RR}$ and $\gamma_1$), but still the spectrum will depend on too many free parameters to allow for a 
complete analytical analysis:
\be \label{eq:params}
\alpha, \ \beta_1, \ \beta_2, \ \gamma_2 ,\ \xLL , \ \mu_d, \ \mu_s, 
\ee
plus the angle $\theta$. 
In the next Section~\ref{sec:scan} we will probe the full parameter space by employing a numerical scan.
Here, we will focus on a special case, i.e. the limit where the masses are approximately custodial 
invariant. In practice, this means that all the states in the same multiplet of the custodial $\SU(2)_C$ are (nearly) degenerate. This limit is beneficial as 
it ensures that loop corrections from the pNGBs to electroweak precision tests remain small~\cite{Gunion:1990dt,Englert:2013zpa,Blasi:2017xmc}.

\begin{figure}[h]
\centering
\includegraphics[width=.45\textwidth]{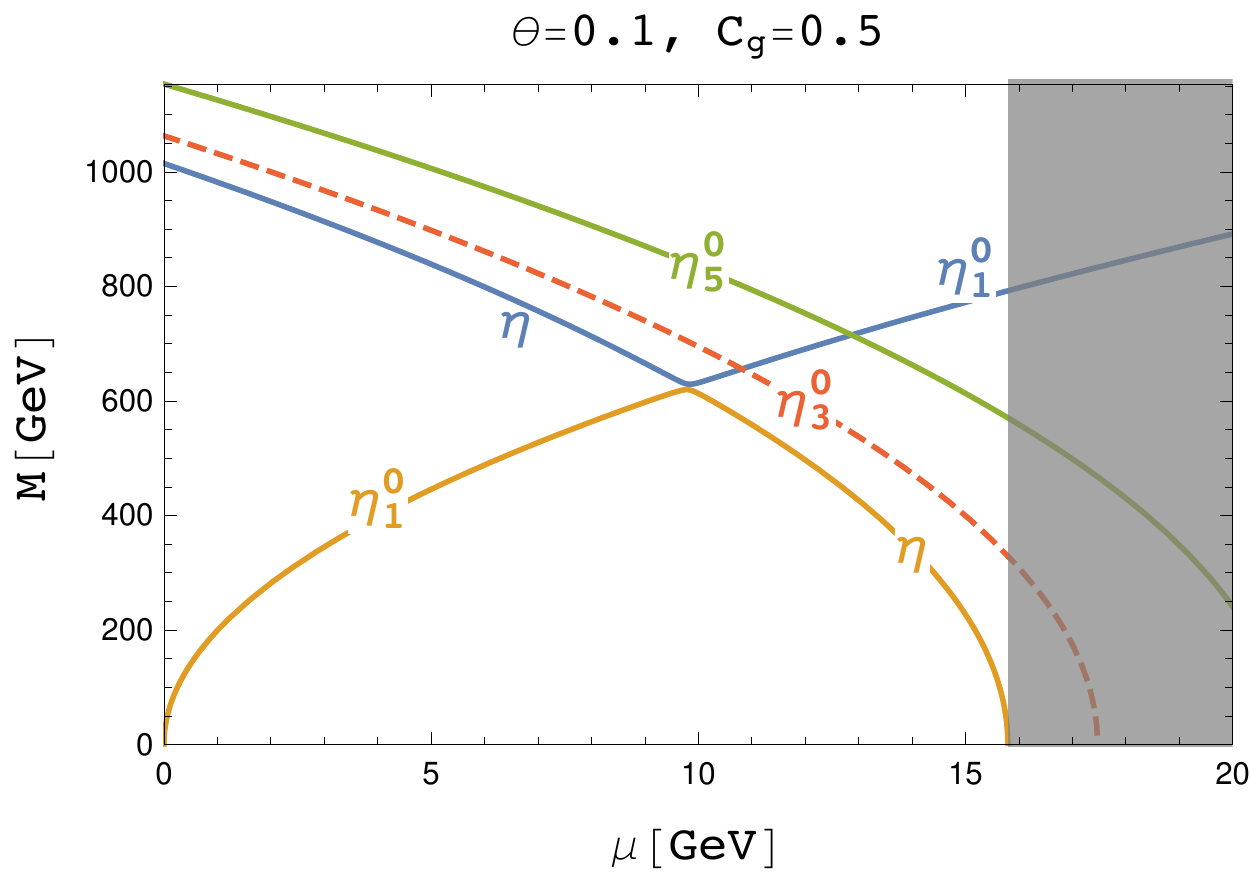}
\includegraphics[width=.45\textwidth]{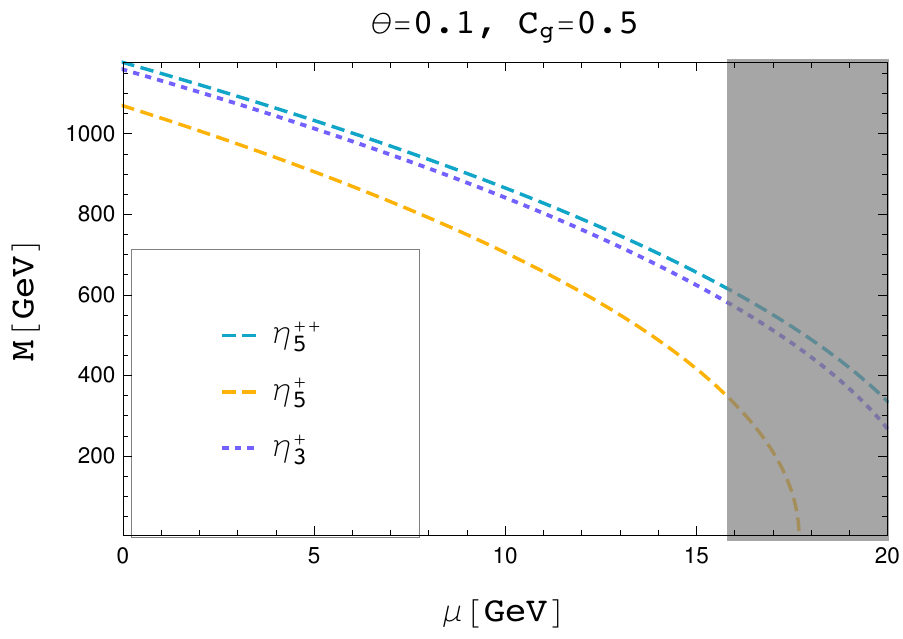} \\
\includegraphics[width=.45\textwidth]{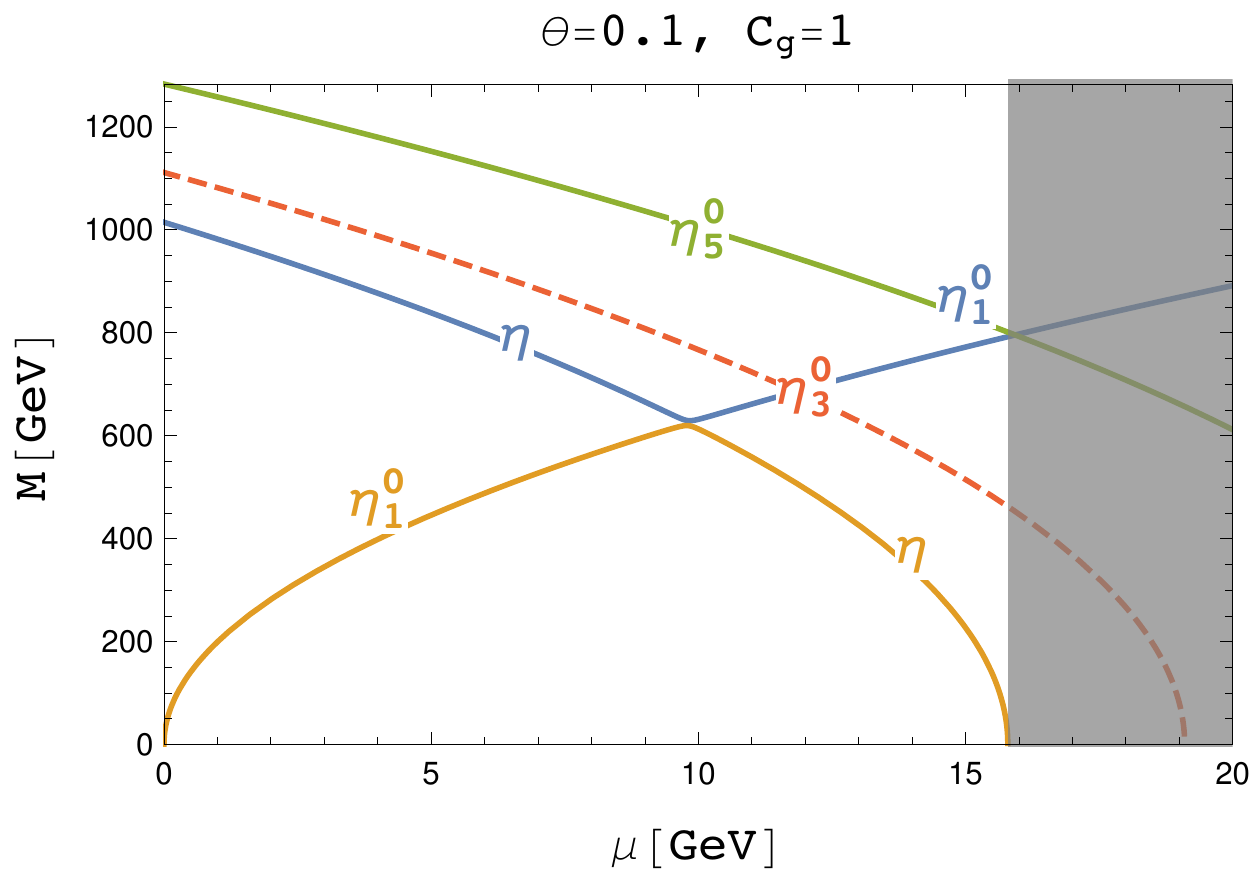}
\includegraphics[width=.45\textwidth]{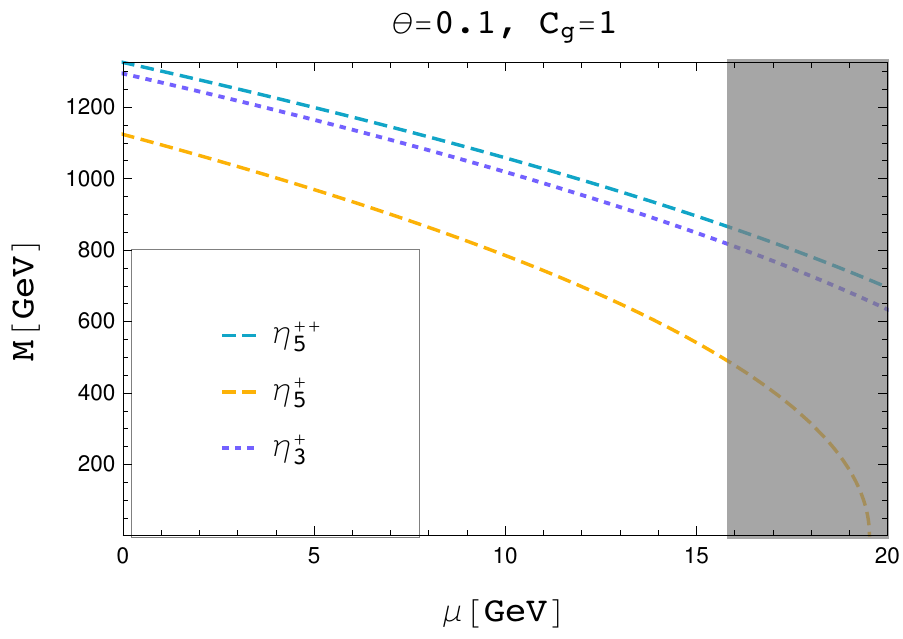}
\caption{Plots of the $A_L$-$A_R$ spectra at NLO, in the `custodial'  case described in the text, including the gauge term of Eq.\eqref{eq:vgauge}. Note that the triplet and five-plet are squeezed towards the singlet as $C_g \to 0$, while the masses of the gauge and custodial singlets, respectively $\eta$ and $\eta_1^0$, do not depend on $C_g$ in this custodial scenario.}
\label{fig:CustPlot}
\end{figure}

In order to find the custodial limit, we first impose that the mixing between states in different multiplets vanishes. This translates into the following 
two different conditions for the neutral and charged sectors, respectively: \footnote{For simplicity, in the explicit computation presented here, we neglect the gauge contribution \eqref{eq:vgauge}. However, we will consistently take it into account in the spectrum presented in Fig.~\ref{fig:CustPlot}. Although the presence of $C_g \neq 0$ leads to a modification of Eq.\eqref{custodial}, the main features that will be outlined in the following remain unchanged.}
\begin{align} \label{custodial}
\text{Neutral}& \quad \sim  \hspace{.5cm} \left(4C_{LL,1}+C_{LL,2} + C_{LL,3}\right) + 8\, C_{RR,1}\,, \nonumber \\
\text{Charged}& \quad \sim  \hspace{.5cm}  \left(4C_{LL,1}+C_{LL,2} + C_{LL,3}\right) + 8\, C_{RR,1}- 8 (C_{RR,1}+C_{LL,1}) \st^2 \,.
\end{align}
We see that the difference between the two conditions is a term proportional to $\st^2$, so if we set to zero one of them, the custodial violation in the other sector will be small for small $\st$.
We thus choose to enforce custodial invariance in the neutral spectrum by implementing the following condition, in terms of the ratios defined in Eq.~\eqref{eq:beta}:
 \be \label{eq:cust1}
 \xLL=-\frac{8}{4+ \beta_1 + \beta_2}\,, \quad \mbox{with} \;\; C_{RR,1} + C_{LL,1} = C_{RR} \frac{-4 + \beta_1 + \beta_2}{4 + \beta_1 + \beta_2}\,.
 \ee
Remarkably, the above condition can be achieved by tuning the value of the pre-Yukawa $y_L$.
 
An interesting fact is that, once \eqref{eq:cust1} is implemented, the neutral spectrum does not depend anymore on $\beta_1$ and $\beta_2$. Also, 
$\gamma_2$ does not appear \footnote{We recall that $\gamma_1$ has already been eliminated.} and therefore the neutral spectrum only depends 
on the left-over parameters $\mu_{d,s}$ and $\theta$ ($C_m$, introduced in Eq.~\eqref{eq:cm}, always multiplies the masses $\mu_{d,s}$, thus it can be reabsorbed in their definition).

The masses of the neutral components of the triplet and five-plet are respectively given by
\begin{align} \label{eq:triplet}
 m_{\eta_3 ^0}^2\  &= \ \frac{2}{3} \frac{m_h^2}{\st^2} \ctt - \frac{16 C_m \, v}{3 \st} \left( \mu_d + 4 \mu_s - 4 (\mu_d+ \mu_s)  \st^2 \right) \,, \\ 
 \label{eq:fiveplet}
 m_{\eta_5^0} ^2  \ &= \ \frac{2}{3} \frac{m_h^2}{\st^2}  \frac{\ctt^2}{\ct^2}\,  - \frac{16 C_m v}{3\st}\, \left( \mu_d + 4 \mu_s - 8(\mu_d+\mu_s) \st^2\right)\,.
\end{align} 
As already mentioned, the dependence on $\beta_{1,2}$ has completely dropped out, furthermore the two masses are equal up to corrections of order $\st^2$.
The remaining two neutral states $\eta_1^0$-$\eta$, being both custodial singlets, have a residual mixing. The mass matrix is given by
\begin{multline}\label{eq:mix}
M^2_{\eta_1^0 - \eta}\ =  \frac{m_h^2}{\st^2 \ct^2}  \left(
\begin{array}{cc}
 \frac{2}{3} \left( 1-\frac{5}{4} \st^2 \right)^2 & - \frac{1}{2} \sqrt{\frac{5}{3}} \left( 1-\frac{5}{4} \st^2 \right) \st^2 \\
- \frac{1}{2} \sqrt{\frac{5}{3}} \left( 1-\frac{5}{4} \st^2 \right) \st^2 & \frac{5}{8} \st^4 \\
\end{array}
\right) + \\ \frac{16 C_m v}{3 \st}  \left(
\begin{array}{cc}
 -(\mu_d + 4 \mu_s + (\mu_d + \mu_s) \st^2)  &  2 \sqrt{\frac{3}{5}} (\mu_d + \mu_s) \st^2 \\
2 \sqrt{\frac{3}{5}} (\mu_d + \mu_s) \st^2 & \frac{3}{5} (\mu_d + 4 \mu_s + (\mu_d + \mu_s) \st^2 )\\
\end{array}
\right)\,.
\end{multline}
We note that the first term, proportional to the Higgs mass, has zero determinant, meaning that a massless eigenvalue is present in the spectrum for 
vanishing $\mu_d = \mu_s = 0$, and therefore non-vanishing HF masses are necessary in order to avoid it. We also recognise the same 
pattern seen for the other two masses: the mixing is suppressed by $\st^2$ and, in the small $\theta$ limit, the mass of $\eta_1^0$ approaches the mass 
of the other two neutral states, while the singlet $\eta$ mass only comes from the HF masses.

Among the charged pNGBs, the doubly charged state $\eta_5^{++}$ mass reads:
\be \label{eq:fiveplus}
m_{\eta_5 ^{++}}^2 \  = \ \frac{2 m_h^2}{3 \st^2} \frac{\ctt}{\ct^2} \left(1 - \frac{ 2 \,\stt^2}{4 +\beta_1+\beta_2} \right) - \frac{16 C_m\, v}{3\st} \bigg( \mu_d 
+ 4\mu_s  - \frac{16 (\mu_d + \mu_s) \stt^2}{4+\beta_1 + \beta_2} \bigg)\,.
\ee
Contrary to the neutral case, it depends on $\beta_{1,2}$, nevertheless it coincides with the mass of the neutral component for small $\st$. 
The system of the two singly-charged states, $\eta_3^+$ and $\eta_5^+$, is more complex due to the residual mixing that arises at order $\st^2$. 
The mass matrix is given by
\begin{multline}\label{eq:mix}
M^2_{\eta_3^\pm - \eta_5^\pm}\ =  \frac{m_h^2}{\st^2 \ct^2}  \left(
\begin{array}{cc}
 \frac{2}{3} \left( 1- \frac{12 + \beta_1 + \beta_2}{2(4 + \beta_1 + \beta_2)} \st^2 \right) & \quad \frac{-4+\beta_1+\beta_2}{3(4+\beta_1+\beta_2)} \ctt \ct \st^2 \\
 \frac{-4+\beta_1+\beta_2}{3(4+\beta_1+\beta_2)} \ctt \ct \st^2 &  \frac{2}{3} \left( 1- \frac{12 + \beta_1 + \beta_2}{2(4 + \beta_1 + \beta_2)} \st^2 \ct^2 \right) \\
\end{array}
\right) -
\frac{16 C_m v}{3 \st} (\mu_d + 4 \mu_s)\, \mathbb{1}_2  + \\
\frac{32 C_m v}{3 \st}   (\mu_d + \mu_s) \st^2 \left(
\begin{array}{cc}
 \frac{12+\beta_1 + \beta_2}{4+\beta_1 + \beta_2}   & \quad -\frac{-4+\beta_1+\beta_2}{3(4+\beta_1+\beta_2)}  \ct \\
 -\frac{-4+\beta_1+\beta_2}{3(4+\beta_1+\beta_2)} \ct  &  \frac{(20+3\beta_1 + 3\beta_2) \st - (12+\beta_1 + \beta_2) \stt}{4+\beta_1 + \beta_2}   \\
\end{array}
\right) 
\,.
\end{multline}
We see that the off-diagonal terms are suppressed by $\st^2$ and consistent with what we found in Eq.~\eqref{eq:cust1}, and that for small  $\st$ 
the masses are degenerate with the other corresponding states.

At the leading order in  $\st$, all the pNGB masses only depend on the combination  $\mu_d + 4 \mu_s$, while the dependence on 
 $\mu_s + \mu_d$ and on $\beta_1 + \beta_2$ arises at order $\st^2$.
Thus, to study the spectrum it is enough to study the dependence on the masses while fixing the value of their ratio, $\mu_d/\mu_s$.
For instance, the results for $\mu_d = \mu_s = \mu$ ($\SO(5)$ 
invariant case) is shown in Figure~\ref{fig:CustPlot}, where we also fixed $\beta_1 + \beta_2 =4$, corresponding to the limit where, for $C_g=0$ and small $\theta$, both the triplet and the
five-plet are exactly degenerate multiplets (i.e., custodial symmetry is fully restored at the quadratic order in the Lagrangian). 
Similarly to previous cases, we see that pNGB masses below $1$ TeV are always present, even though the compositeness scale is high, $f \approx 2.4$~TeV. Furthermore, the five-plet tends to be the heaviest multiplet.

\subsection{Numerical scan of the parameter space}\label{sec:scan}

Since there are in total 7 free parameters, see Eq.~\eqref{eq:params}, a good way to visualise the most general spectrum is to perform a scan over a relevant region 
of parameter space, i.e. over a region where the theory makes consistent predictions. The only requirement of consistency that we impose is the absence of negative-valued mass squared (tachyons) appearing in the spectrum. We have already shown in Section \ref{spectrum}, in a simplified case, that some states 
receive a negative contribution to their mass squared from  $\mu_{d}$ or $\mu_s$, thus tachyons should appear when they exceed some specific value. 
It is not easy, however, to identify such region because of the mixing and the large number of free parameters.
We will also keep track of the value of the Yukawa-independent ratios defined in Eq.~\eqref{eq:ratioAA}, as they indicate departure from NDA when they 
are much larger or smaller than unity. This is not {\it per se} an issue of consistency of the theory, as some form factors may be smaller than what naively 
expected due to the symmetries of the strong sector.

For the scan, we have fixed $\theta = 0.1$, and generated a random sample of $10^4$ points in the parameter space, with the parameters varying 
within the following ranges:
\begin{align*}
0.1 \ &< \ \alpha < 1\, , \quad
0.1 \ < \ \beta_{1,2}  \ < 1 \, , \quad
0.1 \ < \ \gamma_2 \  < 1 \,,  \\
0 \ & < \ \mu_d  \ < 40~\mbox{GeV}  \, , \quad 
0 \  < \ \mu_s  \ < 20~\mbox{GeV} \,, \quad
-10 \  < \ \xLL  \ < 10\,,
\end{align*}
where we absorb $C_m$ in the value of the masses (i.e., fix $C_m = 1)$. Moreover, the parameter $C_g$ has been set to zero:  we have verified that non-tachyonic states are almost all found in a region where $C_g \ll 1$ (and positive), while the other parameters vary within the intervals specified above.

\begin{figure}[t]
\centering
\includegraphics[width=.9\textwidth]{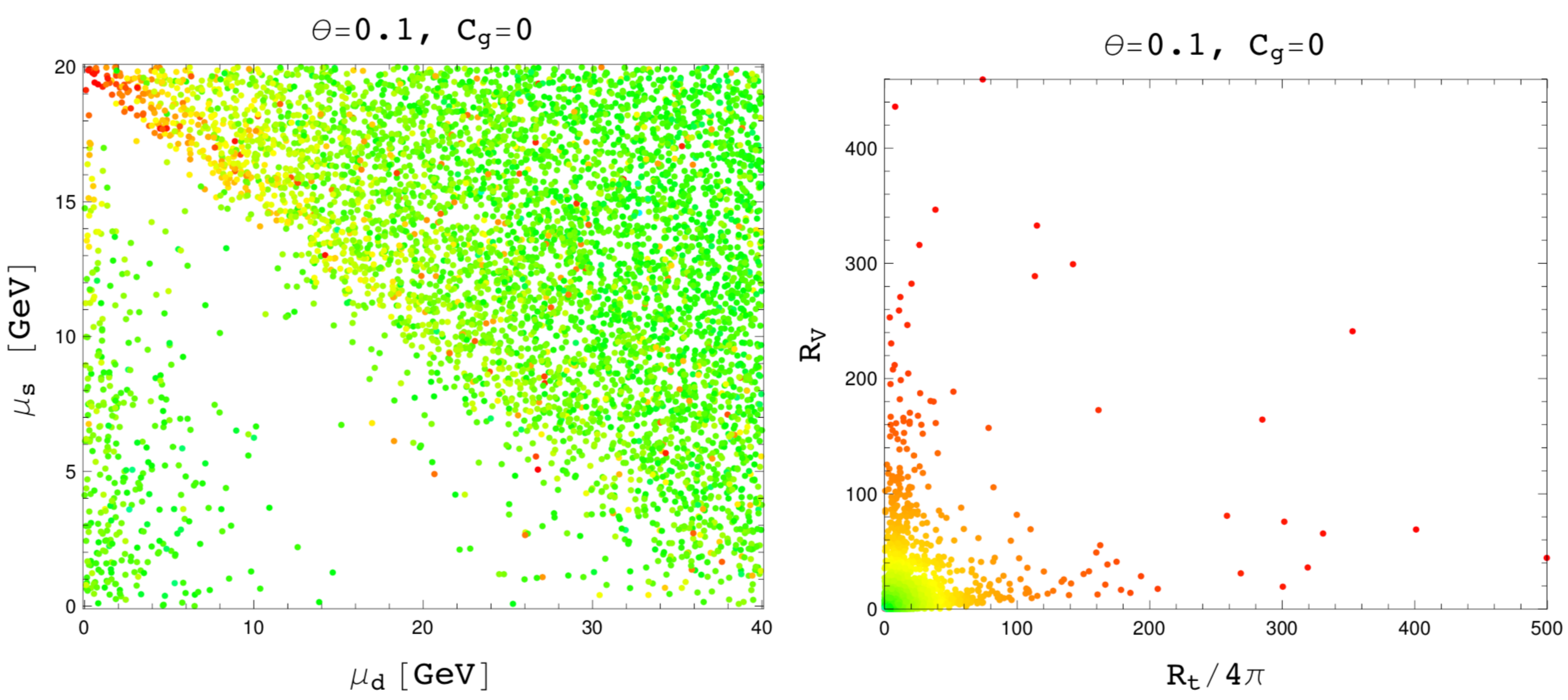}\\
\includegraphics[width=.9\textwidth]{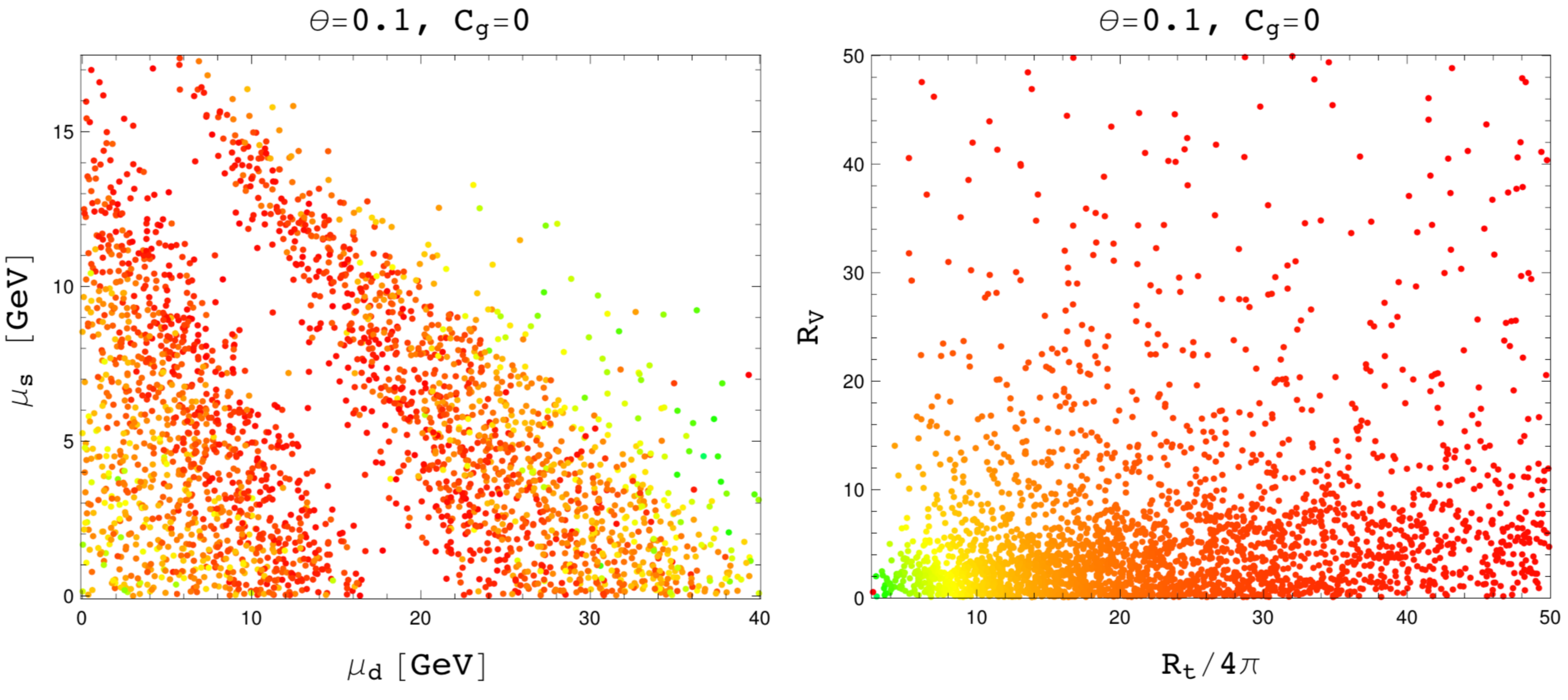}
\caption{\emph{Upper plots}: points corresponding to choices of parameters that yield at least one tachyon in the spectrum (left), with the corresponding values of $R_V$ and $R_t/(4\pi)$ (right).\\
\emph{Lower plots}: regions allowed by the no-tachyon requirement. The colours correspond to different values of $\sqrt{(R_t/(4\pi))^2+ R_V^2}$. We have also required that $R_V < 50$ and $R_t/(4\pi) < 50$.}
\label{fig:tachyons}
\end{figure}

In the upper panels of  Fig.~\ref{fig:tachyons}, we show the distribution of 
points corresponding to spectra with tachyons. The colours correspond to different values of the ratios $R_V$ and $R_t$, defined in Eq.~\eqref{eq:ratioAA}, which are shown in the right panels. We remark 
that most of the tachyonic points lie above a line in the $\mu_d$--$\mu_s$ parameter space, roughly given by $\mu_d + 2 \mu_s = 40$~GeV, and near $\mu_d \sim 0$.

The points corresponding to spectra that are free of tachyons are shown in the lower panels of Fig.~\ref{fig:tachyons}, where we also restricted 
\begin{eqnarray} \label{eq:bound}
0.1 \ < & \ R_V \ &< 50 \,, \nonumber \\
0.1 \ < & \ R_t/(4\pi) \ & < 50   \,.
\end{eqnarray}
The colours discriminate points depending on how close they are to small ratios, or more precisely we give the same colours to points with the same value of 
$\sqrt{R_V^2 + (R_t/(4\pi))^2}$. We see that, as expected, most of the points fill the triangle below the line we identified before. The white band 
that is missing in the middle, however, is not ruled out by tachyons but by our requirement of small ratios, i.e. it's a region where NDA is 
significantly violated. In fact most points are red around this band.

\begin{figure}[!htb]
\centering
\includegraphics[width=.99\textwidth]{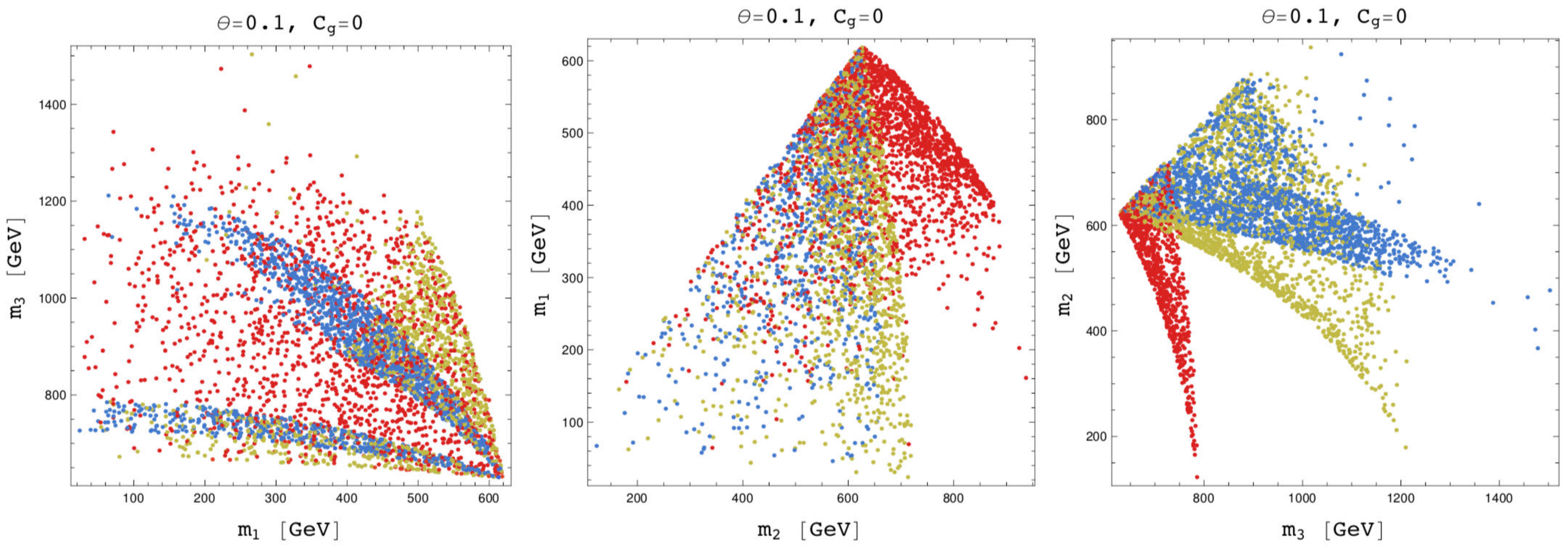} \\
\includegraphics[width=.99\textwidth]{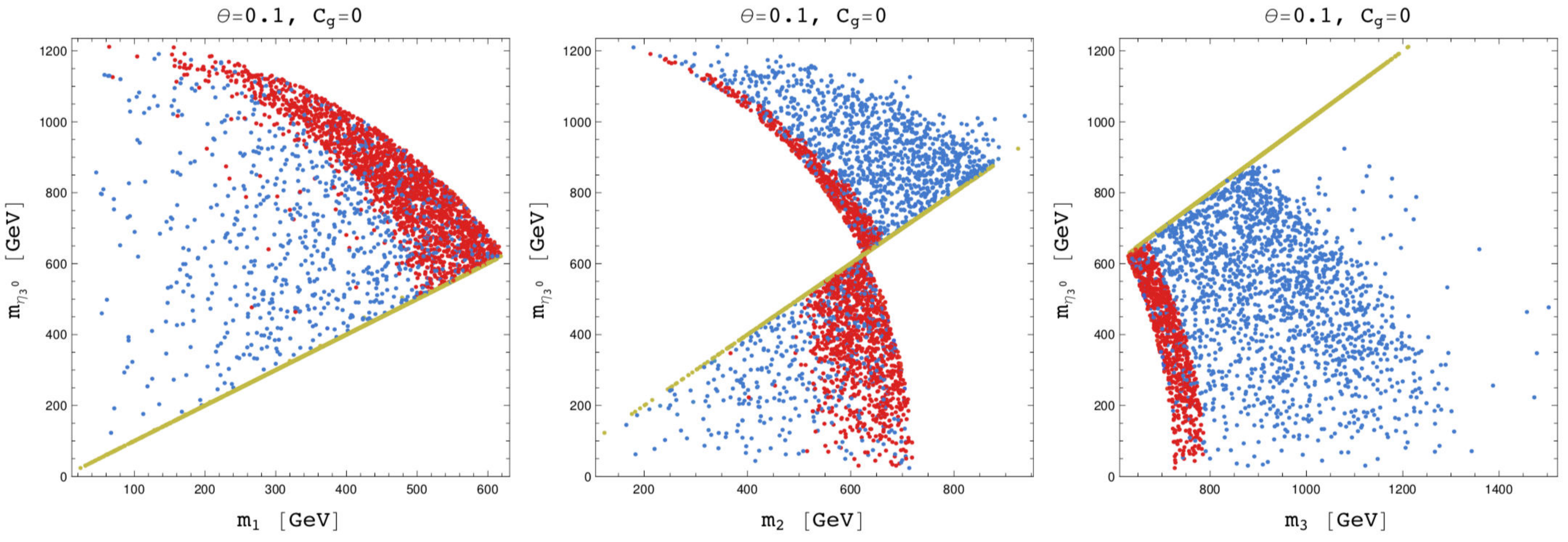}\\
\includegraphics[width=.99\textwidth]{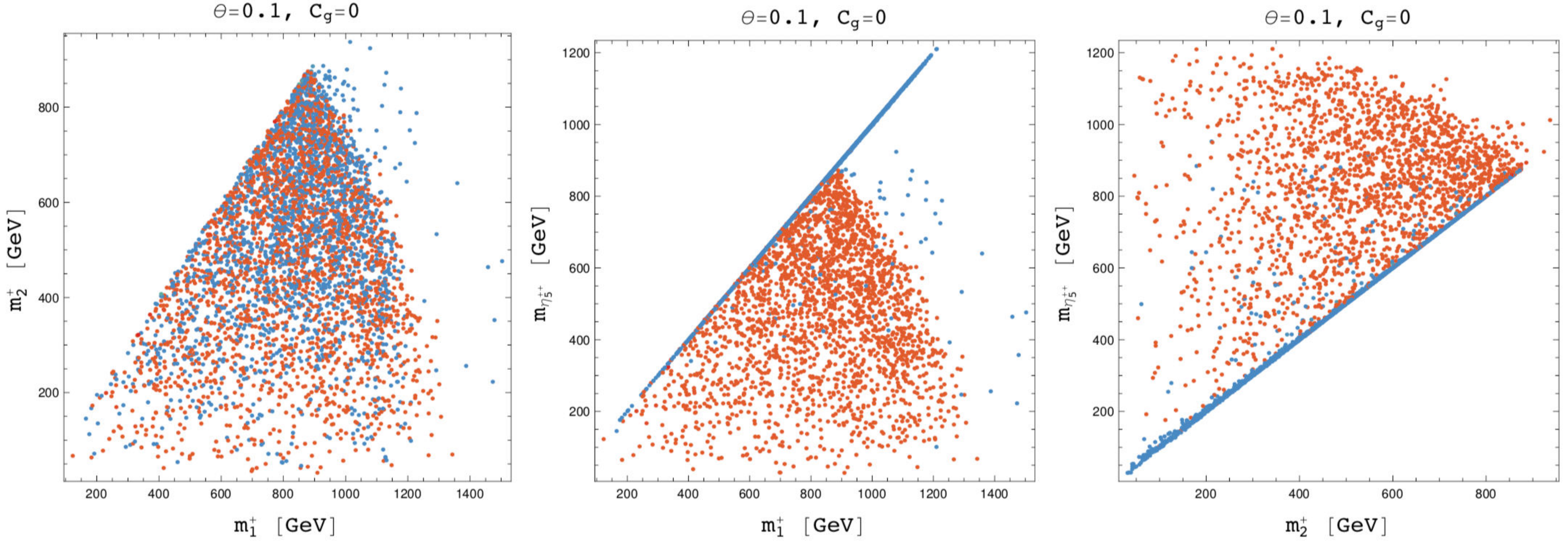}
\caption{Correlations among the pNGB masses. The colours represent the dominant fraction of gauge eigenstate in the state quoted on the horizontal axis: for the neutral pseudo-scalars (first two rows) red stands for the singlet $\eta$,  yellow for $\eta_1^0$ and blue for $\eta_5^0$; for the charged (last row) blue for the triplet $\eta_3^\pm$ and orange for $\eta_5^\pm$.
}
\label{fig:specall}
\end{figure}

Finally, in Fig.~\ref{fig:specall} we show correlations between the various mass eigenstates, $m_{1,2,3}$ for the pseudo-scalars $\eta$--$\eta_1^0$--$\eta_5^0$, $m_{\eta_3^0}$ for the scalar triplet, $m_{\eta_5^{++}}$ for the doubly charged and $m_{1,2}^+$ for the singly charged. The colours keep track of the dominant components of the state in the horizontal axes in terms of the gauge eigenstates, as indicated in the caption. From the top row, relative to the three pseudo-scalars, we see that all the states are lighter than $800$~GeV when they are dominantly singlet, in particular the heaviest one, $m_3$, is always in the range $600$--$800$~GeV. This implies that the whole spectrum is light as long as the singlet is the heaviest state.
The second row of plots, showing the correlation between the pseudo-scalars and the neutral scalar $\eta_3^0$, highlights a more interesting correlation: the neutral triplet is always degenerate with the mass eigenstate that is dominantly custodial singlet $\eta_1^0$ in yellow.  Note that the exact equality we observe here is mainly due to our choice $C_g=0$ in the scan, as a non-zero value of this parameter will induce a splitting due to gauge loops. This effect can also be seen in Fig.~\ref{fig:CustPlot} for the custodial case, where all the masses (except the singlet $\eta$) would become equal in the $C_g=0$ limit. A looser correlation can be observed  between the triplet and the mostly singlet mass eigenstate (red).
The bottom row shows correlations among the charged states: we see again a nice correlation between the doubly charged state and the singly charged eigenstate that is mainly from the triplet $\eta_3^\pm$ in blue, even though there are points that lie outside of the $m_{\eta^{++}_3} = m^+_i$ line. 

In general, a remarkable feature of this model is the presence of light states, whose mass can be as low as $100$~GeV, even though we have fixed $f \approx 2.4$~TeV in the scan. This situation makes the model very promising for searches of new light scalars at the LHC in direct production, or via decays of the top partners~\cite{Bizot:2018tds}.

\section{General vacuum} \label{sec:vacuum2}

The remaining option to eliminate the tadpole of the $\eta_3^0$-field is to  allow for a more general 
vacuum than the one defined in Section~\ref{sec:model}. Following this approach, one is not forced anymore to require a precise cancellation among the parameters 
in the potential in order to set the tadpole to zero. Thus, instead of a relation between the parameters, one of them is simply traded with an additional 
misalignment angle, that we shall denote as $\zeta$.  Let us remark that, contrary to what is usually done in elementary models containing triplets, such 
as the Georgi-Machacek model \cite{Georgi:1985nv}, it is not possible here to generate a custodial symmetric VEV for the triplets (i.e. misalignment). Firstly, this is 
due to the intrinsic CP-parity carried by the scalars, as discussed in Section~\ref{disc}: only the custodial triplet contains a CP-even neutral scalar along which a misalignment can be generated without violating CP. We stress again that a misalignment along any other direction, corresponding to a custodial singlet, would break CP spontaneously.  Secondly, a misalignment along the custodial singlet $\eta_1^0$ automatically generates a 
tadpole for the five-plet field $\eta_5 ^0$, which can only be removed by giving it a VEV (barring exact cancellations between different operators). 

The misalignment matrix $\Omega$, defined in Eq.~\eqref{eq:omega} needs to be generalised as follows
\be
\Omega(\theta) \ \to \ \Omega(\theta_1, \theta_2) \ = \ e^{2 i \theta_1 X^h + 2 i \theta_2 X^{\hat \eta}}\,,
\ee
where the two generators are given explicitly in Appendix~\ref{app:vacuum}.
However, a more convenient parametrisation can be used by observing that
the two generators corresponding to the Higgs and neutral triplet directions, together with their commutator, form a $\SU(2)$ algebra~\cite{Cai:2018tet}:
\begin{align}\label{eq:algebra}
[X^h, X^{\hat\eta}] \ = \ 2i\,  T^7 \,, \quad [X^{\hat \eta},  T^7] \ = \ 2 i \, X^{\hat \eta }\,, \quad [ T^7, X^{\hat \eta}] \ = \ 2 i \, X^h \,,
\end{align}
where the unbroken generator $ T^7$ has the following matrix representation:
\be
 T ^7 \ = \ \frac{1}{2 \sqrt{2}}  \left(
\begin{array}{ccccc}
 0 & 0 & 0 & 0 & 0 \\
 0 & 0 & 0 & 0 & 1 \\
 0 & 0 & 0 & 0 & 1 \\
 0 & 0 & 0 & 0 & 0 \\
 0 & 1 & 1 & 0 & 0 \\
\end{array}
\right)\,.
\ee
The generators $T^{8, 9, 10}$, given in Appendix~\ref{app:gens}, together with $T^7$ and the six generators $\{T_L ^i, T_R ^i \}$, provide a complete
adjoint representation of the unbroken $\SO(5)$ group on the vacuum $\Sigma_0$.
As a consequence of the group structure elucidated above, the vacuum alignment matrix $\Omega$ can be conveniently re-parameterised as
\be
\Omega(\theta_1, \theta_2)  \ \equiv \ R(\zeta) \cdot\Omega(\theta) \cdot R(\zeta)^\dagger \  \,, \quad R(\zeta) = e^{2 i \zeta T^7}\,,
\ee
where
\be
\theta = \sqrt{\theta_1^2 + \theta_2^2}\quad \mbox{and} \;\; \tan \zeta = \theta_2/\theta_1\,.
\ee
The explicit form of $R_\zeta$ is given in Appendix~\ref{app:vacuum}. Thus, the linearly transforming pion matrix reads~\cite{Cai:2018tet}
\be
\Sigma_\zeta (x) = R(\zeta) \cdot \Sigma (x) \cdot R^T (\zeta)\,,
\ee
where $\Sigma(x)$ is defined on the Higgs vacuum, Eq.~\eqref{goldstone}.
As discussed in Ref.~\cite{Cai:2018tet}, the rotation $R_\zeta$ acts on the pion matrix $\Pi (x)$ as a mere change of basis, due to the fact that the generator $T^7$ is not broken by the vacuum $\Sigma_0$. Here, we choose to work in the basis where the pion matrix is given by
\be
\Pi' (x) = R_\zeta \Pi (x) R_\zeta^\dagger\,,
\ee 
with $\Pi (x)$ given in Eq.~\eqref{eq:Pimatrix}.  In this basis, the field we call $h$ has couplings close to the ones of the doublet, while $\eta_3^0$ matches approximately the custodial triplet. We remind the reader that the misalignment by the angles $\theta$ and $\zeta$ induces a mismatch between the pNGBs and gauge eigenstates.

The masses of the EW gauge bosons have now the following expressions
\begin{align}
m_W ^2 \ &= \ \frac{ g^2  f^2}{4} \st^2 \left(1+s^2_\zeta \right)\,, \quad
m_Z ^2 \ = \ \frac{\left(g'^2+g^2\right) f^2}{4} \st^2 \left( 1+ (1+ 2\, \ctt) \, s^2_\zeta \right)\,,
\end{align}
which gives a tree-level correction to the $\rho$ parameter (the Weinberg angle $\theta_W$ is not affected at tree-level):
\be \label{eq:rho}
\delta \rho \ \equiv \ \frac{m_W^2}{m_Z^2 \cos^2 \theta_W} -1 \ =  \ - \frac{2 \ctt (1-c_{2 \zeta} )}{3 - c_{2 \zeta} + 2 \ctt (1-c_{2\zeta})}\,,
\ee
which vanishes either for $\zeta =0$ or for $\theta = \pi/4$. Also, we remark that $\delta \rho < 0$ for $0<\theta < \pi/4$ and $\delta \rho > 0$ for $\pi/4 < \theta < \pi/2$, and that it is symmetric under change of sign of $\zeta$.
The mass of the top quark is also modified in the new vacuum, but its explicit form depends on the chosen spurions. Here we will focus on the anti-symmetric irrep of $\SU(5)$ as in Section~\ref{sec:antisym}, for which it reads:
\be
m_t \ = \ \frac{C_{t, A_L A_R}}{4\pi} f \st c_\zeta \left( \ct - \st s_\zeta\right)\,.
\ee

Besides the tree level correction to the $\rho$ parameter, the new vacuum alignment also modifies the couplings of the would-be Higgs to $W$, $Z$ and tops, in a non-custodial way. We define the modified couplings with respect to the SM ones as
\be
\kappa^\phi_V = \frac{v}{2 m_V^2} g_{\phi VV}\,, \quad \kappa^\phi_t = \frac{v}{m_t} g_{\phi \bar{t} t}\,,
\ee
where $v$ is defined by matching the expression of the $W$ mass to the SM formula, yielding
\be
v = f \st \sqrt{1+s_\zeta^2}\,,
\ee
and $\phi = h, \eta_3^0$. For the $W$, we obtain
\be
\kappa_W^h = c_\zeta \frac{\ct (1+s_\zeta^2) - s_\zeta^2}{\sqrt{1+s_\zeta^2}}\,, \quad \kappa_W^\eta = - s_\zeta \frac{\ct (1+s_\zeta^2) + c_\zeta^2}{\sqrt{1+s_\zeta^2}}\,.
\ee
The corresponding expressions for $Z$ and top are more complicated, and we report them in Appendix~\ref{app:vacuum}. In the limit $\zeta \to 0$, where the vacuum is misaligned along the Higgs direction only, we recover the familiar expression $\kappa_W^h = \ct$ while the coupling of the triplet vanishes. On the contrary, for $\zeta = \pi/2$, where the vacuum is only misaligned along the triplet, $\kappa_W^\eta = - \sqrt{2} \ct$ while the coupling of $h$ vanishes. This confirms our expectation that, in the chosen basis, $h$ matches the doublet and $\eta_3^0$ the custodial triplet.

From Eq.~\eqref{eq:rho} we observed that $\delta \rho$ vanishes for $\theta=\pi/4$, besides the obvious region near vanishing triplet misalignment $\zeta = 0$. It is, therefore, tantalising to think that such a large $\theta$ region may be still allowed by the Higgs data and by precision tests.
The most recent determination of the $\rho$ parameter gives $\rho = 1.0005 \pm 0.0005$~\cite{pdg2018}, where the fit is marginalised on contributions to other oblique observable such as the $S$ parameter. Thus, we will impose a bound on $\delta \rho$ at $3\sigma$, which gives the following range: $-0.1\% < \delta \rho < 0.2\%$.  This is but a rough estimate of the impact of electroweak physics in this model, because at one loop level additional contributions will emerge coming from the modifications of the Higgs couplings and from loops of heavier resonances. For examples of this kind of calculations in other models, we refer the reader to Refs~\cite{Carena:2006bn,Barbieri:2007bh,Arbey:2015exa}. A detailed analysis of this issue is, however, beyond the scope of this paper, because it heavily relies on the details of the model.
One interesting feature is that corrections to the $S$ parameters are typically positive~\cite{Peskin:1990zt,Hirn:2006nt,Contino:2010rs,Panico:2015jxa}, and due to the correlation in the EW fit, a positive contribution to $\delta \rho$ (i.e. $T$) is welcome as it may push the model back into the allowed region~\cite{Grojean:2013qca}. 
In the model under study, this happens for $\theta > \pi/4$. It is the deviation on the couplings of the Higgs that will tell us if such a region is still viable. 

\begin{figure}
\centering
\includegraphics[width=.32\textwidth]{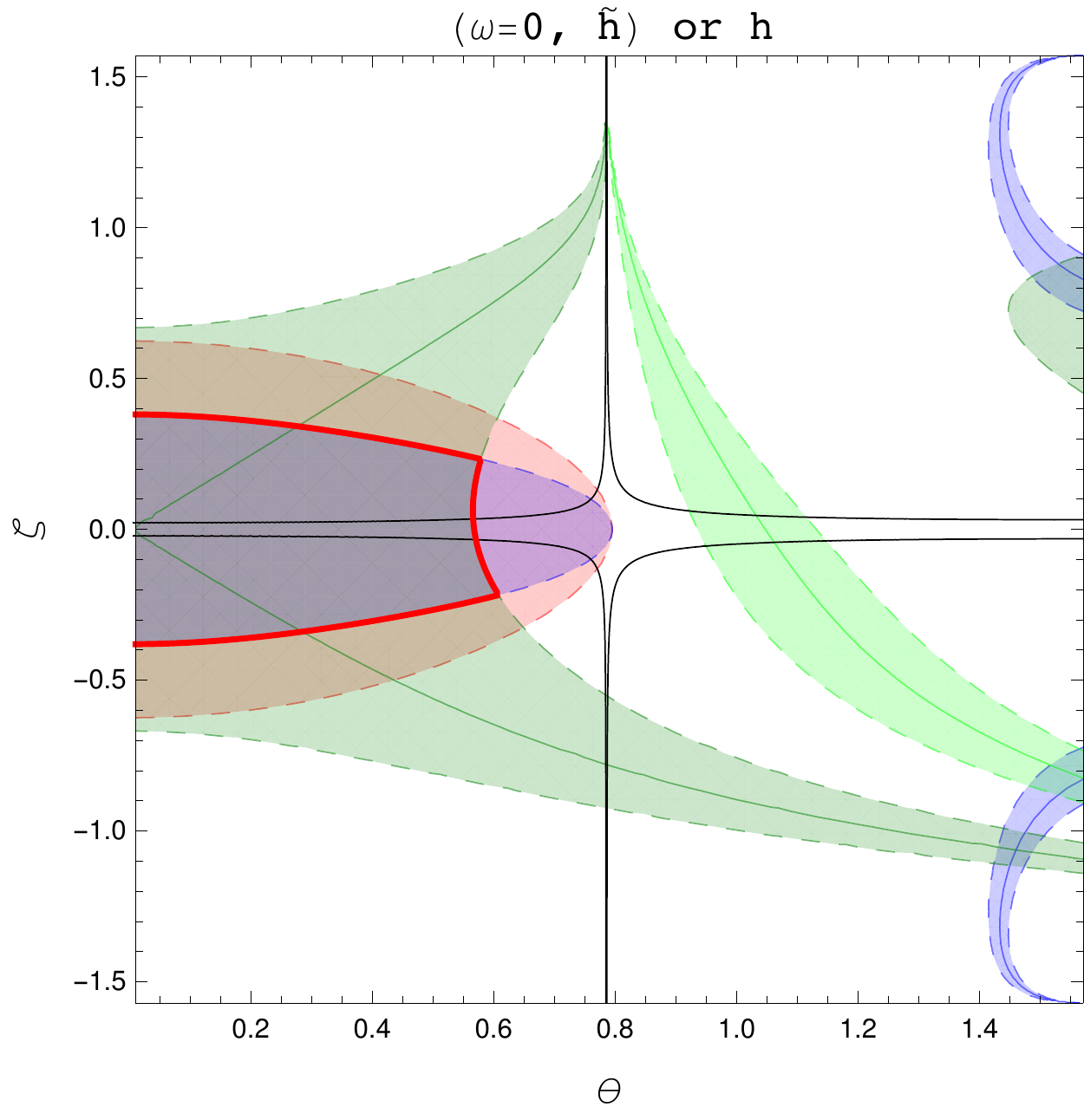}
\includegraphics[width=.32\textwidth]{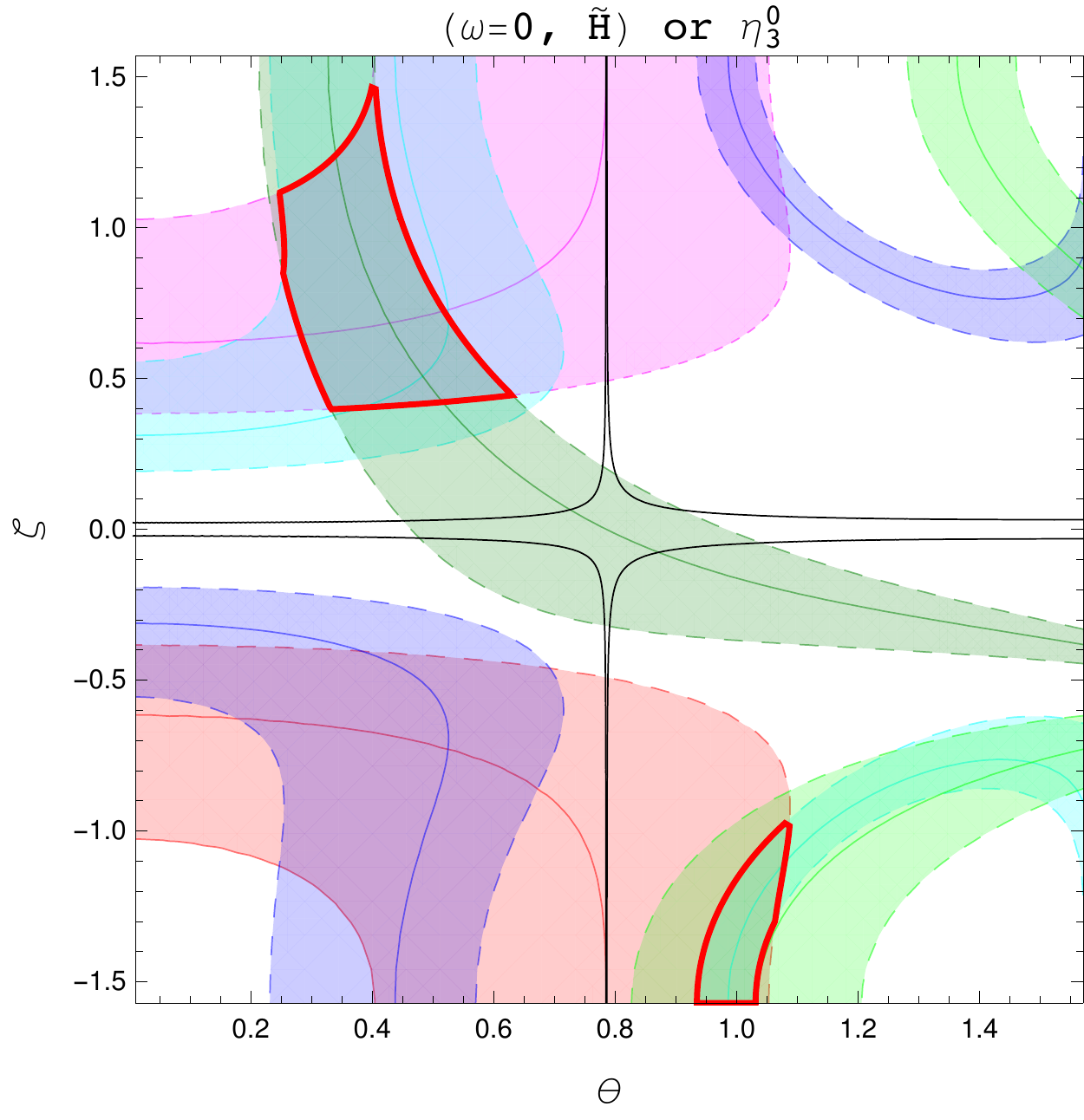} 
\includegraphics[width=.32\textwidth]{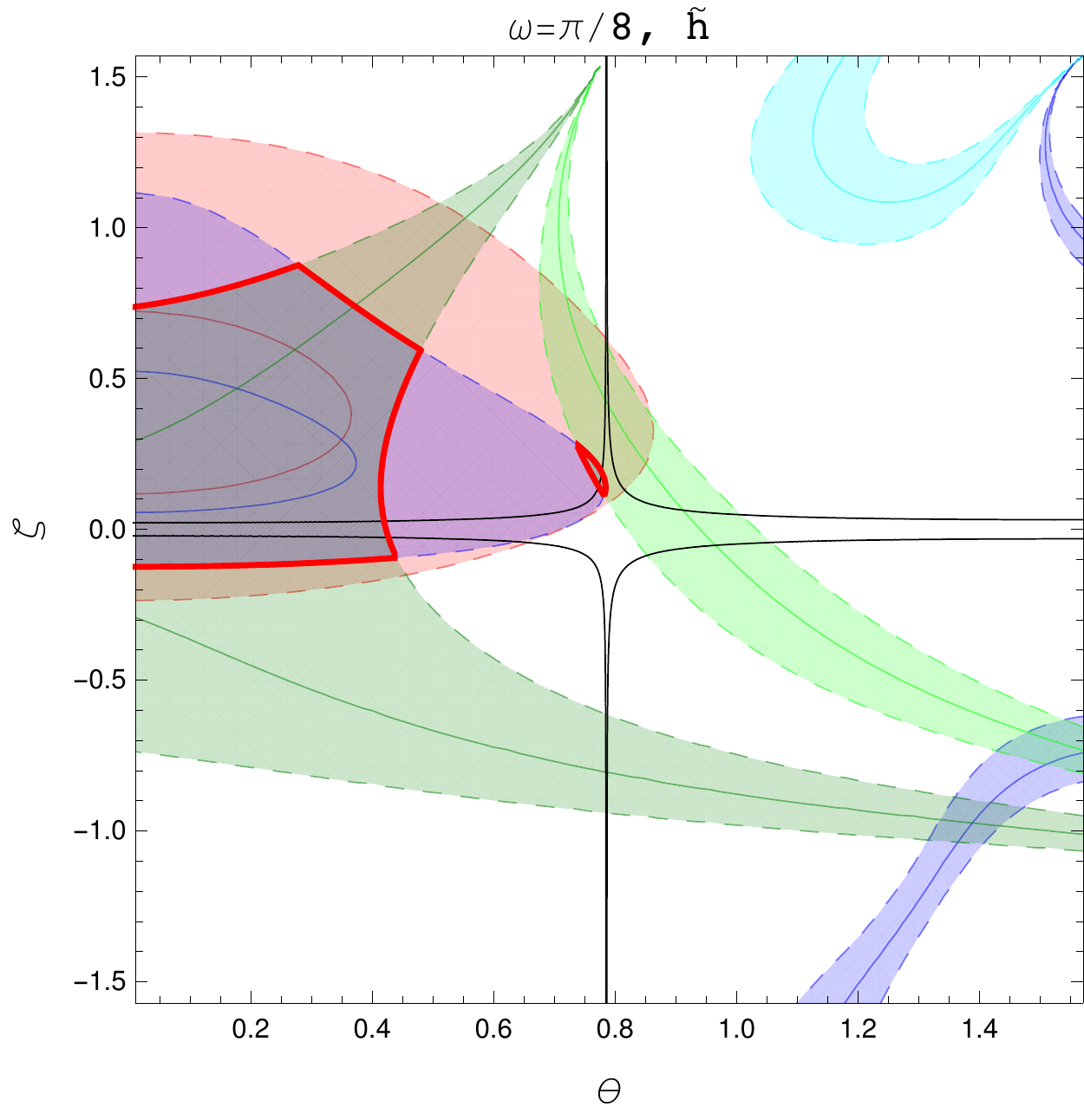} 
\caption{Regions in the $\theta$--$\zeta$ parameter space with scalar couplings within $30\%$ of the SM value to $W^+W^-$ (red) and $ZZ$ (blue), and within $50\%$ to $t\bar{t}$ (green). The region in lighter colours (magenta for $W^+ W^-$, cyan for $ZZ$ and light green for $t\bar{t}$) correspond to flipped sign couplings. The red band encircles the areas where al the couplings are within the stated limits. With black lines we show the $3\sigma$ contours for the $\rho$ parameter, with the allowed region around $\zeta \approx 0$ and $\theta \approx \pi/4$. The three panels correspond to different choices: $\omega = 0$ for the first two, which show the couplings for the doublet $h$ (left) and triplet $\eta_3^0$ (middle), and $\omega = \pi/8$ for $\tilde{h}$ (right).}
\label{fig:coups1}
\end{figure}

\begin{figure}
\centering
\includegraphics[width=.32\textwidth]{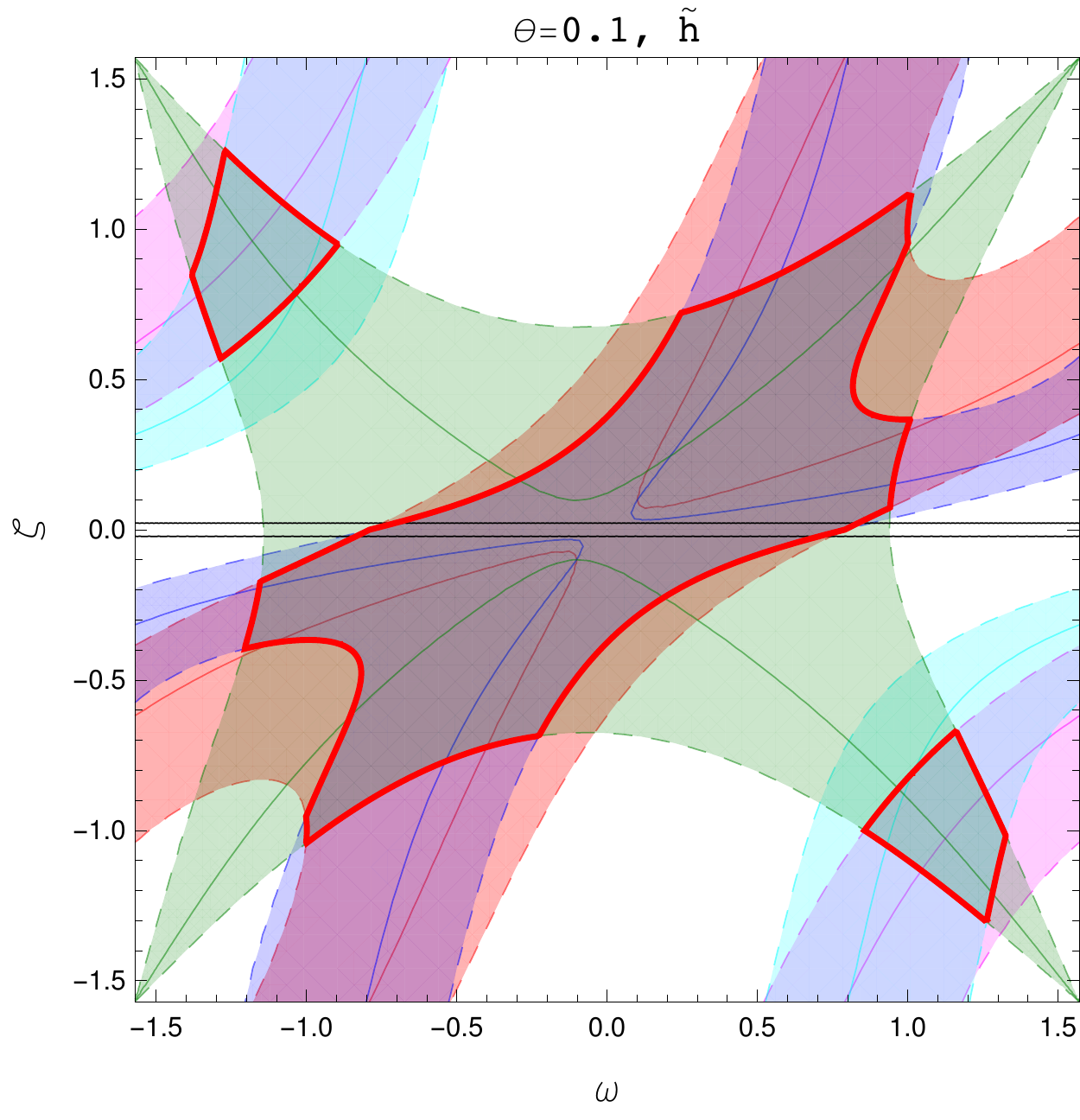}
\includegraphics[width=.32\textwidth]{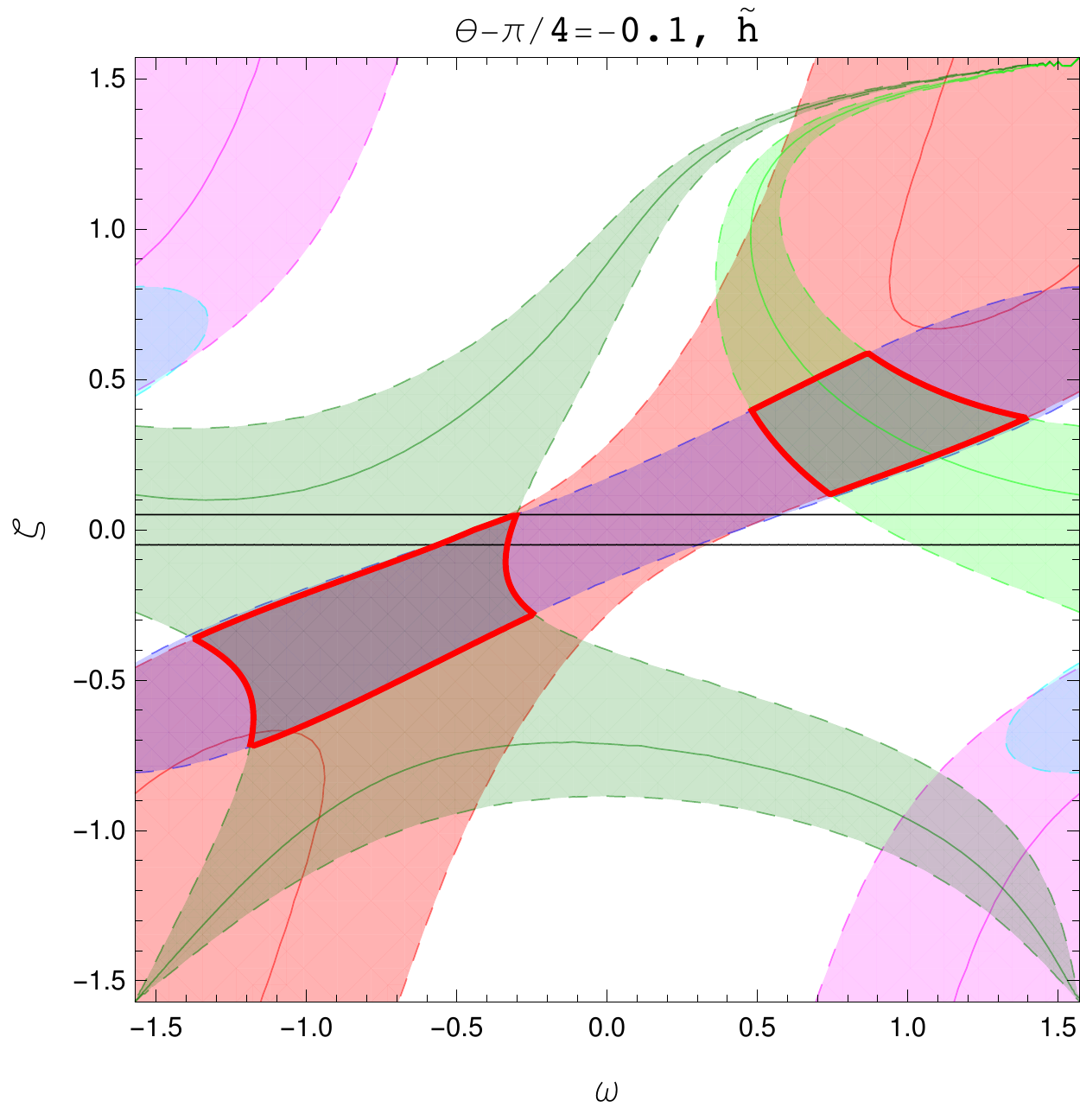}
\includegraphics[width=.32\textwidth]{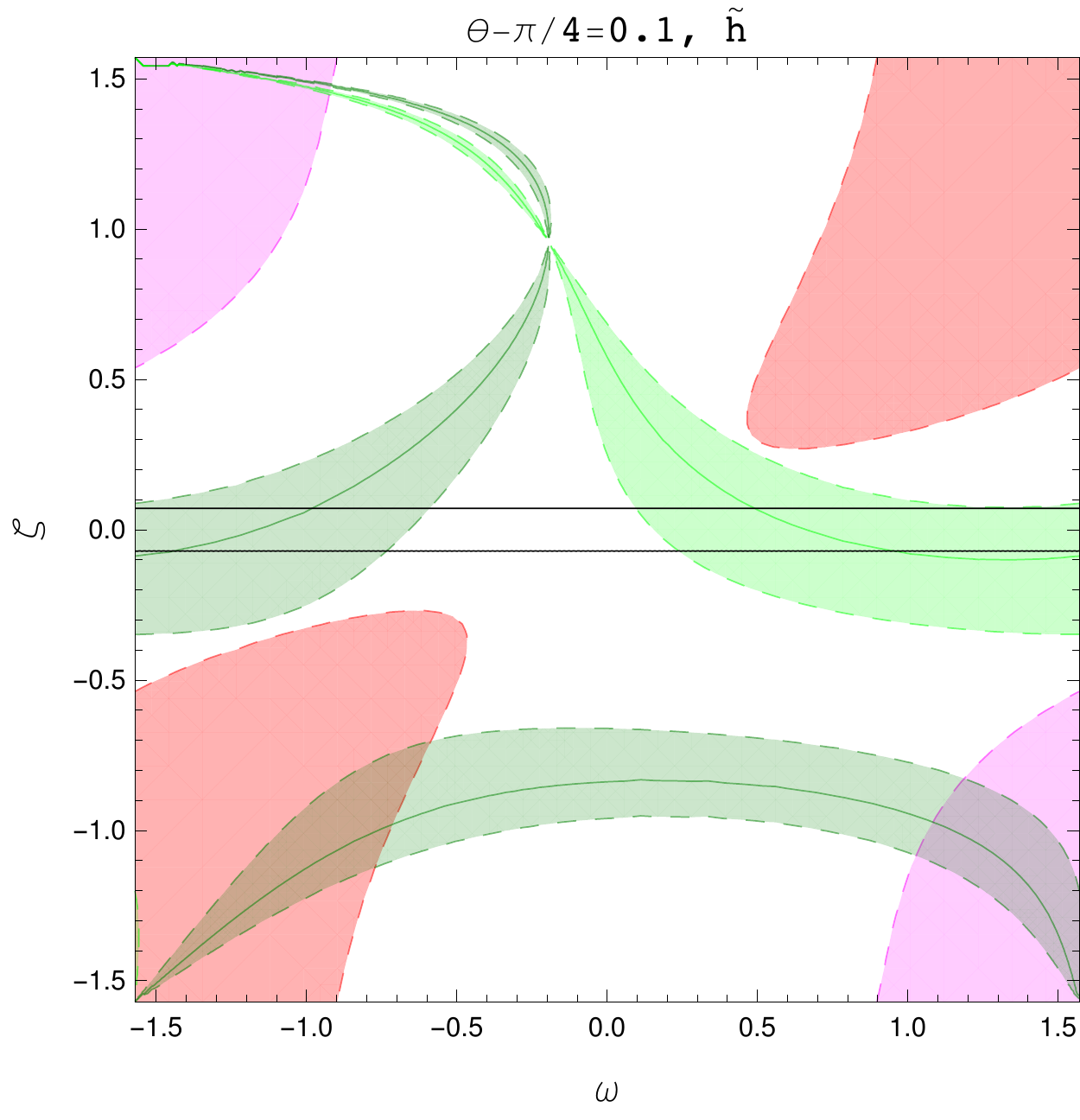}
\caption{Same as Fig.~\ref{fig:coups1} in the plane $\omega$--$\zeta$ and for fixed values of $\theta$. The three panels correspond to the couplings of $\tilde{h}$ for $\theta = 0.1$ (right), $\theta = \pi/4-0.1$ (middle) and $\theta = \pi/4+0.1$ (right).}
\label{fig:coups2}
\end{figure}

A mixing between the two scalars $h$ and $\eta_3^0$ is always generated, while the details depend on the representation of the top spurions and on the free parameters in the pNGB potential. As an example, in Appendix~\ref{app:genvacpot} we briefly discuss the case of the anti-symmetric. 
Here, however, we simply introduce an angle $\omega$ that rotates the basis $(h, \eta_3^0)$ into the mass eigenstate one $(\tilde h, \tilde H)$, defined as follows
\be \label{eq:rot}	
\left( \begin{array}{c}  \tilde h \\ \tilde H \ 
\end{array} \right) = \left( \begin{array}{c c} c_\omega & -s_\omega \\ s_\omega & c_\omega \end{array} \right) 
\cdot \left( \begin{array}{c} h \\ \eta_3 ^0 \end{array} \right)\,,
\ee
where we allow $\omega$ to span $[-\pi/2, \pi/2]$. Note that both mass eigenstates, $\tilde{h}$ and $\tilde{H}$, can, in principle, be identified with the $125$~GeV scalar observed at the LHC.
The modifications to the Higgs coupling depend on three parameters: the two misalignment angles $\theta$--$\zeta$ and the mixing angle $\omega$. 
To compare with the most recent LHC results, we will adopt a very conservative approach and only compare with the most direct available measurements. The reason is that in general additional contribution to the Higgs couplings will arise from loops, which will be especially relevant for the loop induced couplings to gluons and photons. Thus, we need to disentangle the impact of the precise determination of such couplings, via the production rate and the di-photon decay, from the determination of the couplings to $W$'s and tops. 
For the couplings to gauge bosons, the most recent measurements at the LHC provide errors of about $10\%$~\cite{Sirunyan:2017exp,Sirunyan:2018egh,Aaboud:2017vzb,Aaboud:2018puo}, thus we will conservatively allow a $3\sigma$ variation of about $30\%$. For the top coupling, we use the direct measurement coming from the observation of the $t\bar{t}h$ associated production~\cite{Aaboud:2018urx,Sirunyan:2018hoz}, which gives an error of about $15\%$ on the coupling, thus we show a band of about $50\%$ at $3\sigma$.
As a first exploration, we show in Fig.~\ref{fig:coups1} the deviations for the doublet (left plot) and triplet (middle plot) for vanishing mixing, and the same for a small mixing angle $\omega = \pi/8$ (right plot).  We allow couplings with flipped signs, shown with lighter colours in the plot.
The region where all Higgs couplings are compatible with the measurement is thus compared to the most recent best fit of the $\rho$ parameter, with the $3\sigma$ region within the black lines.
 We see that the bound on $\rho$ provides an upper limit on $|\zeta| < 0.023$ for small $\theta$, while an upper bound on $\theta$ around $0.55$ comes from the top coupling. This is an example of the importance of a direct measurement of the top coupling in CH models. As expected, for a pure triplet there is no region where the Higgs couplings can match the experimental results, while the plot on the right shows that increasing the mixing angle tends to shift the region matching the Higgs couplings outside of the band allowed by the $\rho$ parameter.

In Fig.~\ref{fig:coups2} we show the same allowed regions for fixed values of $\theta$. For small $\theta = 0.1$, as shown in the left plot, besides a limit on $|\zeta|< 0.023$ coming from $\delta \rho$, the Higgs couplings to vectors give a bound on the mixing angle of roughly $|\omega| < 1$.
In the middle and right plots, we show values of $\theta$ close to $\pi/4$, where the contribution to $\rho$ is small. For values below $\pi/4$, a small region is still allowed by the Higgs couplings, however remaining marginal as it always entails large deviations to the couplings. It also corresponds to negative mixing angles, and it shrinks to a point for $\theta \to \pi/4$. The right plot shows a value above $\theta = \pi/4$, where a positive contribution to $\rho$ is obtained. However, the values of the would-be Higgs completely exclude this case, as there is no overlap between the allowed regions of the various couplings.
			
\section{LHC phenomenology} \label{sec:pheno}

As we have seen in the previous sections, apart from the singlet, the light scalar content of the $\SU(5)/\SO(5)$ CH model is the same as that of the Georgi-Machacek model~\cite{Georgi:1985nv,Chanowitz:1985ug}, GM for short in the remainder of this section. The LHC phenomenology, therefore, shares some common features to the widely studied one of the GM, see Refs~\cite{Englert:2013wga, Chiang:2015amq, Degrande:2017naf} and references therein.
There are, however, important differences between GM and its composite counterpart: the main one, already highlighted in Section~\ref{disc}, is the fact that the CP properties of the triplets are exchanged. Namely, it is the neutral component of the custodial triplet that is CP-even, while the others (including the singlet) are CP-odd. This property can be determined thanks to the fermionic underlying theory. As a consequence, the custodial singlet VEV of the triplets is forbidden in CP-conserving cases, thus a three level coupling of the new scalars to two massive gauge bosons, $W$ and $Z$, are not generated. 
We remark that the presence of such couplings plays a leading role in the GM phenomenology, as it allows for single production of the scalars via Vector Boson Fusion (VBF) with further decays in di-bosons~\cite{Englert:2013wga}. In particular, a benchmark scenario has been identified, for which the fiveplet is the lightest state of the spectrum, thus providing a motivation for scalar di-boson interpretations of LHC searches~\cite{Logan:2017jpr}. This benchmark, as already stressed, is however not allowed in the CH model, where couplings to a pair of gauge bosons are only generated at one loop level by the WZW anomaly and thus yield negligibly small cross sections.
The leading production mode for the new composite scalars is, therefore, pair production via gauge interactions, which is dominated by Drell-Yan processes~\cite{Han:2007bk, Degrande:2015xnm}.

Couplings to fermions are more model sensitive, as they crucially depend on the origin of the SM fermion masses in the CH model. One difference with respect to the GM model is that couplings of the five-plet to top and bottom quarks are present in some cases, depending on the operator that generates the top mass, while the five-plet is fermiophobic in the GM model. We refer to Section~\ref{sec:topmass} for model details. Note, nevertheless, that all fermion couplings are suppressed by $s_\theta$, thus they tend to be small. There is also a special case, for top partners in the adjoint of $\SU(5)$, where all linear couplings of the pNGBs to the top and bottom vanish. Note that the presence of couplings to tops would induce a coupling to gluons, at one loop level, for the neutral (pseudo-)scalars, thus allowing for single production via gluon fusion. If the couplings of the light quarks and leptons are also generated via FPC, the same rules apply as for the top. There is also the other possibility that their couplings are generated by direct bilinear interactions with the strong dynamics, as in Ref.~\cite{Cacciapaglia:2015dsa}. In this case, the structure of the couplings is more similar to the GM case, as shown in Appendix~\ref{app:effYuk}, as only the five-plet remains fermiophobic. Those couplings would, however, be suppressed by the small Yukawa couplings. 

The presence of a doubly-charged state has been identified as a crucial feature of this model since the beginning~\cite{Georgi:1985nv}, as it potentially leads to low background final states with same-sign leptons~\cite{Englert:2013wga,Chiang:2015amq}. This signature is also present in the composite model, as the doubly charged scalar, $\eta_5^{\pm \pm}$, can only decay in final states that contain two same-sign (virtual) $W$'s. The main decays are thus into a $W^\pm W^\pm$ pair, or into a singly charged state $W^\pm \eta_{1,2}^\pm$. This pattern matches with the GM case~\cite{Chiang:2015amq}. However, the fact that the coupling $\eta_5^{\pm \pm} W^\mp W^\mp$ is induced by an anomaly (thus being loop suppressed) and suppressed by $s_\theta^2$~\cite{Georgi:1985nv} implies that the cascade decays are always preferred, as long as the mass splitting is large enough. Furthermore, the decays of the singly charged and neutral states into gauge bosons are also induced by the anomaly, implying that final states with photons are equally probable as final states with massive gauge bosons, contrary to the GM case where the couplings involving photons are loop induced~\cite{Degrande:2017naf}. \footnote{The high probability of finding photons in the final state is also enhanced by the fact that the only gauge-invariant WZW coupling of the triplets is of the form $\pi_+^i \tilde{B}_{\mu \nu} W_{\mu \nu}^i$, thus it always contains a hypercharge gauge field.}
Finally, it is always possible to add a direct lepton-number violating bilinear coupling of the triplets to leptons, as shown in Appendix~\ref{app:effYuk}. If the vacuum has a very small misalignment along the triplet, as studied in Section~\ref{sec:vacuum2}, this would lead to a composite type-II see-saw mechanism~\cite{Ma:1998dx}. From the phenomenology side, the presence of this coupling is very important, as it would allow unsuppressed tree-level decays of the doubly charged scalar into a resonant pair of same-sign leptons~\cite{Akeroyd:2010ip}.

To summarise, the main differences between the $\SU(5)/\SO(5)$ CH model and GM are:
\begin{itemize}

\item[i)] The main production mechanism is pair production via Drell-Yann, as the couplings of a single pNGB to vector bosons are suppressed, being generated by the WZW anomaly. In the GM model, on the contrary, they are induced at tree level by the custodial-invariant $\SU(2)_L$-triplet VEV. 

\item[ii)] The couplings to fermions depend on the {\it irrep} of the top partners and origin of light fermion masses. In general, the couplings are suppressed by $s_\theta$, and they vanish for top partners in the adjoint. The five-plet is not always fermiophobic.

\item[iii)] Decays to a pair of gauge bosons, of which at least one is a photon, are abundant, as all such couplings are generated by loops. Thus, final states with photons are common for any mass of the scalars. In GM, couplings to photons are loop induced thus relevant only for very low masses below $m_W$ or $m_Z$.

\end{itemize}

\begin{figure}[]
\centering
\includegraphics[width=.49\textwidth]{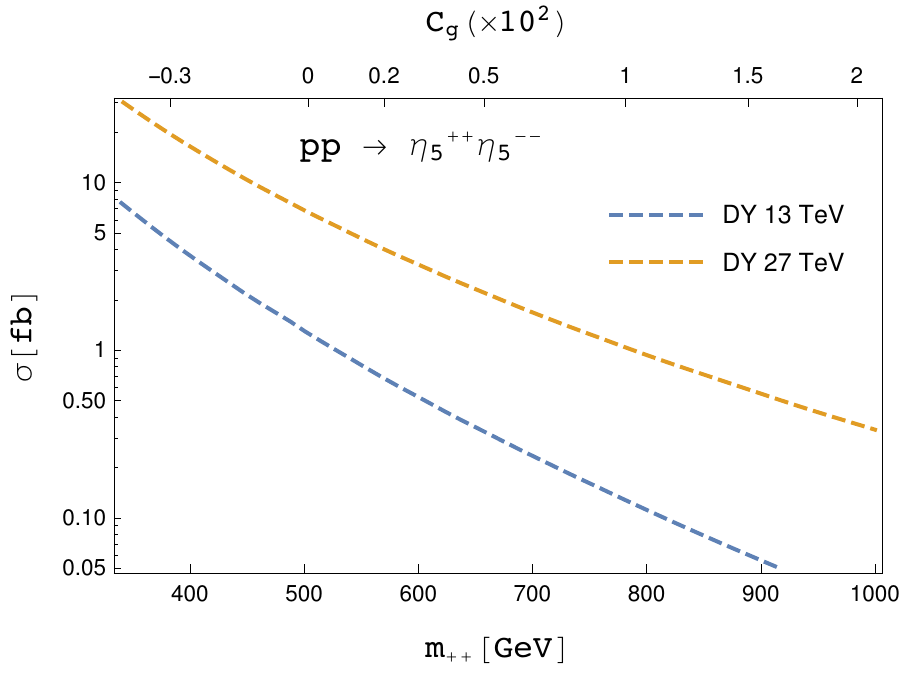}
\includegraphics[width=.49\textwidth]{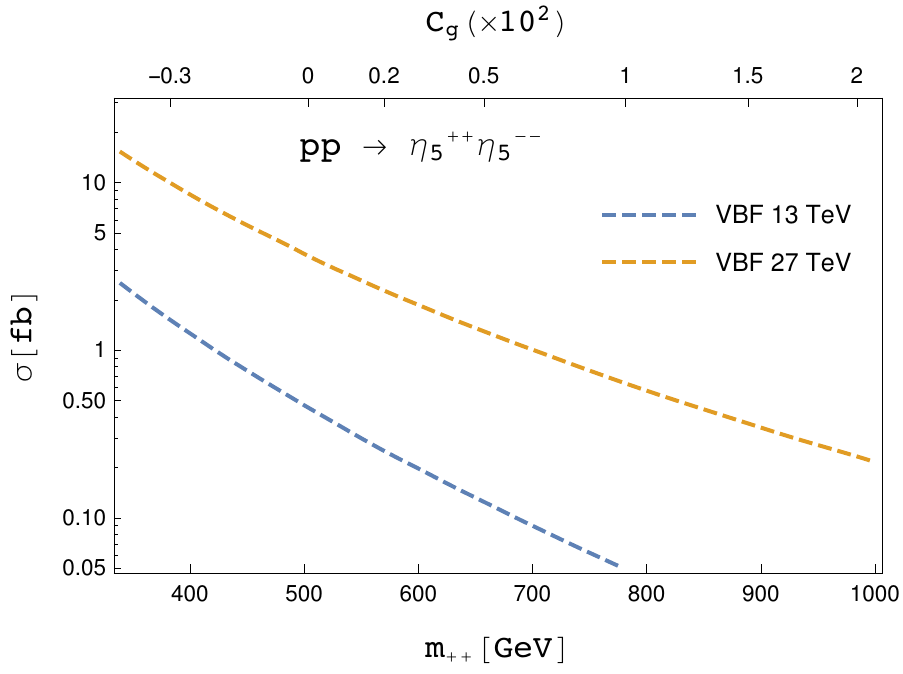}  \\
\includegraphics[width=.49\textwidth]{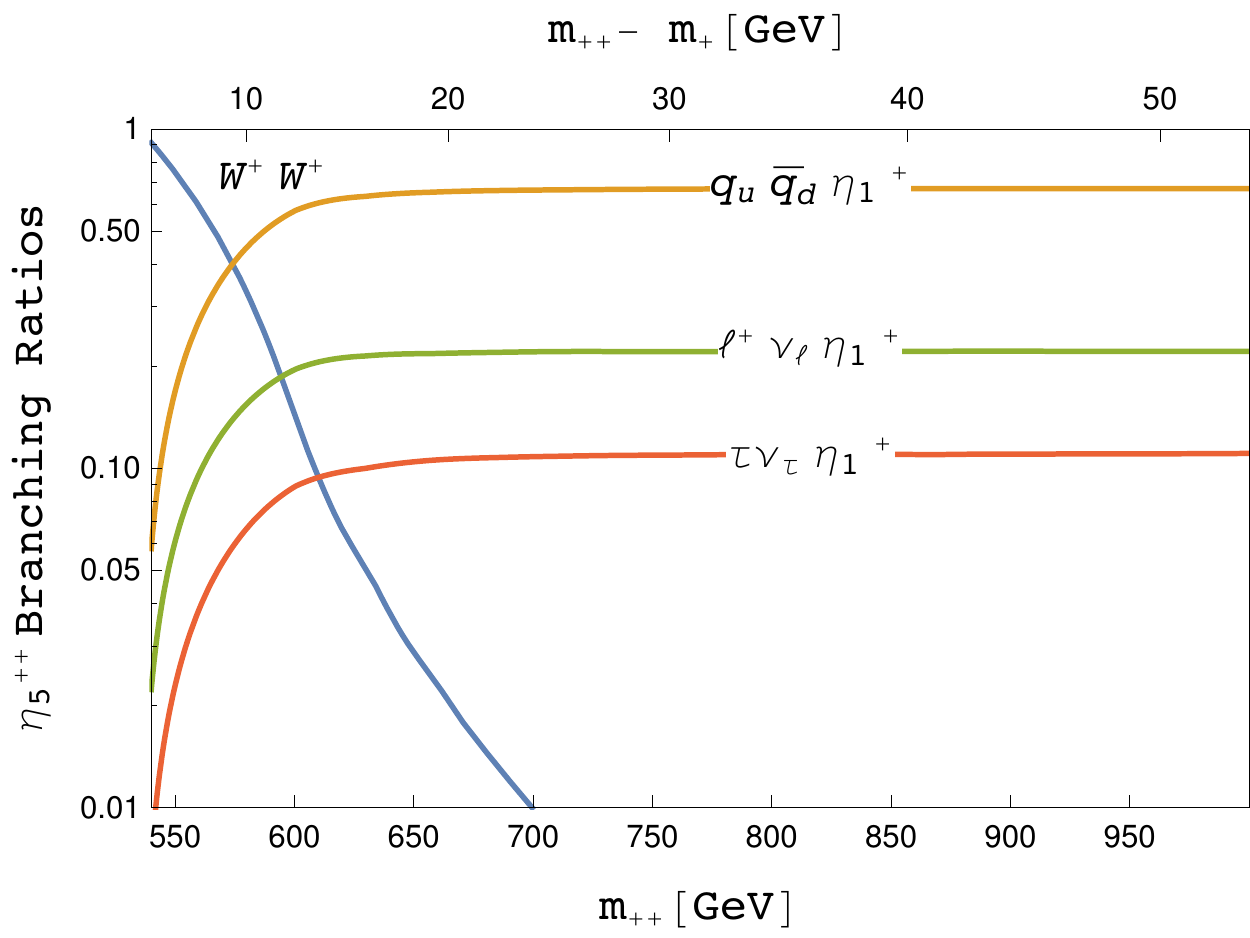}
\includegraphics[width=.49\textwidth]{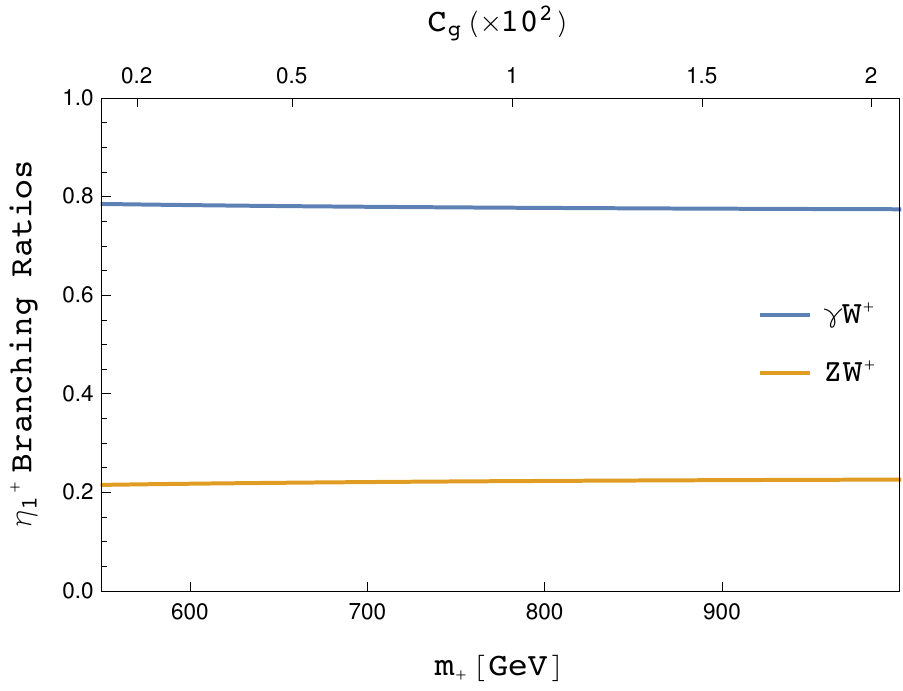}
\caption{\emph{Upper plots}: Cross-section for Drell-Yann (left) and VBF (right) pair production of the doubly-charged pNGB $\eta_5^{++}$, for $13 \TeV$ and $27 \TeV$ (HE-LHC). \\
\emph{Lower plots}: Branching ratios for the $\eta_5 ^{++}$ and $\eta_1^+$ pNGBs, as functions of their masses. For the BRs of $\eta_5^{++}$, $\ell$ stands for both $\mu^+$ and $e^+$.
}
\label{fig:BR}
\end{figure}

To quantify the above statements, we studied in some more detail a benchmark scenario, which gives rather typical predictions. As we already stressed, the spectra follow a universal pattern, independently of the specific top partner representation, if we limit ourselves to viable models with LO potential. We thus choose to study the model with adjoints, for which the couplings to tops vanish for all pNGBs. The spectrum is shown in Fig.~\ref{fig:spectrumDD}, as we also chose $\theta = 0.1$ in our numerical study. We computed the pair production cross sections via Drell-Yann using our own implementation of the model within the \texttt{Feynrules} package~\cite{Alloul:2013bka}, and using the \texttt{UFO} output to generate diagrams and cross sections via \texttt{MADGRAPH5\textunderscore aMC@NLO}~\cite{Alwall:2014hca}. A more accurate determination of the cross sections, including QCD NLO effects, can be found in Refs~\cite{Han:2007bk, Degrande:2015xnm}. In the top row of Fig.~\ref{fig:BR} we show the result as a function of the mass for the dominant pair production of doubly charged states, at the LHC Run-II energy and for the projected energy of a future high-energy LHC option.  We recall that in this model the mass depends on the parameter $C_g$, the coefficient of the gauge contribution to the pNGB masses.
The doubly charged scalar can decay either into same-sign $W$'s, or chain decay to a singly-charged scalar plus a virtual $W$. The branching ratios (BRs) are shown in the lower row of Fig.~\ref{fig:BR}, which shows how the $W^\pm W^\pm$ final state dominates only for small masses below $600$--$650$~GeV.
The singly charged state further decays to gauge boson pairs via the anomaly, with a probability of about $\cos^2 \theta_W \approx 78\%$ -- approximately flat in the considered range of $C_g$ -- of having a photon in the final state.

The same-sign dilepton (SSL) rate of the doubly charged scalar can be estimated to be $\mbox{BR} (\eta_5^{++} \to l^+ l^+ + X) \approx 6\%$, where $l = e, \mu$ (and we include leptonic decays of the tau). Thus, roughly 12\% of the pair production events yield SSL final states, for cross sections of roughly $0.1$~fb for the smallest mass in our benchmark scenario. The low mass range may, therefore, be accessible to LHC Run-II searches based on SSL, like for instance in Refs~\cite{CMS:2015loa,Sirunyan:2017uyt,Aaboud:2017dmy,Aaboud:2018doq}. At higher masses, above $\approx 650$~GeV, decays involving photons become very abundant: using the BRs from Fig.~\ref{fig:BR}, we see that pair production events with two photons in the final state occur around 60\% of the times.
This means that a new search based on hard photons may be more effective in covering this mass range.\footnote{For an analysis of final states featuring several photons using LHC data, see Ref.~\cite{Barducci:2018yer}.} We leave a detailed study of the SSL recast and photon signature for future work. Finally we should mention that, in presence of a Majorana coupling to leptons, resonant decays $\eta_5^{++} \to l^+ l^+$ are also possible, leading to a well-studied signature~\cite{Akeroyd:2010ip}. 
To conclude the discussion, in Table~\ref{tab:xsec}, we provide some specific numbers for two benchmark points, based on $C_g = 0.001$ and $0.01$.

\begin{table} \centering
\begin{tabular}{c  c | c | c | c} \hline
\multirow{2}{*}{Process} &  \multicolumn{2}{c|}{$\sigma$@13TeV (fb)}  & \multicolumn{2}{c}{$\sigma$@27TeV (fb)} \\
 &   $C_g = 0.001$ & $C_g = 0.01$ & $C_g = 0.001 $ & $C_g=0.01$ \\  \hline \hline
$pp \to \eta_5 ^{++} \eta_5 ^{--}$ &  0.91 & 0.12 & 5.1 & 0.16       \\

$ pp \to \eta_1^+ \eta_1^- $ & 0.10 & 0.014 & 0.58 & 0.12 \\
\hline \hline
Masses (GeV) &  &  & & \\ \hline 
$m_{\eta} $ & 1112.2  & 853.6  & -- & -- \\
$m_{\eta_3^0} $& 538.0  & 783.2 & -- &-- \\
$m_1$ & 532.2 & 742.2 &-- &-- \\
$m_2$ & 538.0 & 782.8  & -- & -- \\
$m_{\eta_5 ^{++}}$ & 537.8 & 781.6 & --  & -- \\
$m_{\eta_1^+} $ & 533.2 & 749.1 & -- & --  \\
$m_{\eta_2^+} $ & 538.7 & 787.6 & -- & -- \\ \hline \hline
BR  & & & & \\ \hline 
$\eta_{5}^{++} \to W^+ W^+$  & 0.95 & 0.003 & -- & -- \\ 
$\eta_5 ^{++} \to  q \bar q \,\eta_1^+ $ & 0.03 & 0.67 & -- & -- \\
$\eta_5^{++} \to  l^+ \nu_l \, \eta_1 ^+ $ & 0.01 & 0.22 &  --  & --\\
$\eta_5^{++} \to \tau^+ \nu_\tau  \, \eta_1 ^+ $ & 0.004  & 	0.11  & -- & --\\
$\eta_1^+ \to  \gamma W^+$& 0.79 & 0.78 & -- & -- \\
$\eta_1^+ \to Z W^+ $ & 0.21 &  0.22 & --  & --
\end{tabular}
\caption{Leading production cross-sections at the LHC for $\sqrt{s}=13$  and $27$ TeV for  two benchmark points with $C_g = 0.001$ and $0.01$ ($l=e, \mu$). The BRs refer to the $\eta_5^{++}$ decays, including chain decays into $\eta_1^+$.}
\label{tab:xsec}
\end{table}

\section{Conclusions and outlook} \label{sec:concl}

In this work, we illustrated some key features of the $\SU(5)/\SO(5)$ composite Higgs model, which was first introduced 
in \cite{Dugan:1984hq}. The pNGB spectrum resembles that of the Georgi-Machacek model~\cite{Georgi:1985nv,Chanowitz:1985ug} by the presence of 3 
triplets related by the custodial symmetry. We characterise the model by studying not only the embedding of the electroweak 
symmetry and its breaking, but also the explicit breaking of the $\SU(5)$ symmetry using the spurion formalism. For the top quark we used 
the paradigm of partial compositeness and considered embeddings transforming under complete two-index representations of $\SU(5)$.
This choice is backed up by realistic models based on gauge-fermion underlying theories. For light quarks and leptons, the mass can be generated either by partial compositeness or by direct bilinear couplings, generated at a higher energy scale than the top ones. We classified all the operators at leading
order in the chiral expansion contributing to the top mass (Yukawa) and to the pNGB potential. We then studied their effect on the
vacuum stability and pNGB spectra, thus largely extending the work of Ref.~\cite{Golterman:2017vdj}. We  present the first
comprehensive analysis of the properties of this class of models.

We obtained the most general form of the Yukawa interactions between the pNGBs and the top and (left-handed) 
bottom quarks, showing that, for general embeddings, the couplings of the Higgs and non-Higgs pNGBs are highly correlated.
The linear couplings, in fact, are governed in some cases by only one, two or three coefficients, depending on the operator(s)
generating the top mass.

We then performed the same general analysis for the potential, and analysed the effect of each operator in the vacuum that has been
misaligned along the Higgs direction only. We showed that a misalignment along any custodial singlet direction is forbidden by
CP invariance, thus proving that the custodial preserving triplet VEV allowed in the elementary Georgi-Machacek model
is not allowed in the CP-conserving composite version. The general analysis of the LO potential led us to identify a few viable scenarios, satisfying the requirements of vacuum 
misalignment together with the absence of a tadpole for the CP-even pNGB $\eta_3^0$. The misalignment along the $\eta_3$ triplet needs to be suppressed in order to 
avoid dangerous corrections to the value of the $\rho$ parameter, and also for consistency with the choice of the Higgs vacuum. The three models 
correspond to embedding both the left- and right-handed tops into the adjoint of $\SU(5)$, or embedding the left-handed in the anti-symmetric or the
right-handed in the symmetric (and the other in the adjoint). In all other cases, a triplet tadpole is inevitable, thus inducing a more general vacuum misalignment
that is highly constrained by electroweak precision tests.
In studying the scalar spectra for the three allowed cases, we found a universal pattern related to the transformation properties of the pNGBs under the custodial symmetry. 
Remarkably, the gauge singlet $\eta$ always receives a negative contributions from loops of gauge bosons as opposed to the rest of the scalars, except in the one case outlined at the end of Sec.~\ref{sec:spectra}. 
Moreover, the presence of tachyons often largely restricts the available parameter space, signalling a wrong choice for the minimum. 

The case of the anti-symmetric representation offers an example of NLO potential, and we have shown that a limit exists in which the custodial 
symmetry is restored, at least at the level of the spectrum. The pNGB masses show, however, a different pattern than in the LO cases, which we studied in detail. A common feature, very promising for the LHC phenomenology, is that pNGB masses below 1~TeV are always allowed, even when the condensation scale $f$ is in the multi-TeV range.

Finally, we also briefly discuss the LHC phenomenology of the non-Higgs pNGBs, highlighting the differences with respect to the Georgi-Machacek model. Besides the spectrum, it is remarkable the absence of tree-level couplings of the non-Higgs pNGBs to gauge bosons, which thus reduces single production at colliders. The main production mode is, therefore, production in pairs, via gauge interactions. Furthermore, the fact that couplings to gauge bosons are generated by topological anomalies implies that decay final states with photons are very probable, and even dominant, thus providing a new signature with hard photons. Couplings to fermions are model dependent, and, contrary to the elementary case, they can be set to zero in the adjoint-adjoint case. In other cases, including direct bilinear couplings to the strong dynamics, they are always generated. It turns out that there exist cases where the five-plet is not fermiophobic.
We leave a more detailed study of the LHC phenomenology for future work. Besides the direct pair production that we discuss here, another interesting production mode derives from the decays of the top partners, if they are light enough to be produced at the LHC, or at future higher energy hadron colliders.

\section*{Acknowledgements}

We thank Daniele Barducci  for suggestions concerning the \texttt{Feynrules} implementation of the model. AD is partially supported by the Institut Univ\'ersitaire de France. AD and GC also acknowledge partial support from the Labex-LIO (Lyon Institute of Origins) 
under grant ANR-10-LABX-66 (Agence Nationale pour la Recherche), and FRAMA (FR3127, F\'ed\'eration de Recherche ``Andr\'e Marie Amp\`ere'').
AA thanks the IPNL for hospitality during the completion of this work.

 \newpage

\appendix
\section*{Appendices}

\section{Details of the model}\label{app:models}

\subsection{Choice of the  vacuum} \label{app:vacuum}

The misalignment rotation, $\Omega(\theta)$ is generated by the generator associated to the VEV of the Higgs, $X^{\hat{h}}$, and equals
\beq
X^{\hat{h}} = \frac{1}{2\sqrt{2}} \left( \begin{array}{cccc|c}
 & & & & 0 \\
 & & & & 1 \\
 & & & & -1 \\  
 & & & & 0 \\ \hline
0 & 1 & -1 & 0
\end{array} \right)\,, \quad \Omega (\theta) = e^{4 i X^{\hat{h}} \frac{\theta}{2}} = \left( \begin{array}{ccccc} \label{eq:omega}
1 & 0 & 0 & 0 & 0 \\
0 & c_{\theta/2}^2 & s_{\theta/2}^2 & 0 & i s_\theta/\sqrt{2} \\
0 & s_{\theta/2}^2 & c_{\theta/2}^2 & 0 & - i s_\theta/\sqrt{2} \\
0 & 0 & 0 & 1 & 0 \\
0 &  i s_\theta/\sqrt{2} & - i s_\theta/\sqrt{2} & 0 & c_\theta
\end{array}\right)\,.   
\eeq
The Higgs vacuum thus reads:
\beq
\Sigma_\theta = \Omega(\theta) \Sigma_0 \Omega(\theta)^T = \left( \begin{array}{ccccc}
0 & 0 & 0 & 1 & 0 \\
0 & -s_{\theta}^2 & -c_{\theta}^2 & 0 & i s_{2\theta}/\sqrt{2} \\
0 & -c_{\theta}^2 & -s_{\theta}^2 & 0 & - i s_{2\theta}/\sqrt{2} \\
1 & 0 & 0 & 0 & 0 \\
0 &  i s_{2\theta}/\sqrt{2} & - i s_{2\theta}/\sqrt{2} & 0 & c_{2\theta}
\end{array}\right)\,.   \nonumber
\eeq

In Section~\ref{sec:vacuum2} we introduced a more general rotation $\Omega(\theta, \zeta)$, which is also misaligned along the generator $X^{\hat \eta}$, corresponding to the neutral custodial triplet field $\eta_3^0$, given by:
\begin{align} 
X^{\hat \eta} &= \left(
\begin{array}{ccccc}
 0 & 0 & 0 & 0 & 0 \\
 0 & 0 & \frac{i}{2} & 0 & 0 \\
 0 & -\frac{i}{2} & 0 & 0 & 0 \\
 0 & 0 & 0 & 0 & 0 \\
 0 & 0 & 0 & 0 & 0 \\
\end{array}
\right)\,.
\end{align}
We found that $\Omega (\theta, \zeta) = R_\zeta \Omega (\theta) R_\zeta^\dagger$, with 
\begin{align}
R_\zeta &= \left(
\begin{array}{ccccc}
1 & 0 & 0 & 0 & 0 \\
0 & c_{\zeta/2}^2 & s_{\zeta/2}^2 & 0 & i s_\zeta/\sqrt{2} \\
0 & s_{\zeta/2}^2 & c_{\zeta/2}^2 & 0 & i s_\zeta/\sqrt{2} \\
0 & 0 & 0 & 1 & 0 \\
0 & i s_\zeta/\sqrt{2} & i s_\zeta/\sqrt{2} & 0 & c_\zeta
\end{array} \right)\,. 
\end{align}

In this new vacuum, the couplings of the scalars $h$ and $\eta_3^0$ to $Z$ and top, normalised to the SM values, read:
\begin{eqnarray}
\kappa_Z^h &=&   \st c_\zeta \sqrt{1+s_\zeta^2} \frac{c_\zeta^2 s_{2\theta} - 2 s_\zeta^2 (s_{3\theta} - s_{4\theta})}{1-c_\zeta^2 c_{2\theta} - s_\zeta^2 c_{4\theta}}\,,    \\
\kappa_Z^\eta &=&     - \st s_\zeta \sqrt{1+s_\zeta^2} \frac{c_\zeta^2 (s_{2\theta} + 2 s_{3\theta}) + 2 s_\zeta^2 s_{4\theta})}{1-c_\zeta^2 c_{2\theta} - s_\zeta^2 c_{4\theta}}\,,    \\
\kappa_t^h &=&    \frac{\sqrt{1+s_\zeta^2}}{2(\cot \theta - s_\zeta) c_\zeta} \left( \cot \frac{\theta}{2} + (1-\ct) s_{3\zeta} - 2 \st -c_{2\zeta} (1+2\ct) \tan \frac{\theta}{2} - (1+\ct) s_\zeta  \right)\,,    \\
\kappa_t^\eta &=&  \frac{\sqrt{1+s_\zeta^2}}{\cot \theta - s_\zeta} \left(  c_\zeta^2 + s_\zeta^2 (2\ct-1) + s_\zeta (2\ct+1) \tan \frac{\theta}{2} \right)\,.    
\end{eqnarray}
In the limit $\zeta \to 0$, we recover the results in the Higgs vacuum, namely $\kappa_W^h = \kappa_Z^h = \ct$ and $\kappa_W^\eta = \kappa_Z^\eta = 0$ for the vectors, and $\kappa_t^h = \ctt/\ct$ and $\kappa_t^\eta = \tan \theta$ for the top. In the limit $\zeta \to \pi/2$, where the vacuum is misaligned along the triplet, the couplings of $h$ to vectors vanish while $\kappa_W^\eta = -\sqrt{2} \ct$ and $\kappa_Z^\eta = - \sqrt{2} \ctt/\ct$. In the same limit, the coupling $\kappa_t^h$ diverges due to the fact that the mass of the top vanishes.


\subsection{$\SO(5)$ generators} \label{app:gens}
We list here the missing $\SO(5)$ generators mentioned in Section~\ref{sec:vacuum2}:
\begin{align}
T^8 \ & = \ \frac{1}{2\sqrt{2}} \left(
\begin{array}{ccccc}
 0 & 0 & 0 & 0 & 0 \\
 0 & 0 & 0 & 0 & -i \\
 0 & 0 & 0 & 0 & i \\
 0 & 0 & 0 & 0 & 0 \\
 0 & i & -i & 0 & 0 \\
\end{array}
\right), \ 
T^9 \   = \ \frac{1}{2\sqrt{2}} \left(
\begin{array}{ccccc}
 0 & 0 & 0 & 0 & -1 \\
 0 & 0 & 0 & 0 & 0 \\
 0 & 0 & 0 & 0 & 0 \\
 0 & 0 & 0 & 0 & 1 \\
 -1 & 0 & 0 & 1 & 0 \\
\end{array} 
\right), \
T^{10} \  = \  \frac{1}{2\sqrt{2}} \left(
\begin{array}{ccccc}
 0 & 0 & 0 & 0 & i \\
 0 & 0 & 0 & 0 & 0 \\
 0 & 0 & 0 & 0 & 0 \\
 0 & 0 & 0 & 0 & i \\
 -i & 0 & 0 & -i & 0 \\
\end{array}
\right)
\end{align}

\subsection{Custodial basis} \label{app:custodial}

The real triplet $\pi_0$ and the complex triplet $\pi_{\pm}$ can be rewritten in terms of the custodial $\SU(2)_C$, which is the diagonal  of $\SU(2)_L \times \SU(2)_R$. The bi-triplet decomposes as $(3,3) \to {\bf 5} \oplus {\bf 3} \oplus {\bf 1}$. The triplet components are mapped to the new basis as:
\bea
& \pi_+^+ = \eta_5^{++}\,, \quad \pi_+^0 = \frac{i \eta_3^+ - \eta_5^+}{\sqrt{2}}\,, \quad \pi_0^+ = - \frac{- i \eta_3^+ - \eta_5^+}{\sqrt{2}}\,, & \nonumber \\
& \pi_0^3 = \frac{\eta_1^0 - \sqrt{2} \eta_5^0}{\sqrt{3}}\,, \quad \pi_+^- = \frac{\sqrt{2} \eta_1^0 + \eta_5^0}{\sqrt{6}} + i \frac{\eta_3^0}{\sqrt{2}}\,. &
\eea
We recall that $\pi_-^- = (\pi_+^+)^\dagger$, $\pi_-^0 = (\pi_+^0)^\dagger$, $\pi_-^+ = (\pi_+^- )^\dagger$ and $\pi_0^- = (\pi_0^+)^\dagger$.


\subsection{Mass term and vacuum alignment} \label{app:massvacuum}

In this paragraph we briefly clarify the link between the choice of the mass matrix for the HFs and the choice of the vacuum, which we mentioned in Section~\ref{sec:HFmass}.
To this aim, consider a toy model where the only contribution to the pNGB potential comes from the HF mass term, like in Eq.~\eqref{eq:cm}. For $\mu_s > - \mu_d$, the minimum is at $\theta=0$, and it generates the following masses for the Higgs doublet, triplets and singlet $\eta$:
\begin{equation} \label{eq:spectrumHF}
m_{\rm Higgs}^2 = 8 C_m f (\mu_d + \mu_s)\,, \quad m_{\rm triplets}^2 = 16 C_m f \mu_d\,, \quad m_{\eta}^2 = 16 C_m f (\mu_d + 4 \mu_s)\,.
\end{equation}
We clearly see that for $\mu_s < - \mu_d$, the Higgs doublet squared mass turns negative, thus justifying the expectation that the minimum of the potential would be moved away fro zero. However, the above expressions show that the singlet $\eta$ becomes tachyonic for less negative values of $\mu_s$, i.e. $\mu_s < - \mu_d/4$. This implies that the vacuum needs to be misaligned along the singlet even before the Higgs direction is destabilised.

We will show now that this destabilisation along the singlet, that occurs for negative $\mu_s$, calls for choosing the second inequivalent EW preserving vacuum
 \be \label{vacuum2}
\Sigma_0'=\left(
\begin{array}{cc|c}
 & i \sigma_2 &  \\
 -i \sigma_2 &  & \\ \hline
  & & -1 
\end{array} \right)\,,
\ee
which, compared to the one in Eq.~\eqref{vacuum}, has a minus sign corresponding to the negative singlet squared mass.
One can go from the vacuum $\Sigma_0$ to the new one $\Sigma_0'$ with a $\SU(5)$ transformation by the generator associated to $\eta$ (times an overall phase shift):
\be
\Omega_s = \left. e^{- i \alpha/2} e^{i \sqrt{10}  X_\eta\ \alpha} \right|_{\alpha = \pi/5} =  \left( \begin{array}{ccccc} 1 & & & & \\ & 1 & & & \\ & & 1 & & \\ & & & 1 & \\ & & & & - i \end{array} \right)\,.
\ee
The relation is, therefore, 
\be
\Sigma_0' = \Omega_s  \Sigma_0 \Omega_s^T\,.
\ee
Let us now consider a theory with an HF mass term with $\mu_d > 0$ and $\mu_s < 0$. If we define the EW vacuum $\Sigma_0'$, all the equations for the vacuum alignment Eq.~\eqref{eq:cm} and \eqref{eq:spectrumHF} would be the same but for a flipped sign in front of $\mu_s$. That is, the results are the same as we would obtain for a theory with $\mu_d>0$ and $\mu_s > 0$ around the vacuum $\Sigma_0$. Thus, also for $\mu_s<0$, once the correct EW preserving vacuum is chosen, the theory is well defined at the minimum $\theta=0$. This analysis proves that the HF term alone cannot trigger EWSB.
As a final remark, we would like to remind that the broken generator bases in the two vacua are not the same, thus one needs to define a new pNGB basis on $\Sigma_0'$, which is related to the basis for $\Sigma_0$ in Eq.\eqref{eq:Pimatrix} by
\be
\Pi' = \Omega_s \Pi \Omega_s^\dagger\,.
\ee

\subsection{Potential in the general vacuum for top anti-symmetric spurions} \label{app:genvacpot}

Restricting to the case of the $A_L$-$A_R$ top partner {\it irrep}, the scalar potential generated by the NLO operators, given in Eq.\eqref{potAA}, takes the following simple form:
\begin{align}\label{eq:newpot}
V_{\text{top}}(\theta, \zeta) \ = \  V_0(\zeta)  + A_1 (\zeta) \ctt + A_2(\zeta) c_{4\theta} + B_1(\zeta) \stt + B_2(\zeta) s_{4\theta}\,,
\end{align}
with 
\begin{align}
V_0(\zeta) \ &= \ r_1 +   r_2 \,c_\zeta+ r_3 \,c_{2\zeta}\,, \quad 
A_1(\zeta) \ = \  c_{\frac{\zeta}{2}}^2 (r_4 + r_5 \, c_\zeta)\,, \nonumber \\
A_2(\zeta) \ & = \ r_6 + r_7\, c_{\zeta} + r_8 \, c_{2\zeta}\,, \quad 
B_1(\zeta) \  = \ r_9\, s_ \zeta ^2\, s_{\frac{\zeta}{2}}^{-1} \,, \nonumber \\
B_2(\zeta) \ & = \  r_{10} \, s_ \zeta ^2 \, s_{\frac{\zeta}{2}}^{-1}\,.
\end{align}
Only five of the ten coefficients given above are linearly independent, being related by the following five linear constraints:
\begin{align}
r_2 - 8 r_8 - r_7 - \frac{r_4}{2} \ &= \ 0\,, \quad r_3 - 3 r_8 \ = \ 0 \,, \quad  r_5 + 16 r_8 \ = \ 0 \nonumber \\
r_6+ r_7 - 7 r_8 \ &= \ 0 \,, \quad 2 r_{10} + r_9 \ = \ 0 \,.
\end{align}
Note that, as opposed to the case of the $\theta$-vacuum,  terms proportional to $\stt$ and $s_{4\theta}$ now also appear in the potential. By looking at the 
explicit form of the matrix $\Omega(\theta, \zeta)$ it can be argued that Eq.~\eqref{eq:newpot} already gives the most general 
expansion in $\theta$, since the operators generating the potential have four insertions of $\Omega$, and the latter only contains trigonometric functions 
of $\theta$ that can be re-expressed in terms of $\ct$ and $\st$ (up to unimportant constant shifts).

\subsection{Effective Yukawa couplings and type-II see-saw} \label{app:effYuk}

For completeness, we list here the effective Yukawa couplings between the SM fermions and the strong dynamics, which may be responsible for generating the masses of the light fermions, while PFC is responsible for the top mass~\cite{Cacciapaglia:2015dsa}. We define the projectors on the pNGB matrix such that
\be
\mathcal{P} (\phi_x) \; : \quad \mbox{Tr} [\mathcal{P} (\phi_x) \Pi \Sigma_0] = \frac{1}{2} \phi_x\,,
\ee
where $\phi_x$ is any of the pNGB fields. This will allow us to define gauge-invariant couplings between the pNGB $\Sigma$ matrix and the SM fermions.

\subsubsection*{Up-type fermions}

For up-type quarks, the effective Yukawa couplings read
\be
\mathcal{L}_{\rm Yuk.} \supset  i f y_u\ \left\{ u_L u_R^c\, \mbox{Tr} [\mathcal{P} (H_0) \Sigma] - d_L u_R^c\, \mbox{Tr} [\mathcal{P} (H^+) \Sigma] \right\}\,,
\ee
where, up to linear pNGB terms, the traces read:
\begin{eqnarray}
- i f \mbox{Tr} [\mathcal{P} (H_0) \Sigma] &=& \frac{s_{2\theta}}{\sqrt{2}} + \sqrt{2} \left( c_{2\theta}\ h - s_\theta\ \eta_3^0 \right) - i s_{2\theta} \left( \frac{3}{2\sqrt{5}} \ \eta + \frac{\sqrt{3}}{2} \ \eta_1^0 \right) + \dots \,, \\
- i f \mbox{Tr} [\mathcal{P} (H^+) \Sigma] &=& 2 s_\theta\ \eta_3^+\,.
\end{eqnarray}
The up-type fermion mass, thus, reads
\be
m_u = \frac{y_u}{\sqrt{2}} f s_{2\theta}\,,
\ee
and couplings to all pNGBs are generated, except for the five-plet. This result matches the ones for the elementary Georgi-Machacek model.

\subsubsection*{Down-type fermions}

For down-type quarks and charged leptons, the effective Yukawa couplings read
\be
\mathcal{L}_{\rm Yuk.} \supset  i f y_d\ \left\{ d_L d_R^c\, \mbox{Tr} [\mathcal{P} (H_0^\ast) \Sigma] +  u_L d_R^c\, \mbox{Tr} [\mathcal{P} (H^-) \Sigma] \right\}\,,
\ee
where, up to linear pNGB terms, the traces read:
\begin{eqnarray}
- i f \mbox{Tr} [\mathcal{P} (H_0^\ast) \Sigma] &=& \frac{s_{2\theta}}{\sqrt{2}} + \sqrt{2} \left( c_{2\theta}\ h + s_\theta\ \eta_3^0 \right) - i s_{2\theta} \left( \frac{3}{2\sqrt{5}} \ \eta + \frac{\sqrt{3}}{2} \ \eta_1^0 \right) + \dots \,, \\
- i f \mbox{Tr} [\mathcal{P} (H^-) \Sigma] &=& 2 s_\theta\ \eta_3^-\,.
\end{eqnarray}
The down-type fermion mass, thus, reads
\be
m_d = \frac{y_d}{\sqrt{2}} f s_{2\theta}\,,
\ee
and, again, couplings to all pNGBs are generated, except for the five-plet.

\subsubsection*{Neutrino masses and type-II see-saw mechanism}

Due to the presence of a charged triplet in the pNGB spectrum, direct couplings to a $\Delta L = 2$ leptonic operator is allowed, in analogy to what is done in type-II see-saw models~\cite{Ma:1998dx}. This operator can be generated either by a direct bi-linear coupling or by integrating out fermionic partners in partial compositeness, thus allowing a direct Majorana mass for the left-handed neutrinos. For the former case, the effective couplings read:
%
\be
\mathcal{L}_{\rm Yuk.} \supset f y_\nu \left\{ \nu_L \nu_L\, \mbox{Tr} [\mathcal{P} (\pi_+^-) \Sigma] +  \nu_L e_L\, \mbox{Tr} [\mathcal{P} (\pi_+^0) \Sigma] + e_L e_L\,  \mbox{Tr} [\mathcal{P} (\pi_+^+) \Sigma] \right\}\,,
\ee
with
\begin{eqnarray}
- f  \mbox{Tr} [\mathcal{P} (\pi_+^-) \Sigma] &=&  -  \frac{\st^2}{\sqrt{2}} + \sqrt{2} \left( c_\theta\  \eta_3^0 - s_\theta c_\theta\ h\right) +   \\ 
&& \qquad  i \left( \frac{3}{2\sqrt{5}} s_\theta^2 \ \eta - \frac{5 + 3 c_{2\theta}}{4 \sqrt{3}}\ \eta_1^0 - \sqrt{\frac{2}{3}}\ \eta_5^0 \right)\,, \nonumber \\
- f  \mbox{Tr} [\mathcal{P} (\pi_+^0) \Sigma] &=& \sqrt{2} \left( c_\theta \ \eta_3^+ + i \ \eta_5^+ \right)\,, \\
- f  \mbox{Tr} [\mathcal{P} (\pi_+^+) \Sigma] &=& -2  i \ \eta_5^{++}\,.
\end{eqnarray}
The constant term in the first line indicates that a mass for the neutrinos is generated, and the proportionality to $\st^2$ suggests that this is equivalent to the dimension-5 Weinberg operator in the SM. If this term were absent, as it may be in case of partial compositeness, a Majorana mass term may be generated by giving a small VEV to the triplet $\eta_3^0$.
Also, this coupling can potentially generate sizeable couplings of the five-plet to leptons, in particular for the doubly charged, thus allowing $\eta_5^{++} \to e^+ e^+$ decays at tree level.

\section{Proof of equivalence between our basis of $\SU(5)$ invariant operators and a basis of $\SO(5)$ invariant ones} \label{app:equivalence}

Instead of constructing explicitly a basis of $\SO(5)$ invariant operators following the procedure defined in Refs~\cite{Contino:2010rs,Mrazek:2011iu}, we will prove the equivalence of the two bases by tracing the contribution of the various $\SO(5)$ components inside the $\SU(5)$ invariant operators defined in the main text.

Firstly, we observe that for each pair of $\SU(5)$ spurions in Table~\ref{tab:topspurions} there exists a single $\SO(5)$ component in common. Thus, operators that contain two different spurions can only correspond to a single operator constructed in terms of $\SO(5)$ invariants. The same is true for operators containing only the anti-symmetric. In the following we will therefore only focus on operators containing two adjoints, or two symmetric (and/or the conjugate) irreps.

For the adjoint, the only LO template operator has the form
\beq \label{eq:templateDD}
\mathcal{O}_{D_1 D_2} = \Tr [D_1^T \Sigma^\dagger D_2 \Sigma]\,.
\eeq
To isolate the contribution of the two $\SO(5)$ components, we first observe that the adjoint spurions $D_i$ can be mapped into a 2-index irrep of $\SU(5)$ by multiplying them by the pNGB matrix $\Sigma$, namely taking the combination $D_i \Sigma$ (and $\Sigma^\ast D_i$ for the conjugate). The symmetric and anti-symmetric components can thus be extracted as follows:
\beq
\left. D_i \Sigma \right|_{\rm sym/asym} = \frac{1}{2} \left( D_i \Sigma \pm (D_i \Sigma)^T \right)\,.
\eeq
We can construct, therefore, two operators from the template in Eq.~\eqref{eq:templateDD}:
\begin{eqnarray}
\mathcal{O}_{D_1 D_2, {\rm sym/asym}} &=& \frac{1}{2}\ \Tr [D_1^T \Sigma^\dagger \left( D_2 \Sigma \pm (D_2 \Sigma)^T \right)] \nonumber \\
 &=& \frac{1}{2} \ \left( \Tr [D_1^T \Sigma^\dagger D_2 \Sigma] \pm \Tr [D_1^T \Sigma^\dagger \Sigma D_2^T] \right) \nonumber \\
 &=& \frac{1}{2} \left( \mathcal{O}_{D_1 D_2} \pm \Tr [D_1^T D_2^T] \right)\,.
\end{eqnarray}
The last line thus shows that the two operators constructed with the symmetric and anti-symmetric components of the adjoint are equivalent, up to a pNGB-independent operator that only contains the trace of the two spurions. Note that such trace vanishes for the top mass operator in Section~\ref{sec:topmass}, while it simply gives a constant term in the potential in Section~\ref{sec:pot}.

For the symmetric spurion, and its conjugate irrep, it is most convenient  to map it into a matrix transforming like the adjoint, $S_i \Sigma^\dagger$ (and $\Sigma S_i^c$). The traceless component of such matrix will contain the $\SO(5)$ symmetric component, as this is the only irrep in common to the adjoint, while the remaining term corresponds to the $\SO(5)$ singlet component. We can thus decompose it as follows:
\beq
S_i \Sigma^\dagger = \left. S_i \Sigma^\dagger \right|_{\rm sym} + \frac{1}{5}\ \Tr [S \Sigma^\dagger]\ \mathbb{1}_{5\times 5}\,,
\eeq
and similarly for the conjugate spurion. From the above decomposition, it is clear that the double-trace operators will only pick up the $\SO(5)$ singlet component of the symmetric (and conjugate) spurions. On the other hand, the single-trace operator gives
\beq
\Tr [S_1 \Sigma^\dagger S_2 \Sigma^\dagger] = \Tr [\left. S_1 \Sigma^\dagger \right|_{\rm sym} \left. S_2 \Sigma^\dagger \right|_{\rm sym}] + \frac{1}{5}\ \Tr [S_1 \Sigma^\dagger ] \ \Tr [S_2 \Sigma^\dagger]\,,
\eeq
and similarly for the operator containing the conjugate. This shows that the operator constructed in terms of the $\SO(5)$ symmetric component is equivalent to a linear combination of the single and double trace operators.

The considerations above show that a basis constructed in terms of $\SU(5)$ invariants at LO is equivalent to a basis constructed in terms of $\SO(5)$ invariants.




\newpage

\bibliography{bibSU5}
\bibliographystyle{JHEP-2-2}
\end{document}